\documentclass[preprint,12pt]{aastex62}

\shorttitle{BHR71}
\shortauthors{Tobin et al.}

\newcommand{\nthp}{\mbox{N$_2$H$^+$}}

\newcommand{\ntdp}{\mbox{N$_2$D$^+$}}
\newcommand{\nht}{\mbox{NH$_3$}}

\newcommand{\cateo}{\mbox{C$^{18}$O}}
\newcommand{\thco}{\mbox{$^{13}$CO}}
\newcommand{\twco}{\mbox{$^{12}$CO}}
\newcommand{\kmspc}{\mbox{km s$^{-1}$ pc$^{-1}$ }}
\newcommand{\kms}{\mbox{km s$^{-1}$}}
\newcommand{\kkms}{\mbox{K km s$^{-1}$}}

\begin{document}

\title{The Formation Conditions of the Wide Binary Class 0 Protostars within BHR 71}
\author{John J. Tobin}
\affiliation{Current Address: National Radio Astronomy Observatory, 520 Edgemont Rd., Charlottesville, VA 22903, USA; jtobin@nrao.edu}
\affiliation{Homer L. Dodge Department of Physics and Astronomy, University of Oklahoma, 440 W. Brooks Street, Norman, OK 73019, USA}
\affiliation{Leiden Observatory, Leiden University, P.O. Box 9513, 2300-RA Leiden, The Netherlands}

\author{Tyler L. Bourke}
\affiliation{SKA Organization, Jodrell Bank Observatory, Lower Withington, Macclesfield, Cheshire SK11 9DL, UK}

\author{Stacy Mader}
\affiliation{CSIRO Astronomy and Space Sciences, Parkes Observatory, PO BOX 276, Parkes NSW 2870, Australia}

\author{Lars Kristensen}
\affiliation{Centre for Star and Planet Formation, Niels Bohr Institute and Natural History Museum of Denmark, Copenhagen University, {\O}ster Voldgade 5--7, DK-1350 Copenhagen K, Denmark}

\author{Hector Arce}
\affiliation{Department of Astronomy, Yale University, P.O. Box 208101, New Haven, CT 06520-8101, USA}

\author{Fr\'ed\'eric Gueth}
\affiliation{Institut de Radioastronomie Millim\'etrique (IRAM), 38406 Saint-Martin d’H\`eres, France}

\author{Antoine Gusdorf}
\affiliation{LERMA, Observatoire de Paris, \'Ecole normale sup\'erieure, PSL Research University, CNRS, Sorbonne Universit\'es, UPMC Univ. Paris 06, F-75231, Paris, France}

\author{Claudio Codella}
\affiliation{INAF, Osservatorio Astrofisico di Arcetri, Largo E. Fermi 5,
50125 Firenze, Italy}
\affiliation{Univ. Grenoble Alpes, Institut de Plan\'etologie et d’Astrophysique de Grenoble (IPAG), 38401 Grenoble, France}

\author{Silvia Leurini}
\affiliation{INAF - Osservatorio Astronomico di Cagliari, Via della Scienza 5, I-09047 Selargius (CA), Italy}

\author{Xuepeng Chen}
\affiliation{Purple Mountain Observatory, Chinese Academy of Sciences, Nanjing 210034, China}

\begin{abstract}
We present a characterization of the binary protostar system 
that is forming within a dense core in the isolated dark cloud BHR71. 
The pair of protostars, IRS1 and IRS2, are both in the Class 0 phase, determined from
observations that resolve the sources 
from 1~\micron\ out to 250~\micron\ and from 1.3~mm to 1.3~cm.
The resolved observations enable the luminosities
of IRS1 and IRS2 to be independently measured (14.7 and 1.7~L$_{\sun}$, respectively), in addition to the bolometric temperatures 
68~K, and 38~K, respectively. The surrounding
core was mapped in \nht\ (1,1) with the Parkes radio telescope, and
followed with higher-resolution observations from ATCA in \nht\ (1,1) and
1.3~cm continuum. 
The protostars were then further characterized with
ALMA observations in the 1.3~mm continuum along 
with \ntdp\ ($J=3\rightarrow2$), \twco, \thco, and \cateo\ ($J=2\rightarrow1$) molecular lines. 
The Parkes observations find evidence for a velocity gradient across the core
surrounding the two protostars, while ATCA reveals more complex velocity
structure toward the protostars within the large-scale gradient.
The ALMA observations then reveal that the two protostars are at
the same velocity in \cateo, and \ntdp\ exhibits a similar velocity structure as
\nht. However, the \cateo\ kinematics reveal that the rotation on scales
$<$1000~AU around IRS1 and IRS2 are in \textit{opposite} directions. 
Taken with the lack of a systematic velocity difference between the pair,
it is unlikely that their formation resulted from rotational fragmentation. We instead conclude that the
binary system most likely formed via turbulent fragmentation of the core.

\end{abstract}

\section{Introduction}
Binary and multiple star systems are a frequent outcome of the star formation 
process at both low and high masses \citep[e.g.,][]{raghavan2010,duchene2013}. While
the statistics of multiple star formation in the main-sequence, pre-main sequence phase, and even
the protostellar phase are becoming much better characterized \citep{kraus2011,raghavan2010,wardduong2015,chen2013,tobin2016a},
the physical processes that lead to multiple star formation are not well-characterized observationally. 
The main epoch of multiple star formation is the protostellar phase \citep{tohline2002,chen2013},
when the protostar(s) are surrounded by a dense infalling cloud(s) of gas and dust. Thus, it is imperative to 
examine the process of multiple star formation in the protostellar phase with a combination of 
continuum and molecular line tracers to reveal the physical processes at work.

While it is essential to examine multiple star formation in the protostellar phase, star formation
typically occurs in clusters \citep[e.g.,][]{ladalada2003,megeath2012} increasing the complexity of
the data, especially for spectral line tracers that reveal the motion of gas along 
the line of sight. Toward small proto-clusters in molecular clouds (e.g., NGC 1333),
complex motions of the surrounding gas can 
complicate interpretation of kinematic data,
making multiple star formation difficulty to characterize.
Thus, stars forming in nearby
isolated clouds, the so-called Bok Globules, can be valuable laboratories for the study of 
multiple star formation. This is especially true for wide multiple star systems, where separations are greater
than 1000 AU, because the kinematics should be less confused along the line of sight. The BHR71 system 
\citep{bourke1995a,bourke2001,chen2008,yang2017} is composed of two Class 0 protostars \citep{andre1993} 
separated by $\sim$16\arcsec\ (3200 AU; 200~pc distance), and it is an ideal system to
study the physics of wide multiple star formation. The fact that both components
of the binary system are in the Class 0 phase means that the protostars are very young,
likely less than $\sim$150 kyr in age \citep{dunham2014}. Therefore, the initial conditions
of their formation are not likely to have been erased by the dynamical interactions of the
protostars and their effects on the surrounding cloud from their outflows \citep{arce2006}.

The formation of wide multiple star systems has been thought to be related to the rotation of
the infalling envelope of the protostar, where greater rotation rates would lead to an increased
likelihood of fragmentation. Many theoretical and numerical studies have 
highlighted the importance of rotation in the formation
of wide and close multiple star systems \citep{larson1972,bb1993,boss1995,boss2002,boss2013}. 
The importance of rotation is typically parametrized as $\beta_{\rm rot}$ which
is the ratio of rotational energy to gravitational potential energy. 
\citet{chen2012} compiled a list of all known single and multiple protostars and 
calculated their $\beta_{\rm rot}$ values, finding that most binary/multiple protostars have $\beta_{\rm rot}$ $>$ 0.01.
However, $\beta_{\rm rot}$ is observationally derived from measurement of velocity 
gradients in the cores/envelopes surrounding
the protostar(s), and \citet{tobin2011,tobin2012a} highlighted the possibility that these velocity gradients
may not truly reflect rotation due to the possibility of large scale asymmetries 
in the envelope structures \citep{tobin2010} and filamentary structure \citep[e.g.,][]{andre2010, hacar2011}. Furthermore, numerical simulations of star formation in turbulent
molecular clouds have shown that the turbulence can lead to the formation of multiple star systems
on $\sim$1000~AU scales \citep{padoan2002,offner2010}, possibly removing the need for bulk envelope 
rotation to initiate multiple star formation. 
Therefore sensitive molecular line data tracing the gas kinematics are required, spanning
orders of magnitude in spatial scales to differentiate between different scenarios of multiple star
formation.

We have obtained a comprehensive dataset on BHR71 spanning the scales of the star forming
core at $\sim$1\arcmin\ (12000~AU) resolution down to 1\farcs5 (300 AU) to examine the full range
of scales within this isolated binary protostar system. With this dataset, 
we aim to determine what physical process(es)
lead to the formation of at least two protostars in this system. The results
will be useful for interpreting observations of more evolved multiple star systems and
other forming multiple systems. The dataset includes \textit{Herschel} photometry, 
Australia Telescope Compact Array (ATCA) continuum and molecular line observations, 
Parkes Radio Telescope \nht\ line mapping, 
and Atacama Large Millimeter/submillimeter Array (ALMA) observations of 
continuum and molecular lines. The paper is organized as follows: the observations 
are presented in Section 2, the observational results regarding classification,  
photometry, and continuum are presented in Section 3, the results from the
and molecular line data are presented in Section 4, the results 
are discussed in Section 5, and our conclusions are presented in Section 6.

\section{Observations}

\subsection{Parkes Radio Telescope}

We observed BHR71 with the Parkes 64m radio telescope on 2012 September 06 (project P825). 
The 13mm receiver was used and we
observed an on-the-fly (OTF) map of BHR71 covering an 8\arcmin$\times$8\arcmin\ 
area surrounding BHR71 in the NH$_3$ (1,1) transition using dual polarization. 
The usable surface of the Parkes dish is 55m at 1.3~cm, resulting in a
beam size of $\sim$60\arcsec, see Table 1 for a summary. Weather was
good and T$_{\rm sys}$ was $\sim$100~K. The observations were 
conducted shortly after the 13mm receiver was mounted and an all-sky
pointing model could not be completed prior to our observations and we instead used a local 
pointing solution on the nearby quasar 1057-797. 

The flux calibration of the data was done using the quasar 1057-797, 
which was the gain calibrator for the ATCA observations 
that were conducted
a few days prior (see below). The data were reduced 
using the \textit{livedata} and \textit{gridzilla} tasks within AIPS++. 
The \textit{livedata} task was used to perform baseline
subtraction and scaling to the proper flux density of each spectrum in the map, while 
\textit{gridzilla} was used to construct a datacube from the 
individual spectra in FITS format. After mapping the data,
we noticed that there was a systematic offset in the central 
coordinates of the map; BHR71 was located $\sim$1\arcmin\ north of the pointing center.
The systematic offset likely resulted from the lack of an all-sky 
pointing solution prior to the observations. To recenter the map, we smoothed the
ATCA data to the same resolution as the Parkes data and 
matched peak emission of the core. Given that some emission is resolved-out in the ATCA
observations, there is likely some residual pointing uncertainty of the 
map on the order of 15\arcsec\ to 20\arcsec. 

The FITS cube was then ingested into CLASS, which is 
part of the GILDAS\footnote{http://www.iram.fr/IRAMFR/GILDAS}
software package  using the CLASS function \textit{lmv}. We then used the built-in
NH$_3$ line fitting functions to construct the centroid velocity maps and the linewidth maps.

\subsection{ATCA Observations}

BHR71 was observed with ATCA during 2012 in 
the EW367 configuration and H214 configuration, project C2665. ATCA consists of 6 antennas that are 22 meters in
diameter. One antenna is fixed on a pad 6~km from the array center, the remaining 5 antennas are reconfigurable
on an east-west track and a shorter north-south track. In both observations, we targeted
the NH$_3$ (1,1), (2,2), and (3,3) inversion transitions at 23.694, 23.722, and 23.870 GHz. The
primary beam of the 22 meter antennas at these frequencies is $\sim$144\arcsec. The observations
are summarized in Table 1.

The Compact Array Broadband Backend (CABB) correlator was used for our observations, 
providing 2~GHz of continuum bandwidth (with 2048 channels each 1~MHz wide) 
and we placed 1~MHz zoom bands, each with 
2048 channels (0.5 kHz channels) at the frequencies of 
the NH$_3$ (1,1), (2,2), and (3,3) inversion transitions. 
We used 6 zoom bands for each of NH$_3$ (1,1) and (2,2), and 4 zoom bands
for NH$_3$ (3,3). We used 1057-797 as the complex gain calibrator,
0537-441 as the bandpass calibrator, and 1934-638 was the 
flux calibrator for both the H214 and EW367 observations.

We observed BHR71 on 2012 April 22 in the EW367 configuration, where the maximum baseline
is 367 meters with all antennas positioned along the east-west track; the shortest
baseline was $\sim$38 meters. The weather conditions were variable during the observation
with T$_{\rm sys}$ typically $\sim$100~K. The observations were conducted for a full 12 hour 
synthesis track. However, there was rain during the last two hours of the observations and data from that
time period were unusable. The observations were conducted as a 7-point mosaic where the
goal was to obtain a constant surface brightness sensitivity over the central 
$\sim$144\arcsec\ primary beam. However, due to an error in the mosaic setup,
the mosaic was not as wide in the right ascension direction as intended.
This did not adversely affect the sensitivity of our map 
because the NH$_3$ emission from BHR71 was adequately
mapped with this mosaic pattern.

BHR71 was also observed on 2012 September 24 with ATCA in the H214 configuration where 5 antennas were
positioned along the north-south track with a shortest baseline length of $\sim$47~meters. The 
observing track length was 8 hours and T$_{\rm sys}$ was $\sim$80~K throughout the observations.
The weather conditions were good and the phase between antennas was stable during the observations. 
These observations were also conducted in a 7-point mosaic, with uniform
coverage of the central $\sim$144\arcsec.

The data were calibrated and edited using the Australian Telescope
National Facility (ATNF) version of the MIRIAD software package \citep{sault1995}. The phases
and amplitudes for each baseline were inspected and edited to remove periods of high phase
noise and/or amplitude deviations. We also flagged the single antenna on the 6~km baseline 
because of high phase noise and lack of uv-coverage at intermediate baselines.

The combined EW367 and H214 datasets were imaged using the \textit{clean} algorithm 
implemented in MIRIAD. We generated both a 1.3~cm continuum
image of BHR71 and channel maps centered on the NH$_3$ (1,1), (2,2), and (3,3) transitions;
however, only the \nht\ (1,1) line was well-detected. We did not combine
the ATCA and Parkes data due to the positional uncertainty in the Parkes data, we instead 
use them independently for the spatial scales they are sensitive to.

\subsection{ALMA Observations}

BHR71 was observed by the ALMA 12m array, the ALMA Compact Array (ACA), and the Total Power
Array (TPA) under the project 2013.1.00518.S. 
The observations were conducted in band 6, with a central
tuning frequency of 225 GHz. In all observations, the correlator was configured to
observe $^{12}$CO ($J=2\rightarrow1$), $^{13}$CO ($J=2\rightarrow1$), C$^{18}$O ($J=2\rightarrow1$), \ntdp\ ($J=3\rightarrow2$), and continuum. The continuum
basebands were located at 232.5 GHz, a region without many 
significant spectral lines, and 218 GHz. The
218 GHz band had significant contamination from spectral lines and was not used for continuum
measurements. The continuum bands were observed in TDM mode, with 128 channels over the 2 GHz
of bandwidth each. $^{12}$CO and \ntdp\ were observed in the same baseband, with
each having 58 MHz of bandwidth and 1920 channels, providing a velocity resolution of $\sim$0.08~\kms.
$^{13}$CO and C$^{18}$O were also observed in the same baseband
with identical bandwidth and velocity resolution. See Table 1 for a summary of the observational
setup.

The 12m array observations were conducted on 2015 January 17 in a 14-point rectangular 
mosaic with 34 antennas operating, covering a
45\arcsec$\times$45\arcsec\ region centered between the two protostars. The precipitable water
vapor (PWV) was 3.64~mm at the time of execution. Each mosaic point was observed
for $\sim$62 seconds and the complete scheduling block was executed in $\sim$44 minutes. 
The observations were
manually calibrated by T. Hunter at the North American ALMA Science Center (NAASC) using CASA 
version 4.3.1 and we used the provided reduction script to generate the calibrated data products. 
We then self-calibrated the data using the observed dust continuum; not all mosaic points contained 
a detection of the continuum source, therefore self-calibration was only applied
to the mosaic points where the continuum was detected. We performed both phase and amplitude
self-calibration on the data. The refined phase and amplitude solutions were then applied to the
spectral line base bands. The data were then imaged using the \textit{clean} task
in CASA 4.4. We used interactive cleaning for the continuum and spectral line cubes. For
the spectral line data, we manually drew masks around the emission in each velocity channel
and refined the mask when needed after each iteration of \textit{clean}.

The ACA observations were conducted on 2014 May 19 in a 5-point mosaic centered between
the two protostars. The observation took $\sim$50 minutes and 9 antennas were operating. Each 
mosaic point was observed for 74 seconds and the science observations had a total
of 12.3 minutes on-source. The PWV during these observations was 0.7~mm. 
These data were calibrated by B. Mason at the NAASC using the ALMA pipeline and CASA
version 4.2.2, and we used the reduction script to regenerate the calibrated data product. We also
self-calibrated these data, all mosaic points contained the continuum source due to the larger
primary beam of the 7m antennas. The refined phase and amplitude solutions from the continuum
were then applied to the spectral line data. The data were imaged using the \textit{clean}
task in interactive mode following the same methodology as outlined for the 12m data.

The TPA observations were conducted on 2015 May 2, in three executions of 
$\sim$57 minutes each observing a 93.6\arcsec$\times$93.6\arcsec\
on-the-fly (OTF) map using 2 TPA antennas. The
PWV at the time of execution was 0.9 to 1.3~mm. We found that the pipeline-reduced data did not adequately 
fit the spectral baseline of the $^{12}$CO data due to the broad linewidth. We 
re-ran the reduction script in CASA version 4.3.1 and manually adjusted the region to 
fit the spectral baseline in order to avoid the high-velocity $^{12}$CO emission.

We then combined the $^{12}$CO, $^{13}$CO, C$^{18}$O, and \ntdp\ data using the methods
outlined in the CASA guide for M100 using 
CASA version 4.6.0\footnote{https://casaguides.nrao.edu/index.php/M100\_Band3\_Combine\_5.1}. 
The 12m+ACA visibilities were combined using the
\textit{concat} task in CASA, with the weights being properly adjusted by CASA to compensate 
for the lower sensitivity of the ACA data. We used the \textit{clean} task to image the combined
dataset in the same manner as the individual datasets. We then followed the CASA guide for M100
to combine the 12m+ACA and TPA data using the \textit{feather} task. The fully combined spectral
line dataset had significantly reduced artifacts for all molecules where there was significant 
extended emission. We make use of the fully combined data (12m+ACA+TPA) in this paper,
except for two \cateo\ position-velocity (PV) diagrams and this is noted in the text and caption.
Furthermore, the continuum maps are only 12m+ACA because total power data are not available.

\subsection{Magellan PANIC Observations}

We observed BHR71 with the Magellan Baade 6.5m telescope located at Las Campanas using the 
PANIC \citep[Persson's Auxilary Nasmyth Infrared Camera;][]{martini2004}
near-infrared imager on 2009 January 17 and 18. The PANIC instrument uses a 1024$^2$ detector with 0\farcs12 pixels, 
providing a 2\arcmin$\times$2\arcmin\ field of view. The observations for BHR71 were previously presented in
\citet{tobin2010a} and we only observed H and Ks bands with this instrument. 
The seeing was exceptional during these observations at $\sim$0\farcs4.

\subsection{Cerro-Tololo ISPI Observations and Photometry}

We observed BHR71 using the Infrared Side-Port Imager \citep[ISPI;][]{vanderbliek2004} on the Blanco 4m telescope
at Cerro-Tololo on 2009 June 11. The ISPI observations and data reduction were described in detail in \citet{tobin2010a},
and the imager has a 10\arcmin\ field of view with a 2048$^2$ detector.
We use the J, H, and Ks-band images from ISPI for near-infrared photometry in this paper 
due to the larger field of view
as compared to PANIC. Also the absolute calibration of the ISPI
data is expected to be more robust due to the larger
number of stars available for calibration against 2MASS \citep{skrutskie2006}.
We measured the photometry in the J, H, and Ks bands using
a circular aperture radius of 30\arcsec\ (6000 AU) for IRS1 and a 5\arcsec\ (1000~AU) radius
for IRS2. IRS2 was undetected in this aperture. The background was measured 
using the median of an off-source patch of sky given the highly extended nature of the scattered light emission.
The data were calibrated using photometry from the 2MASS catalog \citep{skrutskie2006} and 
the photometry are listed in Table 2. The seeing during these observations was $\sim$0\farcs9.

\subsection{\textit{Spitzer} Observations and Photometry}

BHR71 was observed with the \textit{Spitzer Space Telescope} using the 
IRAC \citep{fazio2004}, MIPS \citep{rieke2004}, and IRS instruments \citep{houck2004}. 
The IRAC and MIPS observations were part of the \textit{cores2disks} Legacy program and
the reduced data were obtained from the \textit{cores2disks} data archive.
The IRS observations were obtained as part of the IRS guaranteed time program 
and previously published in \citep{yang2017}.
The MIPS 24~\micron\ photometry for IRS1 and IRS2 were presented in \citet{chen2008}, and 
we re-extracted the IRAC photometry of the extended scattered light nebulae toward
BHR71 IRS1 and IRS2 at 3.6, 4.5, 5.8, and 8.0~\micron\ using same the methods outlined 
for the protostar BHR7-MMS in \citet{tobin2018}. We extracted the photometry from 
a 6000 AU (30\arcsec) radius aperture toward IRS1 and a 1000~AU (5\arcsec) aperture radius 
toward IRS2. The 6000 AU aperture 
for IRS1 includes flux from IRS2, but the contribution is at most $\sim$10\% (see Table 2).
We also extracted the photometry from a 10000 AU radius (50\arcsec)
centered on IRS1. We chose these apertures as they are convenient fiducial values and
radiative transfer models offer the option to extract flux within different apertures
around the source(s) \citep[e.g.,][]{whitney2003a}. The 1000~AU aperture restricts the scattered
light emission to a region close to the source and is less affected by the source structure;
the 6000~AU and 10000~AU apertures are useful characterizing the total flux density of scattered light from the source(s).
Due to the small contribution from IRS2 at wavelengths shorter than 24~\micron, we decided to extract the large apertures
for IRS1, in addition to the small uncontaminated apertures.

\subsection{\textit{Herschel} Observations and Photometry}

We obtained the \textit{Herschel} observations toward BHR71 as part of the program OT1\_jtobin\_1.
BHR71 was observed with the PACS photometer \citep{poglitsch2010} on 2011 July 28. The 100 and 
160~\micron\ maps were obtained simultaneously with a map size of 10\arcmin$\times$10\arcmin, observing two
orthogonal scans for 18.5 minutes each. The 70~\micron\ data were obtained separately, mapping a smaller,
2\arcmin~$\times$~2\arcmin\ region, also observing 2 orthogonal scans for 14.4 minutes each. Both sets of PACS observations 
were made using the medium scan speed (20\arcsec\ per minute) for best image quality.
The SPIRE \citep{griffin2010} observations were obtained as part of the same program as the PACS
observations, on 2011 August 16. We observed a 10\arcmin~$\times~$10\arcmin\ region around the protostars 
with orthogonal scan legs for a total time of 17.25 minutes. 
In our subsequent analyses, we utilized the Jscanam\footnote{This is a port of the Scanamorphos 
map making software \citep{roussel2013} into the \textit{Herschel} Interactive 
Processing Environment (HIPE).} products downloaded from the 
\textit{Herschel} Science Archive.

We performed photometry on the PACS and SPIRE data using both aperture photometry
and point spread function (PSF) photometry. At 70, 100, 160, and 250~\micron\ we used
PSF photometry to measure the flux densities toward both IRS1 and IRS2. The sources
are significantly blended at wavelengths longer than
100~\micron\ and are within the PSF wings
of each other at 70 and 100~\micron.
We used the IDL program \textit{starfinder} \citep{diolaiti2000} to perform PSF photometry. Due to the small mapped region, we could not measure the PSF from the
data. Instead, we used the Vesta PSF measurements from the pointing refined data that we 
then rotated and smoothed to match the data. This enabled us to measure the photometry toward
the two sources, with the positions being automatically fit at 70, 100, and 160~\micron. Due to
the increased blending at 250~\micron, we specified the positions for the two sources
that were identified at shorter wavelengths and extracted the 250~\micron\ photometry. Due
to the lack of other sources in the field, we could not cross-calibrate our PSF photometry with
aperture photometry to verify its accuracy. However, the \textit{Herschel} Orion Protostar Survey (HOPS)
used both aperture and PSF photometry and found close agreement between the two \citep{furlan2016}.
The flux density measured using PSF photometry is expected to be systematically
lower than the aperture photometry because aperture photometry includes
additional extended emission surrounding the sources.
PSF photometry from PACS has been characterized by Ali et al (2018, in prep.) for the 
HOPS protostars \citep[e.g.,][]{furlan2016}, finding that the PSF photometry
either agrees with the aperture photometry or is systematically low due to additional
extended flux in the aperture photometry. \citet{magnelli2013} also used PSF photometry,
but created empirical PSFs from their data which were truncated to not include
the full PSF wings. This resulted in their photometry being systematically low, $\sim$70\% of
the aperture measurement at 160~\micron. However, we used the full Vesta PSF to fit our data
and should not have this systematic effect in our measurements.

To measure the aperture photometry, we used 40\arcsec\ radius apertures at 70, 100, 
and 160~\micron, centered between the two protostars. 
We chose different apertures for the \textit{Herschel} PACS photometry in
comparison to the shorter wavelength photometry because the aperture and
color corrections are tabulated for this aperture.
We also corrected for
the encircled energy fraction as documented in the PACS Data Handbook. The PSF photometry 
of IRS1 and IRS2 added together is about 20\% less than the aperture
photometry at each wavelength,
which is reasonable because the PSF photometry excludes the extended
emission surrounding the sources.

For the SPIRE photometry, we used standard apertures of 
22\arcsec, 30\arcsec, and 40\arcsec\ for the 250, 350, and 500~\micron\ bands, 
respectively. These apertures are again different from other wavelengths because they have their
aperture and color corrections tabulated.
We started with the extended source map and utilized the 
tabulated beam correction, color correction, and aperture correction 
for a source with a spectral index of 3.5 and a temperature of 30~K. These are
reasonable assumptions for a protostar as the SED peaks at $\sim$100~\micron\ and the
dust opacity spectral index will result in a spectral index between 3 and 4 in the 
sub-millimeter range. Our photometry are taken toward the region of compact emission toward 
the protostar locations and not from the entire extended core; thus, we do not need
to worry about the proper zero-point flux calibration as in \citep{sadavoy2018}

\section{Observational Results}

The multitude of infrared and submillimeter data taken toward BHR71 are shown in Figures 1 \& 2.
Figure 1 shows the high-resolution ground-based near-infrared view, 
the \textit{Spitzer} IRAC view, and the mid- to far-infrared view with
 MIPS 24~\micron, \textit{Herschel} 70~\micron, and 100~\micron. 
The PANIC images are shown for H and Ks-bands because of their better image quality, 
while the J-band image is from ISPI. Very little emission is detected 
in the vicinity of IRS2 at wavelengths less than 2.2~\micron,
and the image is dominated by the scattered light nebula
on the blue-shifted side of the IRS1 outflow. The near-infrared images on larger scales
detect the shocked-H$_2$ emission from both outflows \citep{bourke2001}. The 
\textit{Spitzer} IRAC data more obviously show the scattered light emission from both IRS1 and IRS2
in addition to outflow knots, as 
shown previously by \citet{chen2008}. Then the MIPS 24~\micron\ 
and \textit{Herschel} PACS 70 and 100~\micron\ data detect the thermal mid- to far-infrared emission 
from the inner envelopes of both protostars. IRS2 is clearly resolved from IRS1 and also 
appears much redder than IRS1 with substantially less 24~\micron\ emission. The PACS
images were previously published by \citet{yang2017}.

We show a larger field of view around BHR71 from \textit{Spitzer} IRAC and \textit{Herschel} SPIRE in 
Figure 2. The \textit{Spitzer} IRAC data show the envelope surrounding BHR71 in extinction against
the 8~\micron\ Galactic background, in addition to the extended outflow knots from both sources.
The SPIRE maps show the submillimeter emission from the extended BHR71 core surrounding IRS1 and IRS2,
emitting where the 8~\micron\ absorption is apparent. At wavelengths longer than 
250~\micron\ IRS1 and IRS2 are no longer resolved. There is faint 500~\micron\ emission
(appearing red) along the direction of the IRS1 outflow. This may be the result of 
CO ($J=5\rightarrow4$) emission contaminating the longest wavelength 
SPIRE band, similar the known CO contamination to SCUBA-2 maps from 
the ($J=3\rightarrow2$) and ($J=6\rightarrow5$) lines \citep[e.g.,][]{drabek2012}.

\subsection{SED Class of IRS1 and IRS2}

The broad wavelength coverage from the near-infrared to the submillimeter also enables
us to more quantitatively characterize these two protostars. To accurately distinguish the
observational Class for both IRS1 and IRS2, we utilized PSF photometry from 
\textit{Herschel} at 70, 100, 160, and 250~\micron, aperture photometry between 3.6~\micron\ 
and 24~\micron\, and millimeter flux measurements from ALMA, ATCA, SEST, and SIMBA, see Table 2.
All these data are used to construct the SEDs shown in Figure \ref{sed} for IRS1 and IRS2 individually, 
as well as the combined emission of the sources. The bolometric 
luminosity for BHR71 as a whole is 
17.7~L$_{\sun}$. The more luminous source, IRS1, 
has L$_{\rm bol}$=14.7~L$_{\sun}$ and a bolometric temperature (T$_{\rm bol}$) = 68~K, 
while the fainter  source IRS2 has L$_{\rm bol}$ = 1.7~L$_{\sun}$ and T$_{\rm bol}$=38~K. 
Thus, while IRS1 is about 10$\times$ more luminous, both are Class 0 protostars, with
IRS2 possibly being slightly 
less evolved. The relative inclinations of IRS1 and IRS2 could contribute to
their observed T$_{\rm bol}$ difference, but they are both moderately inclined with respect to the
line of sight (see Section 3.3) and it is difficult to determine if this would
fully account for the observed difference. 
We calculate L$_{\rm bol}$ by integrating the
SED using the trapezoidal integration method implemented in the
IDL function \textit{tsum}. Then,
T$_{\rm bol}$ is calculated following \citet{myers1993} (also using \textit{tsum}) in the
calculation of the average frequency of the SED. T$_{\rm bol}$ is the temperature of a
blackbody with the same average frequency of the observed SED \citep{ladd1991,chen1995,myers1998}.

\subsection{ATCA and ALMA Continuum data}

Continuum emission is detected toward both BHR71
IRS1 and IRS2 at 1.3~mm and 1.3~cm with
ALMA and ATCA, respectively, see Figure 4. The ALMA 1.3~mm emission is clearly detected toward both protostars,
with IRS1 being $\sim$12 times brighter at 1.3~mm than IRS2 (Table 2). There is also some extended emission away
from the protostar positions, tracing the envelope, and some tenuous emission detected
in between IRS1 and IRS2. The emission between IRS1 and IRS2 is possibly
analogous to the continuum emission that apparently connects IRAS 16293-2422 A and B \citep{jacobsen2018}.
The well-resolved structure of IRS1 makes Gaussian fitting unreliable for this source. 
The major and minor axis of IRS1 out to the 3$\sigma$ level are $\sim$9\arcsec\
and $\sim$7\arcsec, respectively. IRS2, on the other hand, is more symmetric. The major and 
minor axis to the 3$\sigma$ level are both $\sim$3.9\arcsec\, but Gaussian fitting
yields a deconvolved major axis of 1\farcs69 and a minor axis of 1\farcs23 
with a position angle of 87.5\degr. However, the source does not appear
well-resolved and the source size from Gaussian fitting may not be overly accurate.

To measure the ALMA 1.3~mm continuum flux densities as reliably as possible, we 
generated a sigma-clipped map, where pixels below
the 3$\sigma$ level ($\sigma$=0.68~mJy) are set to zero to avoid including excess noise
in the measurement of the flux density. This method will, however, exclude a small amount
of real emission, but will result in a more repeatable measurement of flux density.
We measured the combined emission from IRS1 and IRS2 in a 35\arcsec\ diameter circle centered
between the sources finding a flux density of 1.41$\pm$0.02~Jy. The flux density of IRS1 alone
was measured within a 25\arcsec\ diameter circle centered on IRS1, measuring a flux density
of 1.28$\pm$0.01~Jy. Finally, the flux density for IRS2 was measured in a 10\arcsec\ diameter aperture,
centered on IRS2, measuring a flux density of 0.12$\pm$0.005~Jy. Note that the
uncertainties on flux densities are statistical only.

The ATCA 1.3~cm continuum data also resolve IRS1 and IRS2, though IRS2 is much fainter; IRS2
is only detected at the 4$\sigma$ level, 12.5 times fainter than IRS1. 
Since the sources appear point-like and not highly-resolved, we measured the flux
densities by simultaneously fitting Gaussians to each source using CASA. The
measured flux densities for IRS1 and IRS2 are 2.7 and 0.22~mJy, respectively.
\citet{chen2008} also 
detected both IRS1 and IRS2 at 3~mm; thus, with the addition of our data we can
examine the radio spectrum toward IRS1 and IRS2, shown in Figure \ref{mm-spectrum}.
The emission from both IRS1 and IRS2 are consistent with dust-only emission out to 1.3~cm, but
with a shallow spectral slope, 2.7$\pm$0.2 for IRS1 
and 2.9$\pm$0.5 for IRS2 (F$_{\nu}$~$\propto$~$\lambda^{-\alpha}$). 
In this formalism, the expected spectral index for dust emission is $\alpha$~=~(2+$\beta$) where $\beta$
is the dust opacity spectral index and $\beta$~$\sim$1.8 is expected for dust grains in
the interstellar medium. Our data indicate $\beta$ of 0.7-0.9, which is consistent with
other studies of dust opacity in protostars \citep{kwon2009,chiang2012}. However, we cannot
rule-out some contribution from free-free emission from either protostar at 1.3~cm. 

We can use the dust continuum data to estimate the inner envelope (and presumed disk) masses
surrounding IRS1 and IRS2. We do this by assuming isothermal and optically thin dust emission, 
adopting a dust-only opacity at 1.3~mm from \citet{ossenkopf1994} 
having $\kappa_{1.3mm}$ 0.899 cm$^2$~g$^{-1}$, and assuming a dust 
to gas mass ratio of 1:100 \citep{bohlin1978}. 
We assume a dust temperature of 20~K for IRS1 and IRS2, comparable to the expected
envelope temperatures at $\sim$1000~AU scales for the luminosities of
IRS1 and IRS2 \citep{whitney2003a}. Finally, employing the equation
\begin{equation}
M_{dust} = \frac{D^2 F_{\lambda} }{ \kappa_{\lambda}B_{\lambda}(T_{dust}) }
\end{equation}
we estimate a total mass (dust+gas) for the inner envelopes/disks of 
both IRS1 and IRS2 to be 1.25~M$_{\sun}$ from measured flux density of 1.41~Jy. 
The flux densities for IRS1 and IRS2 are 1.28~Jy and 0.12~Jy, 
respectively, leading to respective dust+gas masses of 1.13 and 0.11~M$_{\sun}$ 
if the same temperature is assumed for both sources. Since IRS1 is a factor of $\sim$8.6 more
luminous than IRS1, its average dust temperature should be a factor of $\sim$1.7 higher, (34~K)
than IRS2 assuming that T~$\propto$~L$^{0.25}$. With an assumed dust temperature of
34~K for IRS1, the estimated mass becomes 0.59~M$_{\sun}$, making the combined mass 0.7~M$_{\sun}$
considering the likely higher temperature for IRS1.

These masses
are for the inner envelope and disk around the protostars, the larger-scale envelope 
traced by \textit{Herschel} (Figure 2) is resolved-out and not detected in the ALMA observations.
Thus, these mass estimates should be considered lower limits. 

The continuum and photometric data enable us to classify the two protostars and measure the 
amount of mass surrounding the protostars; however, these data do not specifically reveal
what led to the formation of the binary system other than there being a large mass reservoir
in the envelope surrounding BHR71. To better characterize the formation of the binary
system, we must turn to the molecular line data to trace the kinematics of the gas
surrounding the protostars.
But first, we will briefly examine the outflows of BHR71 observed by ALMA.

\subsection{Outflow Morphology from ALMA}

The outflows from BHR71 IRS1 and IRS2 are quite prominent in the scattered light 
images from the near-infrared and \textit{Spitzer}. However, we also traced the 
outflow with \twco\ and \thco\ observations with ALMA (Figure \ref{co-isotopes}). 
The integrated intensity maps of the CO emission from IRS1 and IRS2 (Figure \ref{co-isotopes})
clearly show the red and blue-shifted sides of both outflows. The 
\twco\ maps indicate that two protostars have their outflows inclined in opposite directions.
The blue-shifted side of the outflow from IRS1 is on the south and the red-shifted side is
to the north, while this is opposite in the case of IRS2. This orientation of the outflows
had been previously indicated from the near-infrared and IRAC imaging as well as
from single-dish observations \citep{bourke2001,parise2006}. We also find that the \cateo\ traces
the edge of the outflow cavity walls toward IRS1 in Figure \ref{co-isotopes}. From these
data, we measure outflow position angles of 174\degr\ and -31\degr\ for IRS1 and IRS2, respectively. The
angles are measured east of north and are found by drawing a line that bisects the outflow cavity
and passes through the continuum position.

The outflow from IRS2 has a more narrow opening angle than the outflow of IRS2. The outflow
opening angles of IRS1 and IRS2 are ~55\degr\ and $\sim$47\degr, respectively, measured
 at a distance of 10\arcsec\ from the protostars. Note that these are the full opening angles
of the outflow cavities, radiative transfer modeling studies typically quote the half-opening angles
\citep{whitney2003a}. The \thco\ emission from IRS1 traces
lower-velocity emission better than \twco\ and we measure a slightly larger opening angle
of $\sim$63\degr. The outflow from IRS2 is not well-detected in \thco\ and we are unable
to make a similar measurement. 

The red-shifted side of the outflow toward IRS2 shows peculiar structure at
distances greater than 10\arcsec\ from the protostar. Along the line of sight,
the blue-shifted outflow of IRS1 and red-shifted outflow of IRS2 overlap. The northern
side of the red-shifted outflow
from IRS2 fans out to a much larger opening angle at this point and this continues past
the point where the outflow cavities overlap. This wider feature also appears in the
\thco\ integrated intensity map at lower signal to noise.

A more in depth study
of the outflow system will be presented by Bourke et al. (in prep.).
However, we can make some estimates of the source inclinations using the outflow geometry
in the absence of detailed modeling. This is possible
because there is very limited overlap between the blue- and red-shifted 
sides of both outflows. Assuming that the outflows
are conical, the lack of significant overlap of blue and red-shifted sides, both along
the outflow and toward the protostars, means that the IRS1 outflow must have an 
inclination between 35 and 63\degr, because its opening angle is $\sim$55\degr. 
The outflow is likely closer to 63\degr\ than the other extreme because of 
how extended it is in the plane of the sky.
The IRS2 outflow with a more narrow opening angle of $\sim$47\degr\ 
must have an inclination between 43 and 67\degr. We define 90\degr\ as 
viewing the system edge-on where the outflow is totally in the
plane of the sky, and 0\degr\ is edge-on where the outflow is completely along the line of sight.

\section{Dense Molecular Line Kinematics}

The datasets from Parkes, ATCA, and ALMA (Table 1) combined trace the kinematics of the dense gas
from the scale of the entire BHR71 core to the envelopes surrounding the individual protostars.
It is instructive to start from the largest scales and move inward, thus we will begin 
with a discussion of the Parkes data. To introduce the data, we show the spectrum of
each line toward each source (both sources for Parkes) and plot them in Figure \ref{spectra}.
The spectra are extracted from a circular region matching the size of the beam 
in each observation (see Table 1), except for the ALMA data where we extract the spectra within a 
5\arcsec\ diameter circle. The \twco\ and \thco\ have broad linewidths tracing the outflow and
\cateo\ mainly traces the inner envelope with some outflow contamination.
\ntdp\ traces the cold envelope and this molecule
is mostly destroyed in the 5\arcsec\ region around of IRS1, and \nht\ also traces the cold
gas but on the larger scales probed by ATCA. Given that \nht\ and \ntdp\ have lower intensities
at the protostar positions, we also extracted spectra from the position of the \ntdp\ emission peaks
east of IRS1 and east of IRS2. These positions better show the hyperfine structure of \ntdp\ and \nht.

The largely distinct regions traced by CO and its isotopologues and \nht, \ntdp, and \nthp\
are the result of both the thermal structure of the infalling envelopes and
chemical processes. CO is frozen-out onto dust grains at temperatures
below $\sim$20~K and at typical densities of star forming cores n~$>$10$^5$~cm$^{-3}$
\citep{frerking1982,bm1989}. Observations and astrochemical models show that
\nthp, \ntdp, and \nht\ form abundantly in these cold, dense regions where
CO is frozen-out \citep[e.g.,][]{caselli2002, lee2004, bergin2002}. However, in 
the central regions of the protostellar envelopes, the temperature increases due
to heating from the protostar and accretion which evaporates CO from
the dust grains inside R$\sim$1000~AU (the actual radius depends luminosity). In this
region, there is an extremely efficient reaction of CO with \nthp\ and \ntdp\
that destroys these molecules rapidly. 
Also, \ntdp\ formation is shut off at
T$\sim$20~K because the key molecule in its formation pathway, H$_2$D$^+$ is converted back into
H$_3^+$ and HD \citep[e.g.,][]{herbst1982}.
Then there is also a destruction pathway 
for \nht\ involving reactions with HCO$^+$, which will form in the presence of CO \citep{lee2004};
there is also the possibility that \nht\ will freeze-out onto dust grains before
being completely destroyed by HCO$^+$ at n~$>$10$^6$~cm$^{-3}$ \citep{visser2011}.
This leads to the onion-like chemical structure of protostellar envelopes where
\nthp, \ntdp, and \nht\ exist in gas-phase in the outer regions not in the inner regions,
and CO exists most prevalently in the inner regions without \nthp, \ntdp, and \nht. We
note, however, that these are not the only molecules in the gas phase in these regions.

\subsection{Parkes NH$_3$ Kinematics}

The Parkes data traced the \nht\ (1,1) emission line with 60\arcsec\ (12000~AU) resolution 
out to 8\arcmin\ ($\sim$0.5~pc) scales, 
probing the largest scales available to us in molecular gas. 
\nht\ emission is detected from the entire region where the envelope is seen
in absorption at 8~\micron\ (Figure 2), and extending a bit beyond 
the edge of the map to the southeast
as shown in Figure \ref{parkes-nh3}.
The emission does not likely extend beyond the northern end of the map; the intensity appears
to rise here due to increased noise at the map edge.

Using the \nht\ fitting routines in CLASS, we fit the hyperfine lines in each pixel of the 
map where emission was detected above the 5$\sigma$ level and the line-center 
velocity and linewidth maps that are also shown in Figure \ref{parkes-nh3}.
 The line-center velocity map shows a slight velocity gradient along the long axis of the 
\nht\ emission from southeast to northwest, in the same direction as the extended 8~\micron\ 
extinction. The total velocity change is about 0.15 - 0.2~\kms, corresponding to a gradient
of about 1~\kmspc. We measure the velocity
gradient more quantitatively by fitting the slope of a 1D cut along the envelope orthogonal
to the outflow 
of IRS1 and by fitting a 2D plane to the line-center velocity maps. The resulting
velocity gradient orthogonal to the outflow direction is 1.2~\kmspc\ and the 2D velocity
gradient fit finds a gradient of 1.15~\kmspc\ with a position angle of 101\degr\ east of north.
This velocity gradient could be interpreted as core rotation, as is often done 
\citep[e.g.,][]{goodman1993}, but it was shown in \citet{tobin2011,tobin2012} that
infalling flows along asymmetric envelopes could also produce velocity gradients. If
the velocity gradient of the core reflects rotation, then IRS2 should orbit around IRS1
in the same direction as the core rotation. Furthermore, if IRS1 and IRS2 fragmented
within this rotating core, the individual envelopes should 
have rotation in the same direction
in the case of angular momentum conservation.

The linewidth map in Figure \ref{parkes-nh3} shows some variation
 across the source. The southern-most tip of the map 
has a very broad linewidth, this appears real and could be in an area 
affected by the two outflows. The eastern side of the linewidth map 
is relatively constant at 0.5 to 0.6~\kms\
and on the western side the linewidth dips down to $\sim$0.3~\kms. These linewidths are in excess
of the expected linewidth of \nht\ at 20~K of 0.13~\kms, indicative of an additional non-thermal
component. Some of this non-thermal component could come from unresolved velocity gradients
in the envelope within the 60\arcsec\ beam.

While the Parkes data are quite informative
of the large-scale velocity structure,
they also demonstrate a clear need to follow the
kinematics to smaller scales with interferometric data 
to fully characterize the formation pathway
of IRS1 and IRS2.

\subsection{ATCA \nht\ Kinematics}

The ATCA observations detected significant \nht\ emission near the protostar as
shown in Figures \ref{atca-8micron} \& \ref{atca-nh3}. The emission
morphology is quite filamentary, also concentrated along the extended envelope seen
in extinction at 8~\micron. The strongest emission is found southeast of IRS1 with another
sub-peak located east of IRS2,
between the sources.
The emission west of IRS2 is more diffuse with lower
surface brightness. We also note that the \nht\ emission does not peak on the locations of
the protostars themselves, and the morphology of the emission from the main \nht\ lines is quite similar to 
that of the satellite lines.  We overlay the \twco\ integrated intensity
maps in Figure \ref{atca-8micron} to highlight the regions that the outflows may influence.
At first glance, it does not appear likely that the \nht\ emission east and west of IRS1 is 
significantly affected by the outflows.

The kinematic structure 
derived from the ATCA observations (Figure \ref{atca-nh3}) is also fit 
using the \nht\ fitting routines within CLASS. This was done for each pixel of the 
map where emission was detected above 5$\sigma$, see Appendix A. The ATCA kinematic
structure bears some similarity to the Parkes observations,
but also has some significant departures. The line-center velocity map has features that extend
to both more red-shifted and blue-shifted velocities than were evident in the Parkes map. North
of IRS1, there is a red-shifted feature with velocities greater than 
-4.2~\kms\ and there is a red-shifted feature
directly southeast of IRS1. These features may reflect outflow interaction with the envelope
as evidenced by the shape of the 
conical shape of the red-shifted feature north of IRS1, and the  
position of the red-shifted southeast feature along the edge of the \nht\ emission, bordering the cavity wall of the southern outflow lobe.  
The linewidth map also shows broad linewidth south of IRS1 and to 
the southeast; however, the linewidth is very narrow to the north of IRS1.
 Outflow interactions can produce increased linewidth and/or changes in the 
line-center velocity.

Between IRS1 and IRS2 there is a steep velocity gradient from red-shifted to blue-shifted
in the line-center map that continues past the position of IRS2. In the region
with the steep velocity gradient, there is an accompanying increase in the linewidth
map that results from the velocity gradient being unresolved; such a feature showed up 
several times for other sources in \citet{tobin2011}. This gradient from red to blue is in the same 
direction as the large-scale gradient observed in the Parkes \nht\ map. 
Beyond $\sim\pm$40\arcsec\ from IRS1 along the direction of the extended envelope, the line center
velocities are about equal in both the southeast and northwest sides of the envelope. The 
linewidth northwest of IRS2 is narrow, between 0.2 and 0.3~\kms. To the southeast, the 
linewidth remains above 0.3~\kms\ and may be the result of outflow-envelope interaction, 
as suggested already. However, the increased linewidth feature extends beyond the edge of the 
outflow cavity, yet there is unknown 3D structure along the line of sight.
Depending on the configuration of material, the outflow could affect a larger area
beyond its boundaries, and increased linewidth propagates into the envelope, or
if the envelope is filamentary and curved along the line of sight gas falling-in
toward the protstar could yield an increased linewidth.
To the west of 
IRS2, the linewidth is quite narrow, except for one area that has a linewidth of 0.5~\kms.

We zoom-in on the immediate region around IRS1 and IRS2, the inner 60\arcsec\ of the source,
in Figure \ref{atca-nh3-zoom}, showing the line-center velocity in greater
detail. From east-to-west, crossing the source, the velocity 
goes from -4.55~\kms\ in the eastern-most part of the map to -4.25~\kms\ at IRS1, -4.6~\kms\ at IRS2, 
and down to -4.85~\kms\ west of IRS2. Qualitatively, the gradient goes from blue, to red, and back
to blue; thus, there is not a clear velocity gradient that can be attributed to
rotation around IRS1 and IRS2. Furthermore, the strip of broad linewidth directly north of
IRS2 can be indicative of a rapid velocity change that can result from two velocity components
along the line of sight. This was observed toward several protostars in \citet{tobin2011},
and the higher resolution ALMA \ntdp\ data will be used to investigate this further.

The velocity profiles from the ATCA and Parkes velocity maps
are extracted along lines orthogonal to the outflow directions and shown in Figure \ref{1d-profiles}.
The Parkes data show a velocity gradient comparable to the two-dimensional fit. The ATCA profile
shows significant structure near the protostar and then the velocities move back toward
agreement with the cloud for IRS1. For IRS2, a very steep velocity gradient is derived from
the ATCA data. The fits to the Parkes velocity gradients also find that the core velocity at the 
position of IRS1 is -4.45~\kms.

The velocity and linewidth maps derived from hyperfine fitting and the profiles extracted are useful for providing a
global view of the envelope kinematics. However, they can abstract some more detailed information provided by the line
profiles. We show position-velocity (PV) diagrams for both the main \nht\ lines and satellite 
lines in Figure \ref{atca-nh3-pv}. These are extracted from a 10\arcsec\ strip in the east-west
direction (position angle = 90\degr), centered on IRS1. 
PV diagrams extracted from \nht\ (1,1) datacubes are complex because 
each hyperfine component is a pair of closely spaced lines, so caution must be exercised in 
their interpretation. Figure \ref{atca-nh3-pv} shows that the western portion of the envelope
(containing IRS2) is best described by a single velocity component 
(-4.6~\kms and offsets between -25\arcsec\ to -5\arcsec), 
and this component
shows a slight gradient to more blue-shifted velocities. However, another
distinct velocity component is present on the east side of the envelope 
(-4.0~\kms and offsets between -5\arcsec\ to 25\arcsec),
red-shifted with respect to far west side of the envelope.
This is most evident in the PV diagram of the satellite lines 
where there appear to be three peaks; each velocity component of \nht\ has two hyperfine lines and
the velocity shift is large enough that  one hyperfine from 
the blue and red component are overlapping. This large shift 
appears in the velocity map as the red-shifted component
that is coincident with the red-shifted outflow from IRS1, 
but this red-shifted component extends beyond the 
bounds of the outflow. This could indicate that there is a second velocity component to 
the core, or that the outflow is imparting a bulk redshifted velocity on 
dense molecular gas in the envelope.

In addition to the PV diagrams in the equatorial cuts, we also show a PV diagram extracted
from a larger region along the long-axis of the envelope (a position angle of 123\degr)
in Figure \ref{atca-nh3-pv-diag}. This PV diagram shows that the
features observed in the smaller region are present across much of the envelope with the
red-shifted velocity component appearing throughout most of the East side of the envelope.
The red-shifted velocity component (-4.0~\kms) begins to appear 
just before the PV cut crosses IRS1 (offsets from -5\arcsec\ to 70\arcsec),
indicating that it could be related to the red-shifted outflow from IRS1. However, 
the red-shifted velocities continue as the the cut follows the envelope southeast,
well-beyond the region of influence of the red-shifted outflow from IRS1. The
red-shifted outflow of IRS2 is directed in this region and has substantial \twco\ and
\thco\ emission that fans out toward the east side of the envelope. Thus, it is clear that the eastern
side of the envelope has a second, red-shifted velocity component, but it is unclear
if this is outflow-related or part of the cloud structure itself.

\subsection{ALMA \ntdp\ Kinematics}

With the ambiguity of the \nht\ kinematics near IRS1 and IRS2, higher resolution is
imperative to disentangle the kinematic structure of gas surrounding the binary system. 
The ALMA observations include
\ntdp\ ($J=3\rightarrow2$), the deuterated form of \nthp. \nthp, \ntdp, and \nht\ have been shown
in multiple studies to trace the same (or very similar) 
kinematic and physical structures in both starless and
protostellar cores in single-dish and interferometric studies 
\citep{johnstone2010,tobin2011}. \ntdp\ has the caveat 
that it can be destroyed in the inner regions of
the envelope where \nthp\ and \nht\ can still exist \citep{tobin2013}, but
in the case of BHR71 the \nht\ and \ntdp\ peaks align closely with the
\nthp\ peaks found by \citet{chen2008}.

We show the \ntdp\ integrated intensity maps overlaid on the \textit{Spitzer} 8~\micron\ image, 
the ATCA \nht\ and the 1.3~mm continuum in Figures \ref{atca-alma-8micron} \& \ref{alma-n2dp}.
The \ntdp\ map covers a smaller region than the ATCA \nht\ map; as such, the \ntdp\ does not map
to the 8~\micron\ absorption feature as well as \nht. However, there is good correspondence
to the east of IRS1. The \ntdp\ also tends to avoid the region of the outflow toward IRS1, 
but there is some emission north of IRS1 along the red-shifted outflow. Toward IRS2,
the \ntdp\ does not clearly avoid the outflow, which might be due to the three dimensional
configuration of the envelope. Finally, the \ntdp\ east of IRS2 corresponds well to the
\nht\ map, but east and south of IRS2 the \nht\ intensity is dropping while there is \ntdp\ 
emission surrounding IRS2 on three sides.
We also notice in Figure \ref{alma-n2dp} that the extended 1.3~mm continuum around IRS1 seems to avoid
the brightest areas of \ntdp\ emission. Specifically, the extended emission west of 
IRS1 seems to run through the gap in \ntdp\ toward IRS2.

The kinematic structure is derived from the \ntdp\ data using the CLASS hyperfine
fitting routines and the known hyperfine line positions and relative line ratios \citep{dore2004};
also see \citet{tobin2013}; see Appendix A.
The overall kinematic structure of \ntdp\ is quite similar to \nht\ in the
overlapping regions, except that
some of the velocity gradients are better resolved. There are also still
indications of outflow-envelope interaction with red shifted line-center velocity southeast
of IRS1 and increased linewidth in this region. In the \ntdp\ linewidth map,
the linewidth is remarkably low across the entire envelope. The map is a bit noisy due to the
smaller beam and lower signal-to-noise in some regions. North of IRS1, there is still
the large region of red-shifted velocities that coincide with the likely area of influence
from the outflow. Toward IRS2, the line center velocity map is not highly structured, particularly
away from the region of red-shifted velocities that seem to be associated with 
IRS1. The velocity gradient around IRS2 appears to be more along the outflow direction than
orthogonal to it. Overall, the \ntdp\ line-center velocity near IRS2 is very close the
line-center velocity to the east of IRS1. Thus, it is not clear that the kinematics
strongly indicate rotation across the core on scales comparable to the protostar separation, 
or even around the protostars themselves. The area of large linewidth observed
north of IRS2 in \nht\ with ATCA is not reflected in the \ntdp\ map, likely because this was due to a 
velocity gradient that is resolved in the ALMA map. Given that the velocities of the gas toward
IRS2 and the gas east of IRS1 have comparable velocities, the kinematics surrounding IRS1
and IRS2 do not strongly indicate that they are forming out of kinematically
distinct cores. The only firm conclusion that we can make is that the outflow 
from IRS1 appears to be significantly affecting the kinematic structure to the north.

The velocity profiles from the ALMA \ntdp\ velocity maps are also extracted along 
lines orthogonal to the outflow directions and shown in Figure \ref{1d-profiles}.
Compared to the ATCA \nht\ velocity profiles, the \ntdp\ profiles are significantly more structured
and the velocity away from the protostar position for IRS1 does not as closely
agree with the Parkes velocities. This could result from chemical differences between
\ntdp\ and \nht, or the higher resolution of the \ntdp\ data could be resolving structure
that is smoothed out in the lower resolution ATCA data.

In addition to the line center velocity maps and profiles, we also extracted a PV diagram
from the \ntdp\ emission toward BHR71. This PV diagram is shown in Figure \ref{alma-n2dp-pv}
and is taken in an east-west cut 10\arcsec\ in width, centered on IRS1. The angular resolution of
the \ntdp\ maps is about 5 times finer than the \nht\ map, so some features appear different,
but the major features are similar. The PV cut of \ntdp\ shows very 
little evidence for a systematic velocity gradient across the position of IRS1. There is 
also still a second velocity component evident 
(-4.0~\kms), but this red-shifted component is most evident near the protostar position at an 
offset position of $\pm$5\arcsec. The red-shifted velocity component 
is made less obvious due to the brightest hyperfine lines being spread 
over $\sim$0.5~\kms\ \citep{dore2004}, in contrast to the pairs of ammonia lines that
are less blended. The 
lines also appear slightly more red-shifted than \nht\ because we assigned the rest frequency
to match the brightest \ntdp\ hyperfine line, which is at a higher frequency than the other 
hyperfine lines of comparable relative intensity.

\subsection{ALMA \rm{\cateo} Kinematics}

Inside the \ntdp\ and \nht\ emitting regions, we find that \cateo\ emission peaks directly
on the continuum sources as shown in Figure \ref{co-isotopes}. There is also extended emission around IRS1 and IRS2, but with
most emission concentrated at the location of the continuum sources. Moreover, toward IRS1,
the \cateo\ appears extended along the walls of the outflow cavities. There is also a 
thin ridge of increased \cateo\ brightness extended between IRS1 and IRS2.

We examined the \cateo\ line kinematics for signs of rotation, infall, 
and/or outflow entrainment by 
first examining the integrated intensity, line center velocity, and linewidth
maps shown in Figure \ref{alma-c18o}. These are similar to the maps constructed
for the \nht\ and \ntdp\ data, but instead of fitting the hyperfine lines, we
calculate the standard 1st (centroid velocity) and 2nd (linewidth) 
moment maps because \cateo\ is a single emission line. 
The centroid velocity map shows the signature of the outflow
from IRS1, but there is a very slight `twist' in the velocity map across the source
indicating that there is more going on that just the outflow. Also, the linewidth
map shows a broader linewidth near the position of IRS1, indicative of 
more rapid motion along the line of sight. There is not evidence of enhanced linewidth
toward IRS2, but moment maps are most often dominated by the low-velocity
emission that is strongest, especially in the case of
a map with total power emission included (Figure \ref{spectra})

In order to better examine the motions of the gas near the protostar positions,
we examine integrated intensity maps constructed using only the higher 
velocity blue- and red-shifted emission independently.
Toward IRS1, integrated intensity maps are
shown in Figure \ref{IRS1-c18o} for the red and blue-shifted \cateo\ in 
velocity ranges from low (-5.8 to -4.4~\kms\ and -4.4 to -3.2~\kms), 
medium (-6.2 to -5.6~\kms\ and -3.6 to -2.8~\kms), and high (-6.3 to -5.8 \kms\ and -3.0 to -2.5~\kms). At lower velocities
the red and blue-shifted emission are clearly offset along the outflow direction. At medium
velocities, the emission becomes more compact and the red and blue-shifted emission are oriented
diagonally across the continuum source, possibly tracing both outflow and motion in the equatorial
plane. We also note that velocity gradients in the direction of the outflow may result
from infall motion \citep{yen2013}, but may also be related to outflow-envelope
interaction \citep{arce2006}.

The \cateo\ at the highest velocities has blue and red-shifted emission peaks that are oriented in the
east-west direction (blue on the east, red on the west), with some low-level emission along 
the outflow direction. Thus, this velocity gradient is oriented perpendicular to the outflow, as
expected for rotational motion on $<$1000~AU scales in the inner envelope. Note that the
direction of this velocity gradient is the opposite of what is observed on large-scales
with Parkes and ATCA.

The \cateo\ emission toward IRS2, on the other hand, is found to have 
the red-shifted peak on the east and the blue-shifted
peak is on the west (Figure \ref{IRS2-c18o}). The position angle is not exactly orthogonal 
to the outflow, it is not along the outflow either. Thus, the small-scale velocity 
gradients in \cateo\ appear to be in the opposite directions for IRS1 and IRS2.

To examine the kinematics in more detail, we show PV diagrams of the \cateo\
emission toward IRS1 and IRS2 in Figure \ref{PV-c18o}. These PV cuts are taken
orthogonal to the outflow directions of each source, centered on the continuum source.
For both sources, positive offsets are to the east and negative offsets are to the west. 
The PV diagrams in 
Figure \ref{PV-c18o} are generated using only the data from the ALMA 12m array to reduce confusion with the extended
emission picked up by the ACA and Total Power observations.

The PV diagram for IRS1 shows that at high-velocities near the continuum position, the
blue-shifted side is located to the east and the red-shifted side is located to the west. Toward
lower velocities, the \cateo\ emission is more symmetrically distributed in both position and
velocity. IRS2 shows a considerably different appearance. First, the \cateo\ is 
not as spatially extended, likely because IRS2 is ten times less luminous than IRS1, and
the emitting region is expected to be smaller. Second, the region with the brightest blue-shifted \cateo\ is still on the west
side of the envelope, while the brightest red-shifted \cateo\ is on the east side of the envelope.
Thus, both in the integrated intensity maps and the PV diagrams, the velocity gradients of IRS1
and IRS2 are in the opposite direction.

The PV diagrams also show that the \cateo\ emission of the protostars is slightly different from
the velocity of the core. The \cateo\ emission is best described with a central velocity of
-4.6~\kms\ in contrast to the average core velocity of -4.45~\kms\ from Parkes \nht\ emission. Parts of the
core also show velocities that are even more different, Figure \ref{alma-n2dp} shows that the core near 
IRS2 and east of IRS1 has velocities up to -4.9~\kms, and near IRS1 the velocity can be -4.2~\kms.

To compare the PV diagram of the \cateo\ emission to that of the \nht\ and \ntdp, we use of the 12m+ACA+TP data
to show the PV diagram on comparable spatial scales across both sources and show this larger
scale \cateo\ PV diagram in Figure \ref{PV-c18o-ext}.
The PV diagram shows that there is extended structure from the surrounding core
toward both IRS1 and IRS2. Similar to \nht\ and \ntdp, the \cateo\
shows an apparent second velocity component around -4.0~\kms\ and present from -25\arcsec\ to 
beyond 30\arcsec. However, the protostars themselves seem to be more closely associated
with the blue-shifted velocity component. The blueshifted component is at about -4.9~\kms\ and
the protostars are between -4.7 and -4.6~\kms. We note, however, that the large-scale \cateo\
emision likely reflects the velocity of the gas in the exterior regions of the could because \cateo\
is likely frozen-out onto dusgrams in the higher density interior of the core/envelope where
\ntdp\ and \nht\ are present \citep{frerking1982,bm1989}.

\section{Discussion}

As an isolated core harboring a wide binary system, BHR71 should lend itself to being
an ideal testing ground for wide binary formation. The formation of
such systems have long been thought to be related to the rotation of the core \citep[e.g.,][]{larson1972,boss1979,bb1993,tohline2002}. Thus, our initial expectation was to find 
a classic scenario of clear core rotation, with conserved angular 
momentum leading to increased rotation velocity at
progressively smaller radii. We also expected that a 
difference in radial velocity of IRS2 relative to IRS1 due to orbital motion could be measured because its 
separation, lower overall luminosity, and estimated dust+gas mass. 

\subsection{Kinematic Analysis}

The kinematic structure observed in \nht\ from Parkes and ATCA alone, appeared remarkably
consistent with rotation-induced fragmentation. The Parkes data showed a modest
velocity gradient on arcminute scales, and the ATCA data showed evidence of a larger
velocity gradient across the two protostars. Furthermore, previous ATCA \nthp\
observations toward IRS1 detected a velocity gradient across the main protostar,
but did not detect emission around IRS2 due to the smaller primary beam at 3~mm \citep{chen2008}.
%
%
%
%
%
%
%
%
The velocity gradient observed by Parkes along the major axis of the core as viewed
in \nht\ is 1.2~\kmspc, measured across a diameter of 150\arcsec\ (30000~AU). 
The low-resolution of the single-dish observations smooth out any velocity structure
at small scales and provides a measure of the velocity gradient on the largest scale,
which for the sake of this discussion, we interpret as solid-body rotation. But
once material begins free-fall, rotation is no longer solid-body and conserves 
angular momentum (in the absence of magnetic fields). This 
gradient would correspond to a rotation velocity of 0.087~\kms\ at the core radius of 15000~AU 
and an angular velocity ($\Omega$) of 3.9~$\times$10$^{-14}$~s$^{-1}$. If we assume that 
the entire core at radii less than 15000~AU is in freefall, the infalling gas
should conserve angular momentum. Then,
the rotation velocity should be $\sim$ 0.33~\kms\ at a radius of 4000 AU. At the
separation of IRS2, 16\arcsec\ (3200 AU) the rotation rate would be even
higher, 0.4~\kms. The ALMA \ntdp\ map shows very little evidence for a velocity gradient
of this magnitude from the area directly east of IRS1 to the area surrounding IRS2. 
Furthermore, the \cateo\ velocities of IRS1 and IRS2 are consistent with 
being at the same velocity. 
The lack of observed rotation velocity on 3200~AU scales, however, could be due to the
apparent influence of the outflow on the ambient envelope material, masking hints of 
rotational velocity increases toward the envelope center. But, the largest scales
examined by the ALMA \ntdp\ do not show ordered rotation surrounding
either IRS1 or IRS2.

In the preceding paragraph, we assumed that the entire core was falling in and
that only the velocity gradient on the largest scales reflected solid-body rotation.
If we instead assume that the entire core is not yet collapsing and the 1.2~\kmspc\
velocity gradient reflects solid body rotation
out to a certain radius, then we can calculate different
rotation velocities. Assuming inside-out collapse \citep{shu1977, tsc1984}, the infall radius
must be larger than the companion separation. If we assume an infall radius of 6400 AU, twice the companion
separation, the solid body rotation velocity at this radius would be 0.037~\kms\ with
the 1.2~\kmspc\ gradient 
($\Omega$ = 3.9~$\times$10$^{-14}$~s$^{-1}$), 
below our ability to detect. Furthermore, if angular momentum was conserved from
a hypothetical infall radius of 6400~AU, then the rotation velocity at a radius of 
3200~AU would only be 0.074~\kms. Both the solid-body rotation rate at 
6400~AU and the estimated rotation velocity from conserved angular
momentum at 3200~AU would be below our ability to detect. Nevertheless, 
the core is observed to have velocity structure with significantly larger amplitudes than these extrapolations
from the large-scale core rotation. These inconsistencies can be taken as evidence against ordered
collapse of the system with rotationally induced fragmentation.


Using these observationally measured quantities of the core velocity structure, we can
calculate the estimated stability of the core from the ratio of
rotational energy to gravitational potential energy ($\beta_{\rm rot}$) on the scale of 3200 AU.
While this diagnostic 
assumes solid-body rotation and we use a rotation velocity inferred from the assumption
of constant angular momentum, this analysis enables comparisons to other studies
that determine $\beta_{\rm rot}$ as their key diagnostic derived from the
observations. We follow
the method outlined by \citet{chen2007} which calculates the rotational energy as
\begin{equation}
E_{rot} = \frac{1}{2}I\Omega^2 = \frac{1}{2}\alpha_{rot}MR^2\Omega^2
\end{equation}

and $\alpha_{rot}$ = $\frac{2}{3}(3-p)/(5-p)$, $\Omega$ is the angular velocity derived from the observed
velocity gradient, R is the radius of the envelope, and M is the mass of the envelope. We adopt 
p=1.5 for a spherical envelope in
free-fall. Note that this is not exact because the envelope around BHR71 IRS1 and IRS2 is non-spherical. 
Then the gravitational binding energy is defined from the virial theorem to be
\begin{equation}
E_{grav}=\frac{3GM^2}{5R}
\end{equation}
where G is the gravitation constant; M and R are the same as defined for the previous equation.
Knowing these two equations, $\beta_{\rm rot}$ can be calculated
\begin{equation}
\beta_{\rm rot} = \frac{E_{rot}}{E_{grav}} = \frac{5R^3\Omega^2\alpha_{rot}}{6GM}.
\end{equation}
We can simplify this relationship by multiplying constants through and converting the more natural observed
units such as solar masses, parsecs, and \kmspc, yielding
\begin{equation}
\beta_{\rm rot} = 55.2 \left(\frac{R}{\rm pc}\right)^3 \left(\frac{M_{\rm env}}{1.0\rm\ M_{\sun}}\right)^{-1} \left(\frac{v_{\rm grad}}{\rm km s^{-1} pc^{-1}}\right)^2
\end{equation}
The terms of the equation are now defined as R being the core radius in pc, $v_{\rm grad}$ is 
the velocity gradient in terms of \kmspc\ across this scale, and M$_{\rm env}$ is the core/envelope mass in solar masses.
We are interested in determining the level of rotational support both at a radius of 0.05~pc (10000AU) to assess
the level of rotational support in the envelope as a whole and at a radius of 3200~AU (0.016~pc), 
the separation of IRS1 from IRS2. We first calculate $\beta_{\rm rot}$ for the larger scale using the 
velocity gradient measured across the BHR71 core of $\sim$1.2~\kmspc and an envelope mass 
of $\sim$4.6~M$_{\sun}$ determined for the radius of 0.05~pc from 8~\micron\ extinction 
from \citet{tobin2010}. With these values, we find $\beta_{\rm rot}$ = 0.002 at a radius of 0.05~pc.

To calculate an upper limit of rotational support on the scale of the companion separation (3200~AU; 0.016~pc), we 
assume that angular momentum is conserved and the velocity gradient is more rapid at a radius 0.016~pc. 
Scaling the velocity gradient from the ratio of core to separation/inner envelope radius, the velocity gradient
at a radius of 0.016~pc (3200 AU) is 4.83~\kmspc. We then use the mass measured from the ALMA 
continuum data for both IRS1 and IRS2, which totals 0.7~$M_{\sun}$, factoring in the likely
higher dust temperature toward IRS1. With these numbers, we calculate an upper limit 
to $\beta_{\rm rot}$~=~0.006. Note that 
we call this an upper limit because we assumed that angular momentum was conserved from 
larger scales, but the velocity gradient across the envelope from ALMA \ntdp\ is consistent 
with zero. Furthermore, \citet{tobin2011, tobin2012} argued that for elongated envelopes 
such as BHR71 projected infall motions can contribute to the observed velocity
gradients. Lastly, if the motions do correspond to rotation, inclination is not taken into account,
therefore rotation could be higher if that is the origin of the velocity gradient. 
However, as outlined in Section 3.3 the inclination of the outflows is 
very likely to be greater than 35\degr\, therefore the
correction to the rotation velocity is likely less than a factor of 2.
The $\beta_{\rm rot}$~=~0.006 measurement for BHR71 places this system toward the top end of the distribution
of single systems shown in \citet{chen2012}.
However, this might not be an entirely fair comparison
because those $\beta_{\rm rot}$ values are calculated at a variety of core radii, some 
single-dish and some interferometric. We further emphasize that if we evaluate the velocity
gradient from the ALMA \ntdp, excluding the regions that are likely outflow influenced, 
$\beta_{\rm rot}$ could be much lower. Our calculated value for $\beta_{\rm rot}$ is lower than the
one determined by by \citet{chen2008}, but they probed a different scale focused on
IRS1 using \nthp.

In addition to the apparent velocity gradients, the observations also revealed the presence of
two velocity components of the dense gas to the east of IRS1. The \nht\ and \ntdp\ line center velocities are
distinctly red-shifted in the regions that overlap with the red-shifted outflow lobe of IRS1. Furthermore, 
the PV diagrams shown in Figures \ref{atca-nh3-pv}, \ref{alma-n2dp-pv}, and \ref{PV-c18o-ext}
show that this red-shifted component is also present in equatorial plane
of the envelope. Its presence in the equatorial plane is diminished in the velocity maps that are
derived from hyperfine fitting of the \nht\ and \ntdp\ in Figures \ref{atca-nh3}, \ref{atca-nh3-zoom}, and \ref{alma-n2dp}
because the blue-shifted component has greater line strength.

This additional line component must not be very broad since it can be observed as a distinct component
in the PV diagrams in Figures \ref{atca-nh3-pv}, \ref{alma-n2dp-pv}, and \ref{PV-c18o-ext}. The outflow from 
IRS1 could be inducing bulk motion
in the surrounding molecular gas, similar to the simulations of \citet{offner2014}. Furthermore, the red-shifted
lobe of the outflow from IRS2 could also be contributing to the red-shifted velocities in \nht\ and \ntdp\
on the southeast side of the envelope. The overlap of IRS2 outflow contours in Figures
\ref{atca-nh3}, \ref{atca-nh3-zoom}, and \ref{alma-n2dp} highlight this possibility, despite the region 
also being coincident with the edge of the blue-shifted outflow cavity from IRS1. Alternatively,
the two velocity components could also result from velocity shear that might have been present
during the formation of a core. However, the ambiguity of the formation conditions of the
BHR71 core and its structure along the line of sight makes the origin of the second velocity component and the
interaction of the outflows and dense molecular gas difficult to confidently ascertain.

\subsection{Formation of the Binary System}

Models that consider the formation of multiple star systems from rotating, collapsing envelopes
have difficulty fragmenting for parameters similar to the observed quantities of $\beta_{\rm rot}$ and
$\Omega$ for BHR71. One of
the classic studies of \citet{bb1993} used a rotating collapsing envelope
with an m=2 perturbation to break the symmetry. However, they still needed a fairly rapid
rotation rate for fragmentation on $>$1000~AU scales, using $\Omega$=7.2$\times$10$^{-13}$~s$^{-1}$ resulting in 
$\beta_{\rm rot}$~=~0.16. Thus, these models had over an order of magnitude more rotation in order
to produce fragmentation on the desired scale.

Recent work by \citet{boss2014} explored fragmentation with rotating cloud cores including
magnetic fields. This work did not find fragmentation on scales $>$1000~AU for systems with 
rotation rates $<$10$^{-13}$~s$^{-1}$ nor $\beta_{\rm rot}$~$<$~0.01. The systems that 
most frequently produced fragments on scales $>$1000~AU in those simulations had the
highest rotation rates ($\Omega$=3$\times$10$^{-13}$~s$^{-1}$). Systems still fragmented in the 
presence of magnetic fields, but the models with the strongest magnetic fields did 
inhibit fragmentation. Many other models considering core rotation also had difficulty in
forming multiple systems with $>$1000~AU for rotation rates comparable to or exceeding those found in our
observations \citep[e.g.][]{price2007b,machida2008}.

Putting all this information on the kinematics of BHR71 together
and recent numerical studies of multiple star formation with rotation,
we find it unlikely that
core rotation could have resulted in the formation of the binary system. This is based on the 
reasonable upper limit to the value of $\beta_{\rm rot}$~=~0.006, but in 
reality the core rotation
is nearly zero across the inner envelope, as probed by \ntdp\ and toward the protostars
as measured with \cateo. More to the point, the overall
inner envelope kinematics are not consistent with ordered rotation when viewed with
\nht\ and \ntdp\ at high enough resolution to both resolve the companion and recover
the extended emission of the envelope. However, regardless of the level of true
rotation in BHR71, the evidence for rotation in the opposite direction on scales $<$~1000~AU
cannot be easily reconciled. On the scales of disks, the Hall effect could theoretically
reverse the direction of rotation in the disk \citep{tsukamoto2015,krasnopolsky2010}, but the
motions we observe are on $>$100s of AU scales where the densities should not be high enough
for non-ideal MHD effects to operate efficiently \citep{krasnopolsky2010}.

Due to the observational inconsistencies with angular momentum conservation
and the difficulty of simulations with similar initial conditions to result
in wide companion formation, we must consider alternative scenarios for
the formation of this binary system. One of the leading alternatives is turbulent fragmentation of the 
core \citep{padoan2002,goodwin2004,offner2010}.
The turbulent velocity structure of the simulated molecular clouds will create density perturbations
throughout the cloud. This creation of over-densities can be efficient enough such that these regions of 
locally enhanced density exceed the Jeans mass and collapse to become a protostar. Thus,
a protostellar core or two adjacent cores formed in the presence of turbulence could lack an
ordered rotation pattern and still form a binary system. Simulations find that the initial
separations of these cores are typically $>$500~AU to 1000s of AU \citep{offner2010,bate2012}.

The apparent rotation in opposite directions on small-scales for IRS1 and IRS2 can also
result from turbulent fragmentation. Fragmentation can happen in a turbulent cloud without
rotation of the envelope/core, and the angular momenta of the collapsing regions will be derived from the 
net angular momenta in the turbulent velocities \citep{offner2010,walch2010,offner2016}. The
net angular momenta of the density enhancements that formed IRS1 and IRS2 could have been 
anti-aligned, leading to the opposite rotation directions observed 
on $<$1000 AU scales toward IRS1 and IRS2. Turbulent fragmentation is found to be quite likely
for other widely separated protostellar multiple systems as well \citep{lee2017, pineda2015}.

The disordered velocity structure of the envelope as viewed in \ntdp\ and \nht\ could also
be the result of turbulence in the core, where the bulk motion of turbulent gas could manifest 
itself as a disordered velocity structure. We suggest this because the outflow is unlikely to
produce all the disordered velocity structure that is observed in the core. Specifically, 
we are referring to the area east of IRS1
as shown in Figure \ref{alma-n2dp}, where the velocity along the north-south direction
goes from red-shifted to blue-shifted and back to red-shifted. 

The misaligned angular momentum vectors are not unique to the BHR71 system. The
Class 0 proto-multiple system IRAS 16293-2422
may have misaligned kinematics in its binary system separated by only $\sim$600 AU \citep[e.g.,][]{zapata2013},
the IRAS 04191+1523 system with a separation of $\sim$860~AU has projected angular
momentum vectors that differ by $\sim$ 90\degr\ \citep{lee2017}, and there are more evolved proto-planetary
disk systems that also show similar misalignment \citep{stapelfeldt2003,jensen2014,williams2014,brinch2016}. Thus, the formation of
multiple systems with misaligned or even anti-aligned angular momentum vectors
may be common \citep{lee2016}.

Despite the seeming consistency with turbulent fragmentation, many of the simulations
are focused on fragmentation within a parsec-scale molecular cloud. However, BHR71 is an
isolated core and is not located within a larger cloud \citep{bhr1995}. It is therefore
unclear if the results from turbulent fragmentation are directly applicable. However, \citet{walch2010}
examined the effects of turbulence within isolated cores that are similar in size and mass
to BHR71. They found that the turbulence within the core could lead to non-axisymmetric 
structure and misaligned net angular momenta. But, these particular simulations used
very high levels of initial turbulence, the ratio of turbulent to gravitational binding
energy in the simulated cores was between $\sim$0.3 to $\sim$0.6. Following \citet{walch2012},
BHR71 has a ratio of turbulent to gravitational energy of $\sim$0.04, consistent with
other star forming cores. Therefore, the levels of internal turbulence do not 
appear sufficient to fragment the core in the way that it occurs in the study of 
\citet{walch2010}. \citet{offner2014} also examined isolated, turbulent cores
in their outflow study; however, fragmentation in those simulations was not witnessed on scales
comparable to BHR71 IRS1 and IRS2, but that was also not the focus of their study.


A variation on the theme of turbulent fragmentation is that the shape of the core itself
could result in multiple protostars forming. \citet{bonnell1993} considered the collapse
of cylindrical cloud cores, and due to the shape the ends of the cylinder collapsed first
forming two protostars. These protostars later moved closer together, under their
mutual gravitational attraction, forming a bound system.
The fact that the BHR71 core appears elongated and asymmetric is similar to such a scenario.
Moreover, molecular clouds themselves and even individual protostellar envelopes are often filamentary
\citep[e.g.,][]{andre2010,tobin2010a,looney2007}. Therefore it is possible that cloud
shape played a role in the formation of BHR71 IRS1 and IRS2, but an asymmetric cloud (or portions of it) can only collapse
if they are Jeans unstable.

While turbulent fragmentation can explain many of the features 
we observe in the velocity field, thermal Jeans fragmentation 
is another possible route for the formation of the binary system. If the gas is cold and
dense, it can collapse without rotation or turbulent motions. Such a 
scenario is consistent with the lack of radial velocity shift between the sources. However,
we might still expect the individual protostars to have a common angular momentum 
vector because of the overall velocity gradient observed in BHR71.

Using our knowledge of the envelope
mass, we can calculate the Jeans length for BHR71 to determine if this is a feasible fragmentation mechanism.
The Jeans length is approximately
\begin{equation}
\lambda_J = \frac{c_s}{\sqrt{G\rho}}
\end{equation}
where c$_s$ is the sound speed ($\sim$0.25~\kms\ for gas at 20~K), G is the gravitation constant, 
and $\rho$ is the average density. We calculate the average density from the mass of 4.6~M$_{\sun}$
that is estimated to be enclosed by a radius of 0.05~pc which is $\sim$6$\times$10$^{-19}$~g~cm$^{-3}$ or n$\sim$10$^5$~cm$^{-3}$.
With these values, we estimate the Jeans length to be $\sim$8400~AU, which is much larger than the projected
separation of IRS1 and IRS2. Lower temperatures were certainly possible prior to protostar formation
\citep[e.g.,][]{stutz2010, sadavoy2018} and higher densities as well, but an average density 10$\times$
higher is necessary to make the Jeans length comparable to the current protostar separation.
Thus, thermal Jeans fragmentation is not clearly a favored mechanism on its own to
form the binary system, and turbulent fragmentation remains the 
most likely candidate.

\subsection{Binary Arrangement Along Line of Sight}

The structure of the core and relative positions of IRS1 and IRS2 along the line of sight is a 
source of uncertainty in the interpretation of the kinematics and
companion formation. IRS2 appears to be either behind or at the same distance
with respect to IRS1 because the outflows do not appear to strongly 
interact. Such an interaction is expected to be accompanied by 
a shock, but \textit{Spitzer} H$_2$ line mapping \citep{giannini2011} 
and \textit{Herschel} [O I] mapping \citep{nisini2015} do not
show strong emission at the location where the IRS2 outflow fans out.
Moreover, the blue-shifted lobe of IRS1 exits the globule, as seen in the 
infrared and optical imaging (e.g., Figure 1), so it may be near the front edge of the 
globule (at least, not exactly in the center). Therefore IRS2, if it were even closer 
to us than IRS1, would be really close to the edge, but we see that it is very embedded 
and its blue lobe is not prominent in the near-infrared like that of IRS1.

The red-shifted outflow from IRS2 does fan out on the southeast side of it, but this is 
not what would be expected if the outflow was interacting with the blue-shifted
outflow from IRS1. If the two outflows were interacting, the red-shifted outflow
from IRS2 would be expected to be deflected to the south, the direction of the
blue-shifted flow from IRS1. Since such a deflection is not observed, the arrangement
of the sources makes the possibility of IRS2 being located much in front of IRS1 unlikely.
The extension of the southeastern side of 
the red-shifted outflow from IRS2 may more likely be related to 
how it is interacting with and exiting the core; outflows can also change their orientations
over time as they accrete material with different angular momenta \citep{offner2016}.

The \cateo\ velocities of IRS1 and IRS2 are approximately the same, as are the velocities
of the \nht\ and \ntdp\ gas closely associated with these protostars. Thus, it is
unlikely that they are within separate, but nearby cores along the line of sight, 
despite the second velocity component. With such an
arrangement of IRS1 and IRS2, one could formulate a scenario where the actual 
separation of IRS1 and IRS2 is 2-3$\times$ larger 
(if e.g., IRS2 is located $\sim$6,000 AU behind IRS1),
and assuming that IRS2 still orbits IRS1, the relative velocity 
of the protostars as measured by \cateo\ would be mostly tangential to 
the line of sight making their relative velocities appear smaller. We note, 
however, that regardless how IRS1 and IRS2 are arranged in
three dimensions, the opposite rotation directions for IRS1 and IRS2 are 
still inconsistent with rotation induced fragmentation.


\subsection{Future Observations to Characterize BHR71}

While it is has become somewhat cliche to suggest that further observations are needed
to understand a particular system, we outline here a few specific missing pieces of the 
puzzle that would enable our ideas regarding fragmentation in BHR71 to be confirmed
or ruled-out.

First, the evidence for opposite rotation directions toward IRS1 and IRS2 in \cateo,
which makes rotational fragmentation seem less likely, could be improved. A
factor of 2-4 increases in both sensitivity and resolution would solidify this result.
The present observations only have a resolution of
$\sim$1\farcs5, and the effective time on source for the protostars
was only $\sim$62 seconds on each mosaic point. 

Second, the envelope kinematics on scales between the \ntdp\ and \cateo\ need 
improved resolution to 
better connect the kinematics from $>$1000~AU scales to 
$<$1000~AU scales where \cateo\ is the best tracer.
\citet{chen2008} published 
ATCA \nthp\ observations that had severe limitations of sensitivity and spatial filtering.
However, the regions traced in \nthp\ are coincident with part of the \ntdp\ emission.
Further observations of \nthp\ with ALMA at high-sensitivity and with coverage of both
large and small spatial scales would further illuminate the likely opposite rotation of 
gas near the protostars. While the \nht\ traces down to about the same scale as \nthp\ and \ntdp,
observations of \nht\ cannot currently be obtained at significantly higher resolution 
due to spatial filtering and lower sensitivity of ATCA. Thus, \nthp\ with ALMA is the best
option going forward.

Third, higher sensitivity observations that could
detect the \nht\ (2,2) transition would be helpful in order to determine the gas kinetic 
temperature. However, higher sensitivity and resolution \nht\ observations are 
not trivial because ATCA is currently the only
interferometer in the southern hemisphere that can observe the \nht\ inversion transitions. 
Knowledge of the gas kinetic temperature could enable the susceptibility of the
gas around the protostars to thermal Jeans fragmentation to be assessed. But, such measurements
of kinetic temperature may also be possible using H$_2$CO \citep{mangum1993}, if present in the envelope gas
surrounding the protostars.

\section{Summary and Conclusions}

We have conducted a multi-wavelength and multi-line study of the isolated
proto-binary system located within BHR71, unleashing a large battery of ground-based
southern and space-based observatories on this system. We have resolved the 
two protostars IRS1 and IRS2 in the infrared out to 160~\micron\
using \textit{Herschel} PACS photometry and conducted the most complete 
assessment of the SED toward IRS2. We also fully mapped the BHR71 core
using \textit{Herschel} SPIRE mapping.  We have further characterized the 
continuum and kinematic properties of the envelope surrounding both sources, 
in addition to the individual envelopes around each of IRS1 and IRS2. 
This was done using ATCA and Parkes to observe
\nht\ (1,1) emission from the cold gas, and ALMA to image \ntdp\ and \cateo\, in addition to outflow tracers.

Our main results are as follows:
\begin{itemize}
\item With observations that resolve the protostars out to 160~\micron\ 
and PSF photometry at 250~\micron, we find that both are Class 0 
protostars, as expected. IRS1 has a bolometric luminosity and temperature
of 14.7~L$_{\sun}$ and 68~K, respectively, while IRS2 has values 
of 1.7~L$_{\sun}$ and 38~K. Thus, IRS2 could
be less evolved than IRS1, but there are systematics related 
to source inclination that add uncertainty to T$_{\rm bol}$.

\item We resolve the continuum emission toward each protostar 
at 1.3~mm and 1.3~cm. We find that the radio spectra 
can be consistent with dust-only emission down to 1.3~cm 
for both IRS1 and IRS2, even with the 3~mm point from \citep{chen2008}
included, but we cannot exclude some free-free emission at 1.3~cm.
The total mass
measured around the two protostars 
is 1.25~M$_{\sun}$,
assuming a dust temperature of 20~K for both protostars, and
the masses of IRS1 and IRS2 individually are 1.13~M$_{\sun}$ and 0.11~M$_{\sun}$, 
respectively with an 
assumed dust to gas mass ratio of 1:100 and \citet{ossenkopf1994} dust. 
If the dust temperature around IRS1 is 34~K, accounting for the 
higher luminosity, then the mass of IRS1 would be 0.59~M$_{\sun}$, making
the combined mass 0.7~M$_{\sun}$.
This continuum mass is a lower limit since there
is substantially more cold dust surrounding two protostars 
as viewed by \textit{Herschel}. ALMA simply picks up the
brightest emission close to each protostar. There is also
weak evidence for a bridge of material between IRS1 and IRS2, perhaps
similar to what is observed in IRAS 16293-2422 \citep{jacobsen2018}.

\item The Parkes and ATCA observations seem to indicate a smooth 
velocity gradient across the core that could be interpreted as rotation. 
However, the ALMA \ntdp\ observations at higher resolution reveal a very complex
velocity field that is not smooth. 
Position-velocity diagrams from both
ALMA \ntdp\ and ATCA \nht\ find a second velocity component on the envelope
east of IRS1. Some of this second component is likely to result from the
outflow influence on the envelope, but we also find that there is no
clear velocity gradient across the envelope midplane from IRS1 to IRS2.

\item The ALMA observations also clearly detect \twco, \thco, and \cateo. 
\twco\ and \thco\ mainly trace the  outflow emission from the 
protostars, while \cateo\ traces some emission along the outflow, but also the 
kinematics of the envelopes surrounding IRS1 and IRS2
on $<$1000~AU scales. The \cateo\ 
velocities indicate that IRS1 and IRS2 are at the same velocity 
along the line of sight, so there is no evidence of 
orbital motion in their respective radial velocities.
Moreover,
the apparent rotation of the inner envelopes around IRS1 and IRS2 appear to be in the \textit{opposite} directions!

\item We conclude that the binary formation in BHR71 is 
unlikely to have resulted from rotationally-induced
fragmentation of the core. The upper limit on the ratio of 
rotational energy to gravitational potential energy is low ($\beta_{\rm rot}$~=~0.006).
This value is among the low end of the distribution presented in \citet{chen2012}.
Furthermore, the \ntdp\ velocity field across IRS1 and IRS2, in the equatorial 
plane of the envelope is consistent with no
velocity gradient beyond the separation of the pair. There is no evidence of orbital motion given that
the sources have the same systemic velocities along the line of sight, 
and the apparent rotation of the inner envelopes on scales less than
1000~AU is in the opposite direction for IRS1 and IRS2. Thus, with all the
evidence against ordered rotation leading to the formation of the binary 
system, we conclude that turbulent fragmentation might be the most likely formation scenario, 
given that the envelope has a velocity field that is not well-ordered.
\end{itemize}

\facility{ATCA, Parkes, ALMA, Herschel, Spitzer, CTIO(ISPI), Magellan (PANIC)}

\software{Astropy \citep[http://www.astropy.org; ][]{astropy2013,astropy2018}, APLpy \citep[http://aplpy.github.com; ][]{aplpy}, CASA \citep[http://casa.nrao.edu; ][]{mcmullin2007}, The IDL Astronomy User's Library (https://idlastro.gsfc.nasa.gov/)}

We acknowledge the constructive report from the anonymous referee which 
improved the quality of the manuscript. We also acknowledge useful discussions
regarding BHR71 with M. Dunham and C. Hull, as well as help with \textit{Starfinder} by
N. Murillo.
J.J.T. acknowledges support from NSF grant AST-1814762, the Homer L. Dodge Endowed Chair,
and grant 639.041.439 from the Netherlands
Organisation for Scientific Research (NWO).
H.G.A. acknowledges the support of NSF grant AST-1714710.
X.C. acknowledges support by the NSFC through grant 11473069.
This paper makes use of the following ALMA data: ADS/JAO.ALMA\#2013.1.00518.S.
ALMA is a partnership of ESO (representing its member states), NSF (USA) and 
NINS (Japan), together with NRC (Canada), NSC and ASIAA (Taiwan), and 
KASI (Republic of Korea), in cooperation with the Republic of Chile. 
The Joint ALMA Observatory is operated by ESO, AUI/NRAO and NAOJ.
This work is based in part on observations 
made with Herschel, a European Space Agency Cornerstone Mission with significant 
participation by NASA. Support for this work was provided by NASA through an award 
issued by JPL/Caltech.
The National Radio Astronomy 
Observatory is a facility of the National Science Foundation 
operated under cooperative agreement by Associated Universities, Inc.
This research made use of APLpy, an open-source plotting package for Python 
hosted at http://aplpy.github.com. This research made use of Astropy, 
a community-developed core Python package for 
Astronomy (Astropy Collaboration, 2013) http://www.astropy.org.

\appendix

\section{Hyperfine Fits}
We show examples of the hyperfine fitting from CLASS for \nht\ and \ntdp\
in Figure \ref{nh3-hyperfine} and \ref{n2dp-hyperfine}. The fits
show that there is evidence of a second component in some positions of
the \nht\ and \ntdp\ velocity maps shown in Figures \ref{atca-nh3} and 
\ref{alma-n2dp}, respectively. The east side of the envelope in both
\nht\ and \ntdp\ shows two components in Figures \ref{atca-nh3-pv} and
\ref{alma-n2dp-pv}, respectively. These appear in the hyperfine fits
as broader linewidth because a single velocity component is adopted in
the fitting process.

We further examined the issue of the second velocity component for the fit
to the east side of the envelope, where the second component causes
the most visible change to the linewidth (Figure \ref{nh3-n2dp-hyperfine}). When we include a second component
to the fit, we do find that the linewidths are indeed lower for each velocity component. 
We do not adopt a two component fit for the entire envelope because the 
two component fit often requires fine tuning to obtain acceptable results. Thus a
two-component fit is not feasible to use over the entire spectral cube.

\begin{small}
\bibliographystyle{apj}
\bibliography{ms}
\end{small}

\clearpage

\begin{figure}
\begin{center}
\includegraphics[scale=0.275]{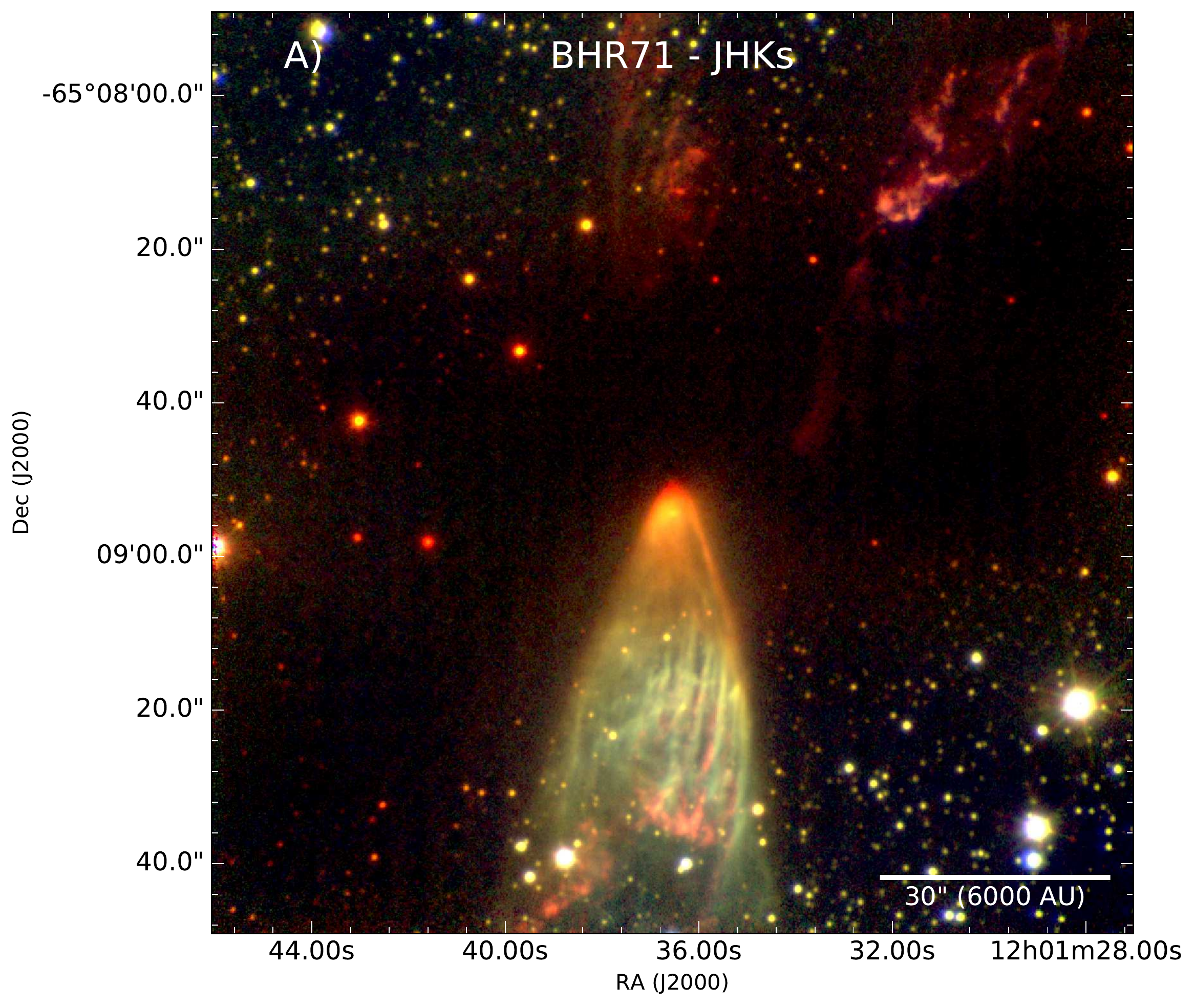}
\includegraphics[scale=0.275]{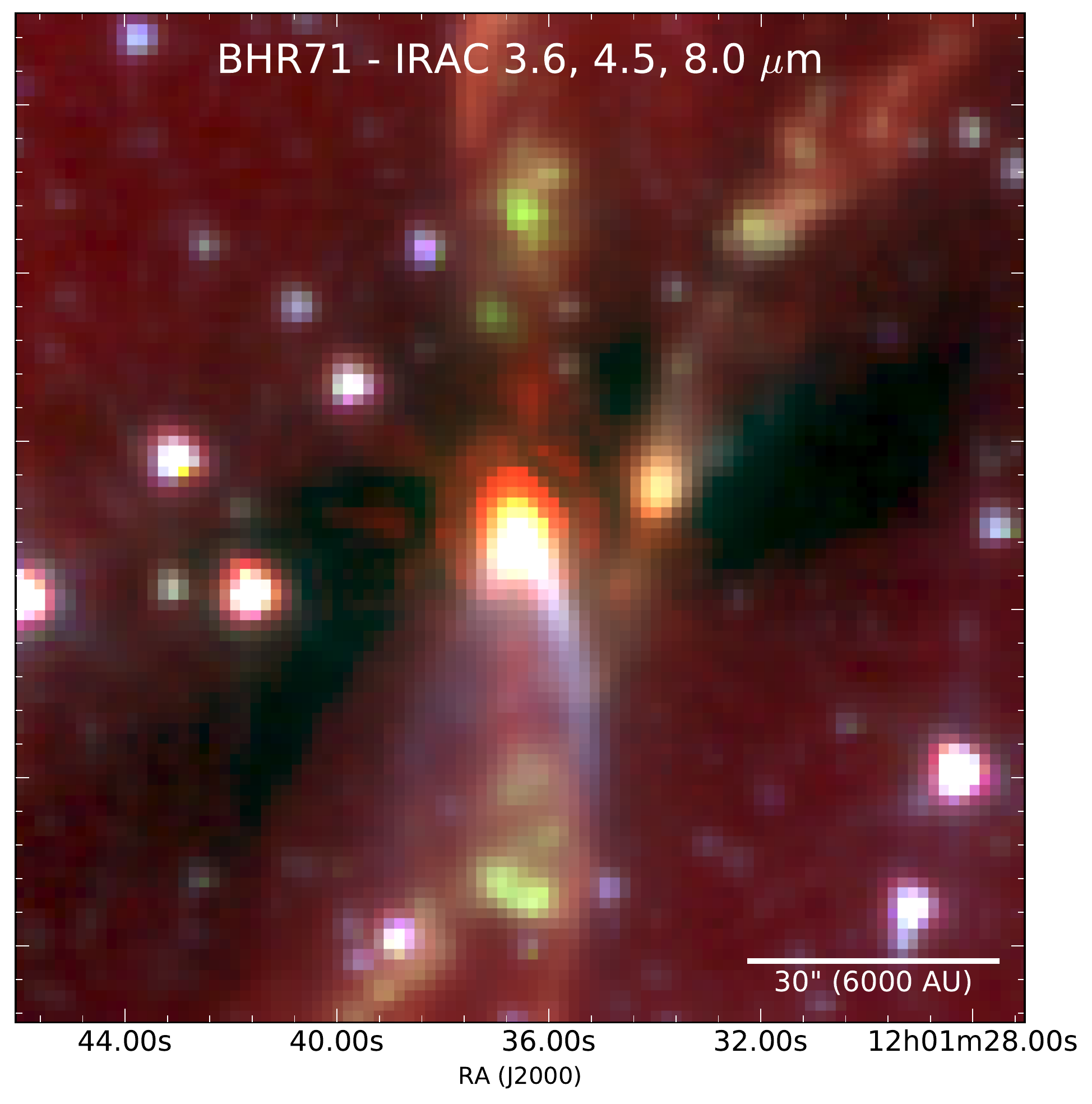}
\includegraphics[scale=0.275]{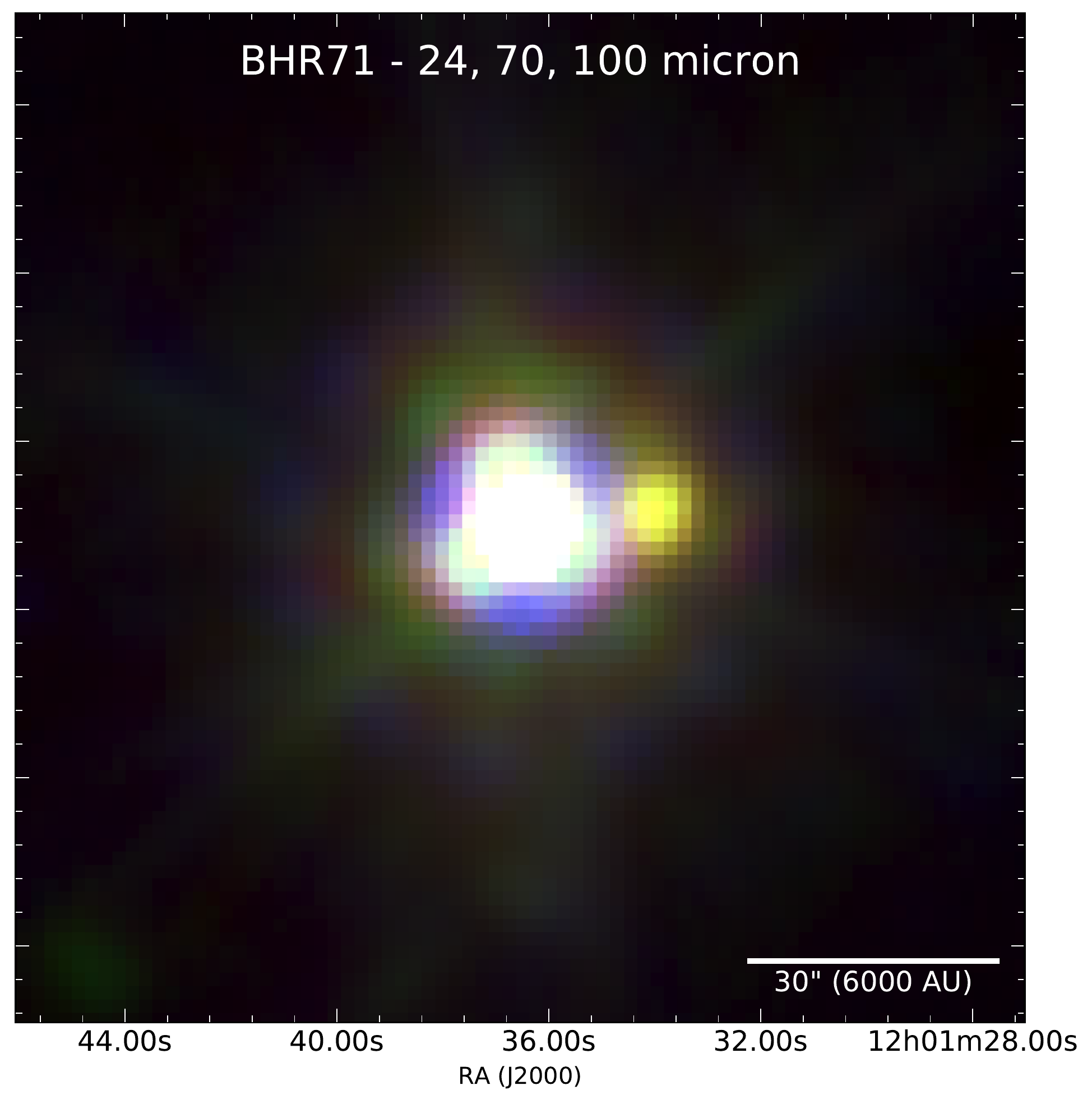}
\end{center}
\caption{False color images of BHR71 at multiple wavelengths. 
The left panel shows BHR71 as viewed in the near-infrared in Ks (PANIC), H (PANIC), and 
J-bands (ISPI). The outflow cavity from the blue-shifted outflow of IRS1 is quite prominent and
the outflow cavity from IRS2 is not distinct, but several knots are 
apparent in the northwest portion of the image. The middle panel 
shows the \textit{Spitzer} 3.6, 4.5, and 8.0~\micron\ images on the 
same scale. IRS1 is still dominant, but IRS2 is much more apparent. 
The right panel shows the \textit{Spitzer} 24~\micron, along with 
\textit{Herschel} 70 and 100~\micron\ imaging that clearly resolve 
the mid and far-infrared emission directly associated
with both IRS1 and IRS2. IRS1 is quite prominent at all three 
wavelengths, but IRS2 is not strongly detected until 70 and 100~\micron\ due to 
its deeply embedded nature. All images map the longest wavelength to 
red, the intermediate wavelength to green, and the shortest wavelength to blue.}
\end{figure}

\begin{figure}
\begin{center}
\includegraphics[scale=0.35]{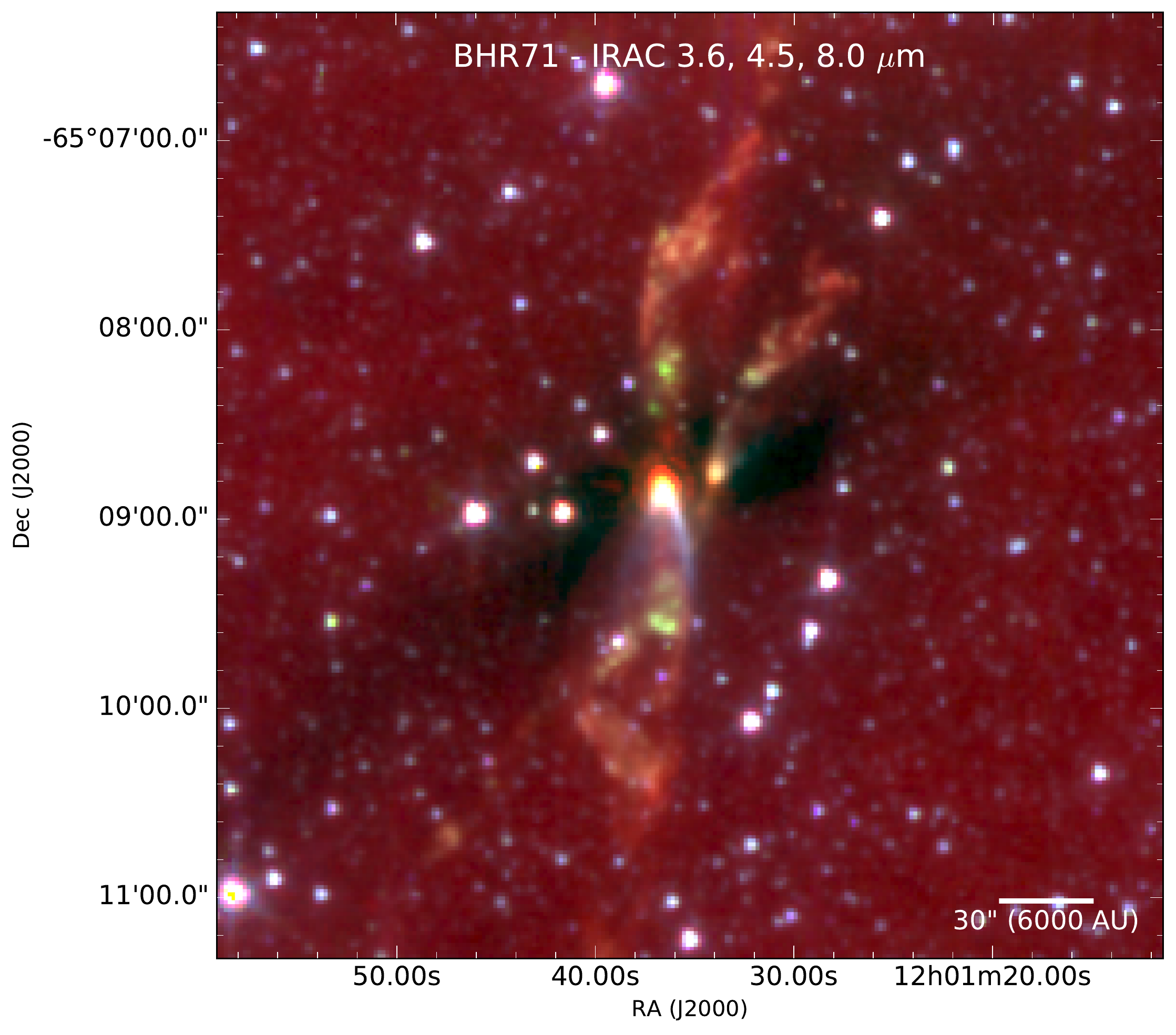}
\includegraphics[scale=0.35]{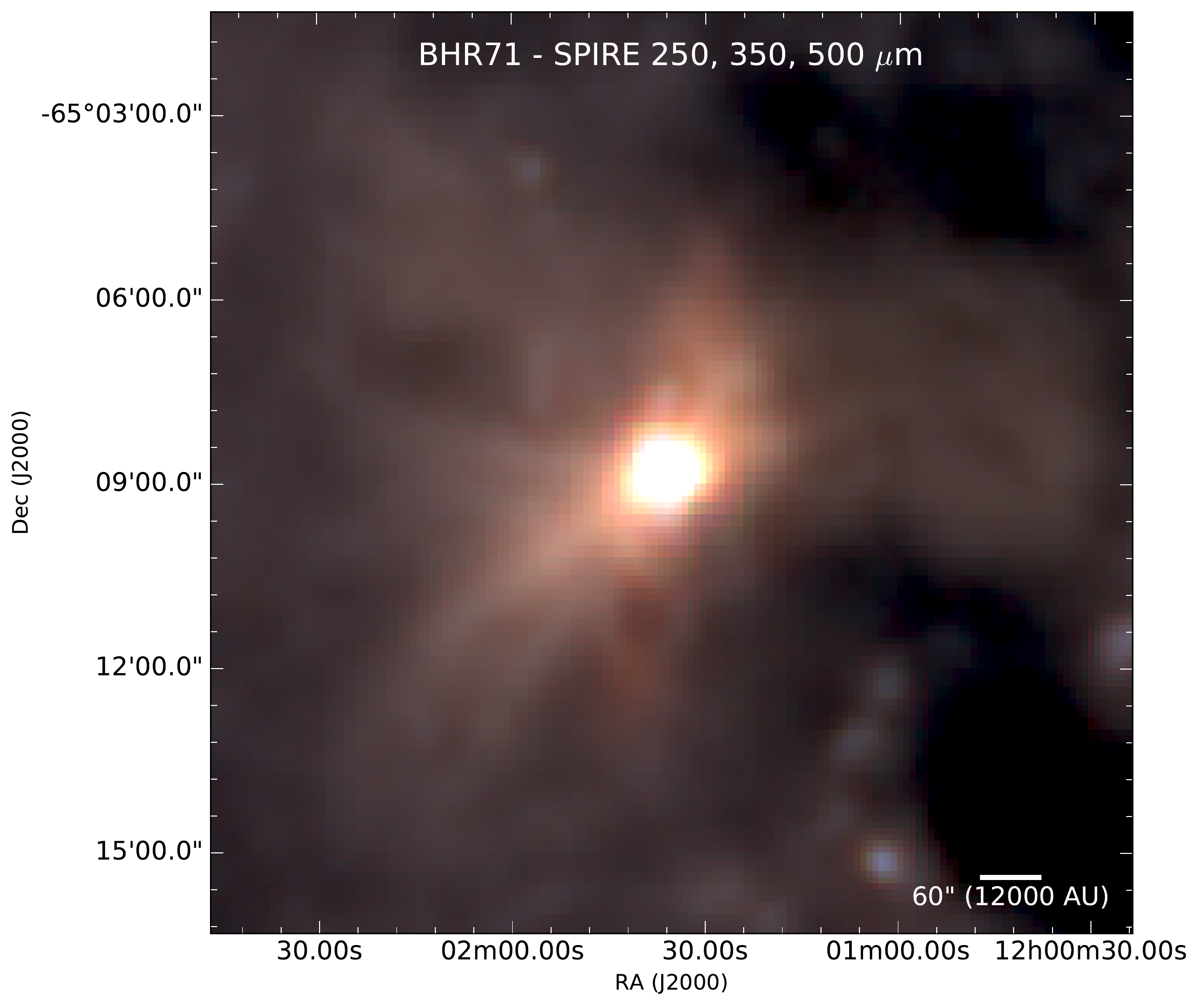}
\end{center}
\caption{False-color images of BHR71 showing a larger-scale view 
from \textit{Spitzer} at 3.6, 4.5, and 8.0~\micron\ (left panel). The 
extended envelope is apparent in absorption against the 8.0~\micron\ 
infrared background. The right panel shows the submillimeter view of 
BHR71 at 250, 350, and 500~\micron\ covering a field of view twice as 
large as in the left image. There is a red feature at 500~\micron\ in the 
outflow direction that is possibly contamination from CO ($J=5\rightarrow4$) 
emission. The core surrounding the protostars is quite extended and well-resolved 
even at the longest wavelengths.}
\end{figure}

\begin{figure}
\begin{center}
\includegraphics[scale=0.8,trim=0.5cm 12cm 2cm 0.5cm, clip=true]{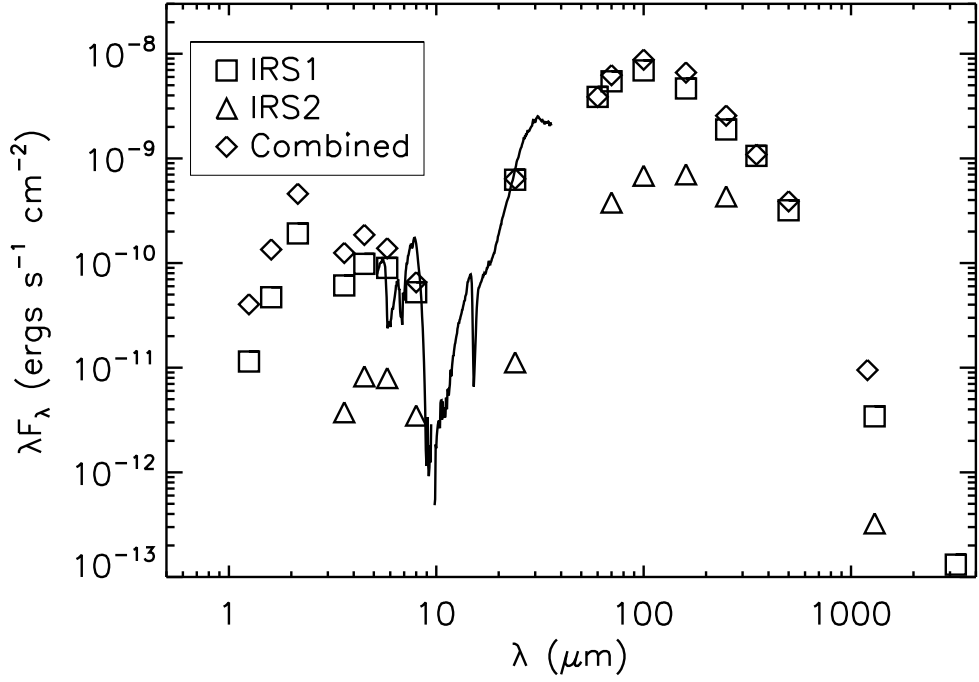}
\end{center}
\caption{Spectral energy distribution (SED) from the near-infrared to the millimeter constructed
from photometry presented in this work, as well as photometry and spectroscopy from the
literature. From the SEDs, IRS1 and IRS2 are measured to have
 L$_{bol}$ = 14.7, 1.7 L$_{\sun}$ and T$_{bol}$ = 68, 38 K, respectively.
The two protostars are well-resolved shortward of 70~\micron\ and at 
wavelengths where they can be observed with interferometry. The fluxes toward IRS1 and IRS2 
are measured with PSF photometry between 70 to 250~\micron. IRS1 is plotted with square 
symbols, IRS2 is plotted with triangles, and the combined SED is plotted with the 
diamonds. IRS1 clearly dominates the SED, but IRS2
has a lower $T_{\rm bol}$, and the SED is seen to peak at slightly longer wavelengths than
IRS1 due to its more deeply embedded nature. The solid line is the \textit{Spitzer} IRS spectrum
from \citet{yang2017}.}
\label{sed}
\end{figure}

\begin{figure}
\begin{center}
\includegraphics[scale=0.5]{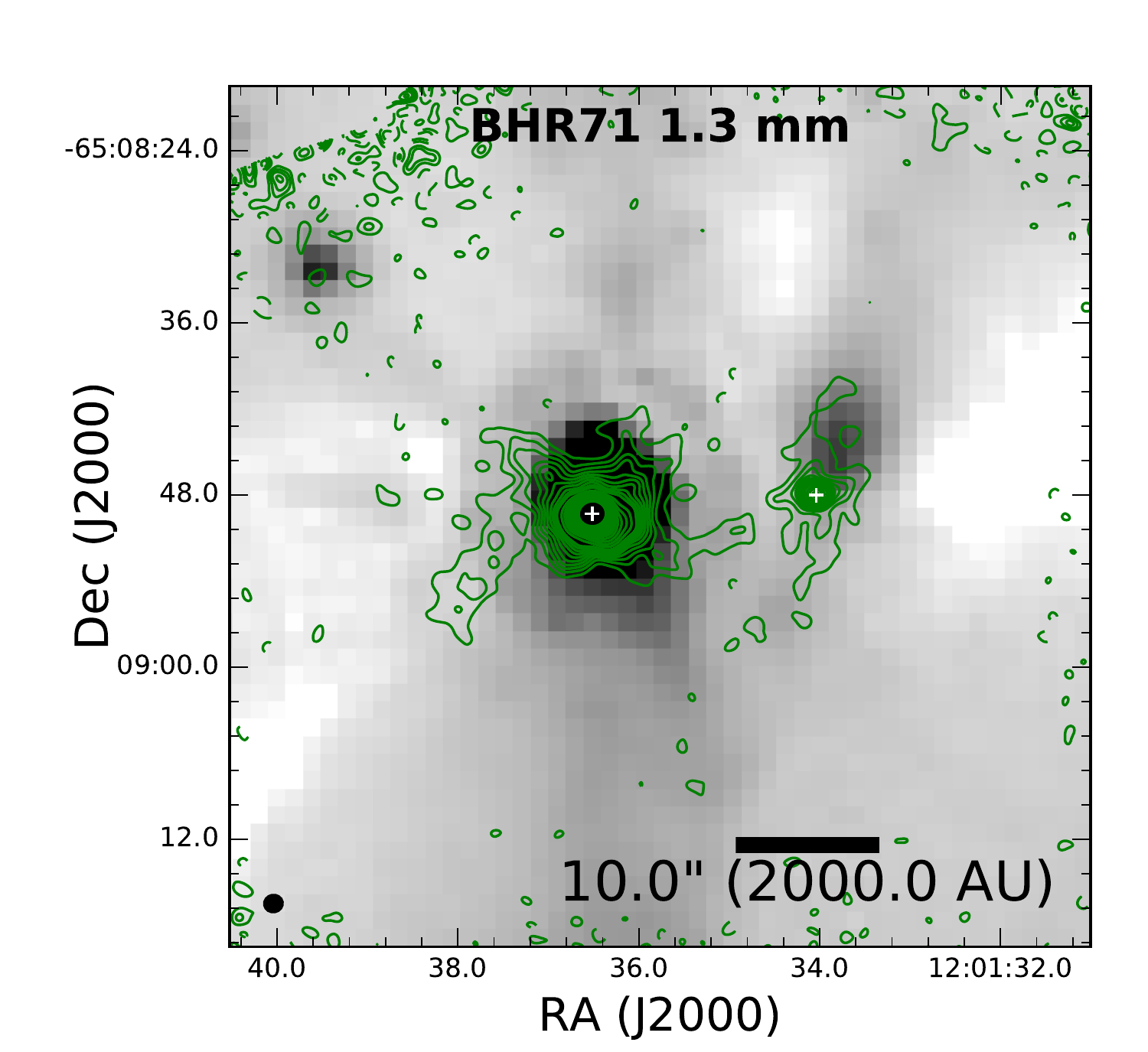}
\includegraphics[scale=0.5]{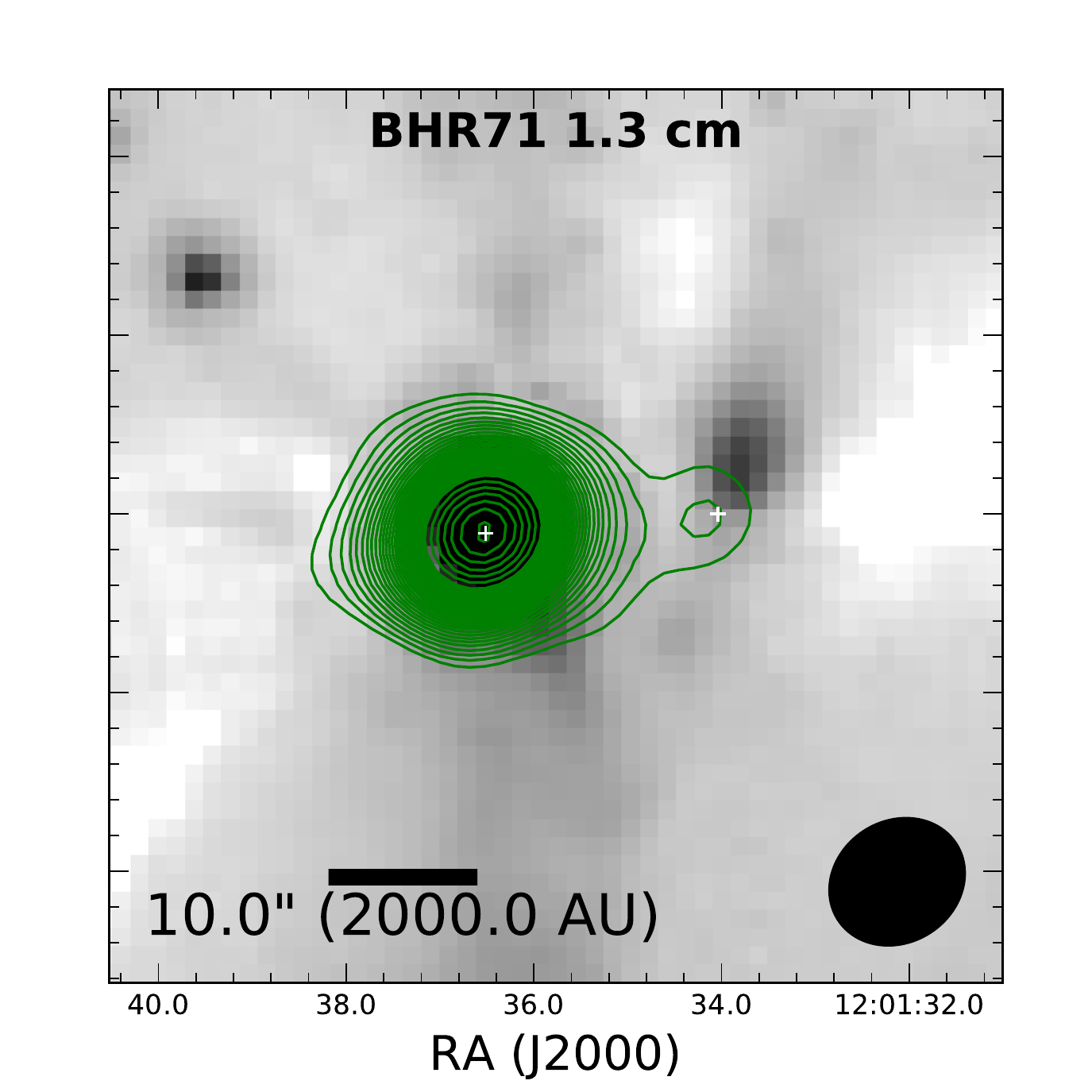}

\end{center}
\caption{ALMA 1.3~mm and ATCA 1.3~cm contours overlaid on the \textit{Spitzer} 8~\micron\
image (grayscale) are shown in the left and right panels, respectively. The ALMA image at higher
resolution shows the dust emission corresponding to the location of the brightest 8~\micron\
emission toward BHR71 IRS1, but toward IRS2, the 1.3~mm emission is located at the base of the
8~\micron\ emission. The same behavior is also shown by the ATCA 1.3~cm image. The ATCA image
might be tracing a combination of both free-free and dust emission. The ALMA 1.3~mm emission
shows extended features; IRS1 shows features extended apparently along the outflow
cavity walls, and there is an extension of emission from IRS1 toward the location of IRS2. The
contours in both panels start at 3$\sigma$ and increase on 3$\sigma$ intervals, 
where $\sigma_{1.3mm}$ = 0.5~mJy~beam$^{-1}$ and $\sigma_{1.3cm}$ = 0.1~mJy~beam$^{-1}$.
}
\end{figure}

\begin{figure}
\begin{center}
\includegraphics[scale=0.5]{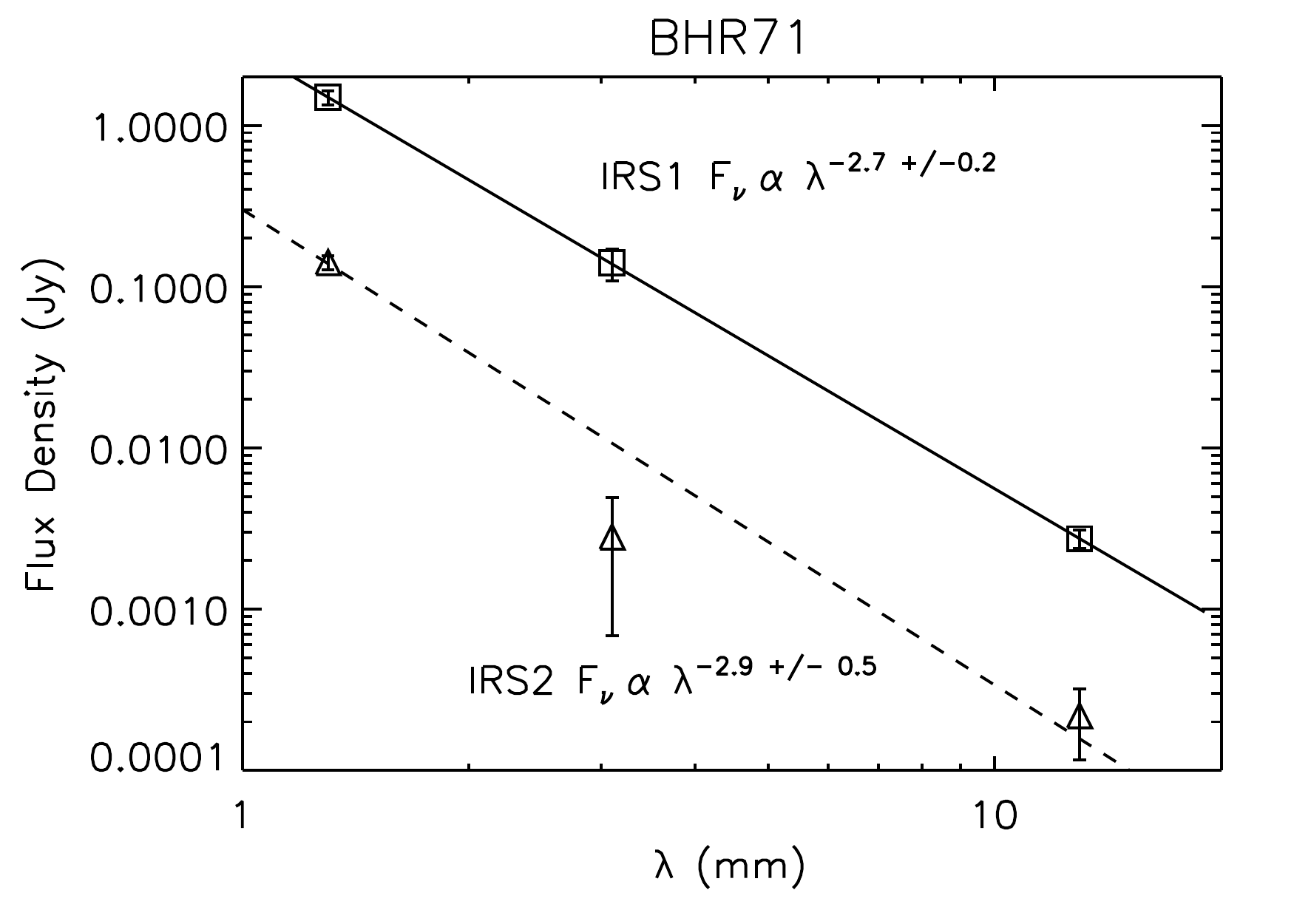}

\end{center}
\caption{Millimeter to centimeter spectrum of BHR71 IRS1 and IRS2. Both are compatible
with dust emission down to 1.3~cm, but we cannot rule out a contribution of free-free
emission at 1.3~cm toward both protostars. The 3.1~mm data point toward IRS2 seems 
quite low compared to the others, given that it would indicate an unphysically steep
spectral slope, but this is also the lowest S/N data point.}
\label{mm-spectrum}
\end{figure}

\begin{figure}
\begin{center}
\includegraphics[scale=0.4]{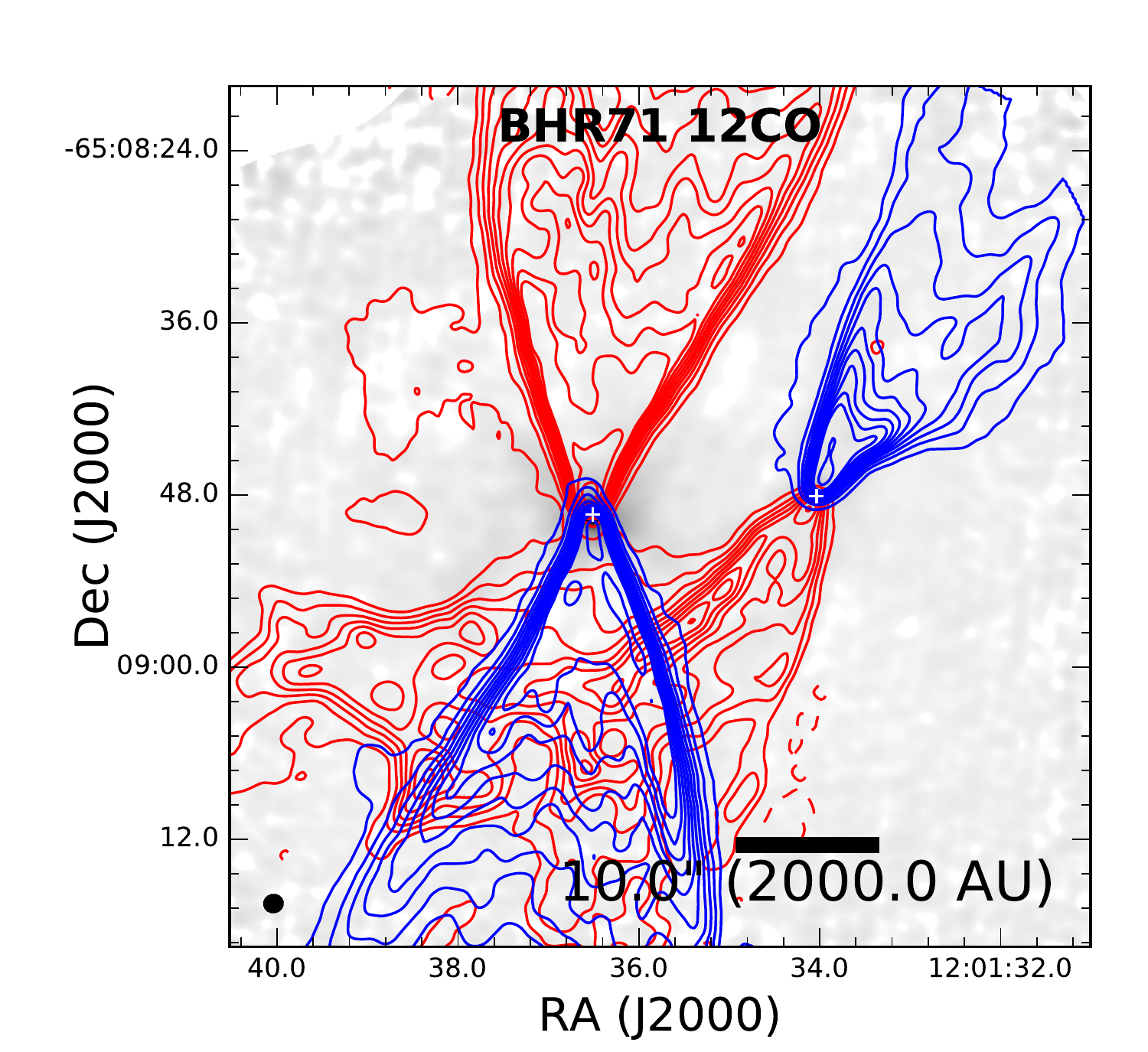}
\includegraphics[scale=0.4]{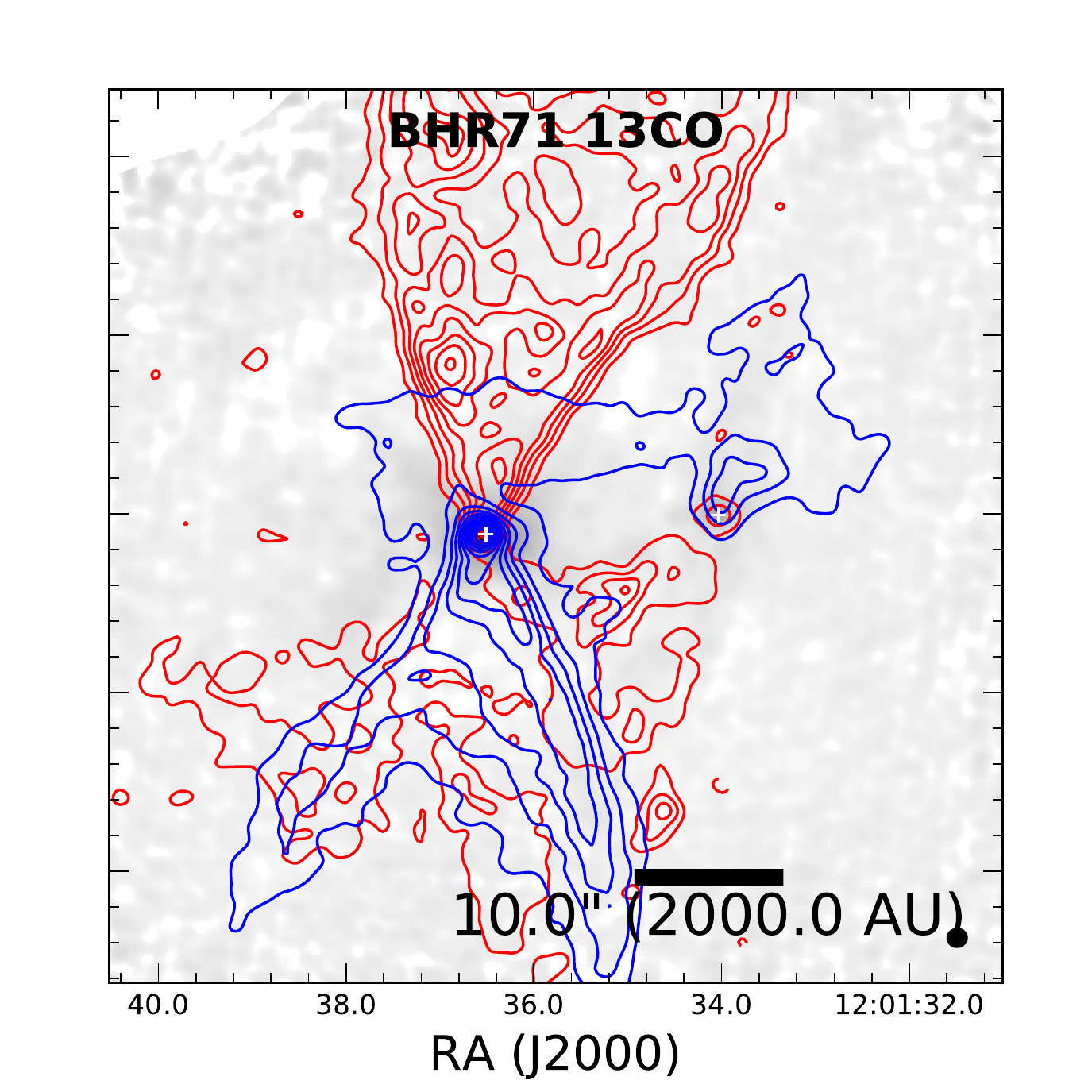}
\includegraphics[scale=0.4]{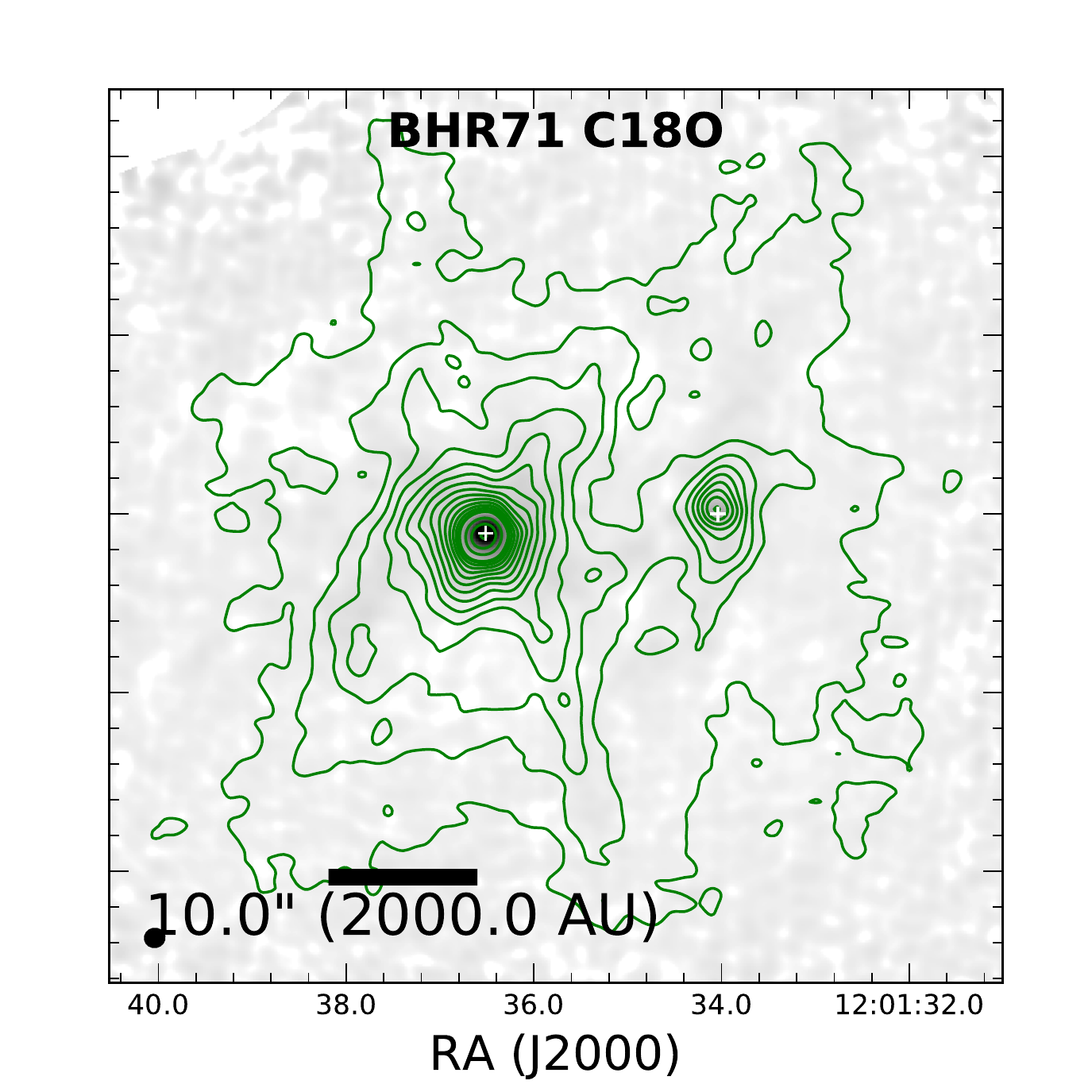}
\end{center}
\caption{Integrated intensity maps of $^{12}$CO, $^{13}$CO and C$^{18}$O ($J=2\rightarrow1$)
overlaid on the ALMA 1.3~mm continuum image (grayscale). The $^{12}$CO and $^{13}$CO
are divided into red and blue-shifted components, with corresponding contour colors; the
C$^{18}$O includes both the blue and red components. The $^{12}$CO and $^{13}$CO clearly show
the outflow emission from the protostars in both lines, with the red-shifted lobe
of the IRS2 outflow being quite broad. The outflow cavities are clearly not axisymmetric, but
the IRS1 cavity especially appears to show reflection symmetry. The C$^{18}$O is
peaked toward both protostars, tracing the warm inner envelopes.
The \twco\ is integrated between -15 to -5.6~\kms\ and -3 to 4~\kms, and
$\sigma_{blue,red}$ = 0.46, 0.4~\kkms, respectively. The blue and red contours 
start 20$\sigma$ and increase at this same interval. The
\thco\ is integrated between -7 to -4.7~\kms\ and -3.2 to -1.5~\kms, and 
$\sigma_{blue,red}$ = 0.13, 0.15~\kkms, respectively, with the blue contours starting
at 50$\sigma$ and increasing on 20$\sigma$ intervals, while the red contours 
start 20$\sigma$ and increase at this same interval. The \cateo\ is integrated between -5.8 to -3.2~\kms,
and $\sigma$ = 0.13~\kkms\ and the contours start at 15$\sigma$ and increase on 10$\sigma$ intervals.
}
\label{co-isotopes}
\end{figure}

\begin{figure}
\begin{center}
\includegraphics[scale=0.5]{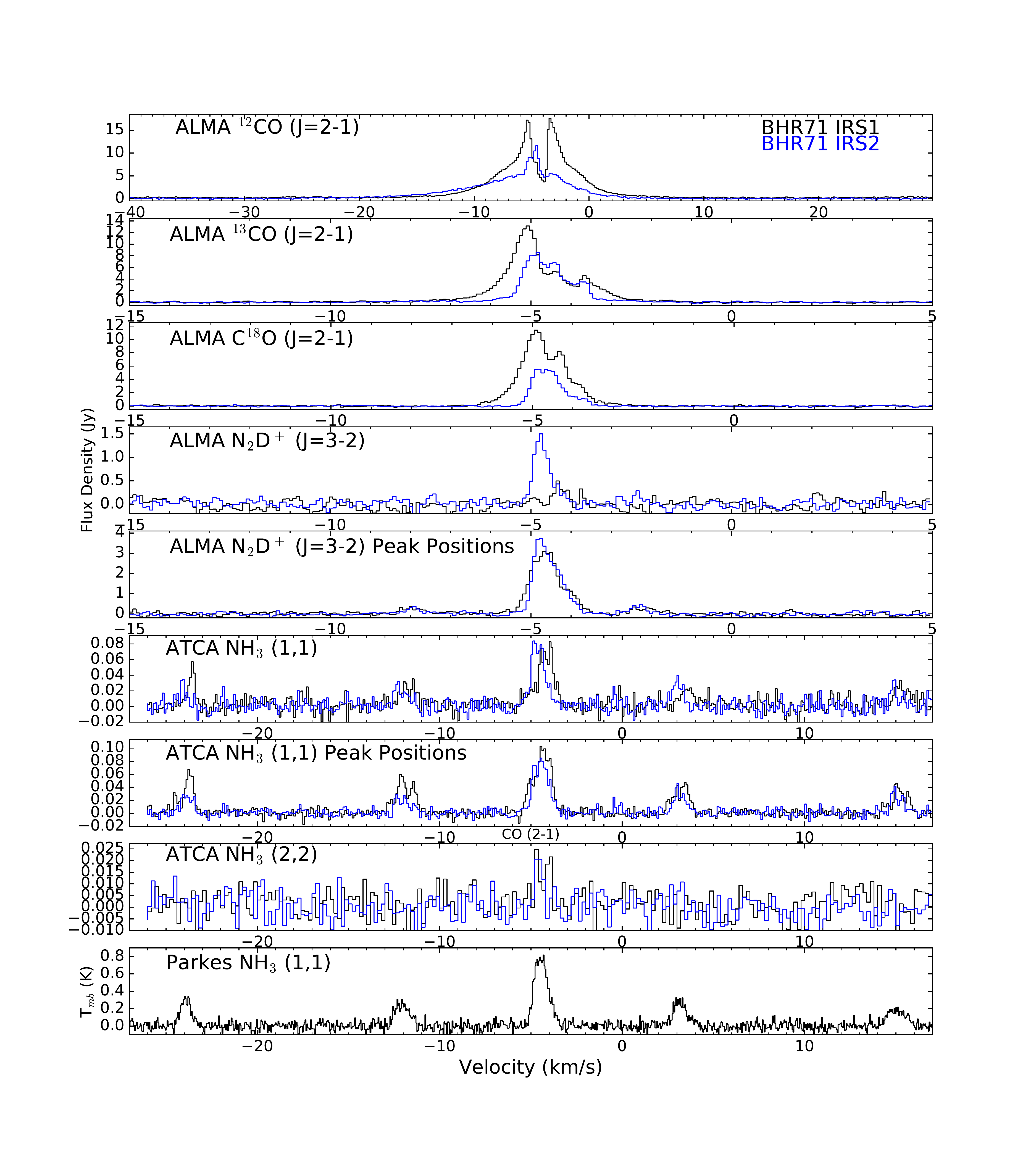}
\end{center}
\caption{Spectra toward the positions of BHR71 IRS1 and IRS2 for each molecule observed. The
ALMA spectra are extracted from a 5\arcsec\ circle centered on the protostars, the ATCA spectra
are extracted from a circle matching the synthesized beam ($\sim$10\arcsec) centered on the protostars,
and the Parkes spectrum is extracted from a 60\arcsec\ circle centered on IRS1. ALMA and ATCA
\nthp\ and \ntdp\ spectra are also extracted from the emission peaks closest to IRS1 and IRS2 in the
panels containing the text `Peak Positions.' Some spectra
are shown in different velocity ranges to highlight different features of the spectra.}
\label{spectra}
\end{figure}

\begin{figure}
\begin{center}
\includegraphics[scale=0.8]{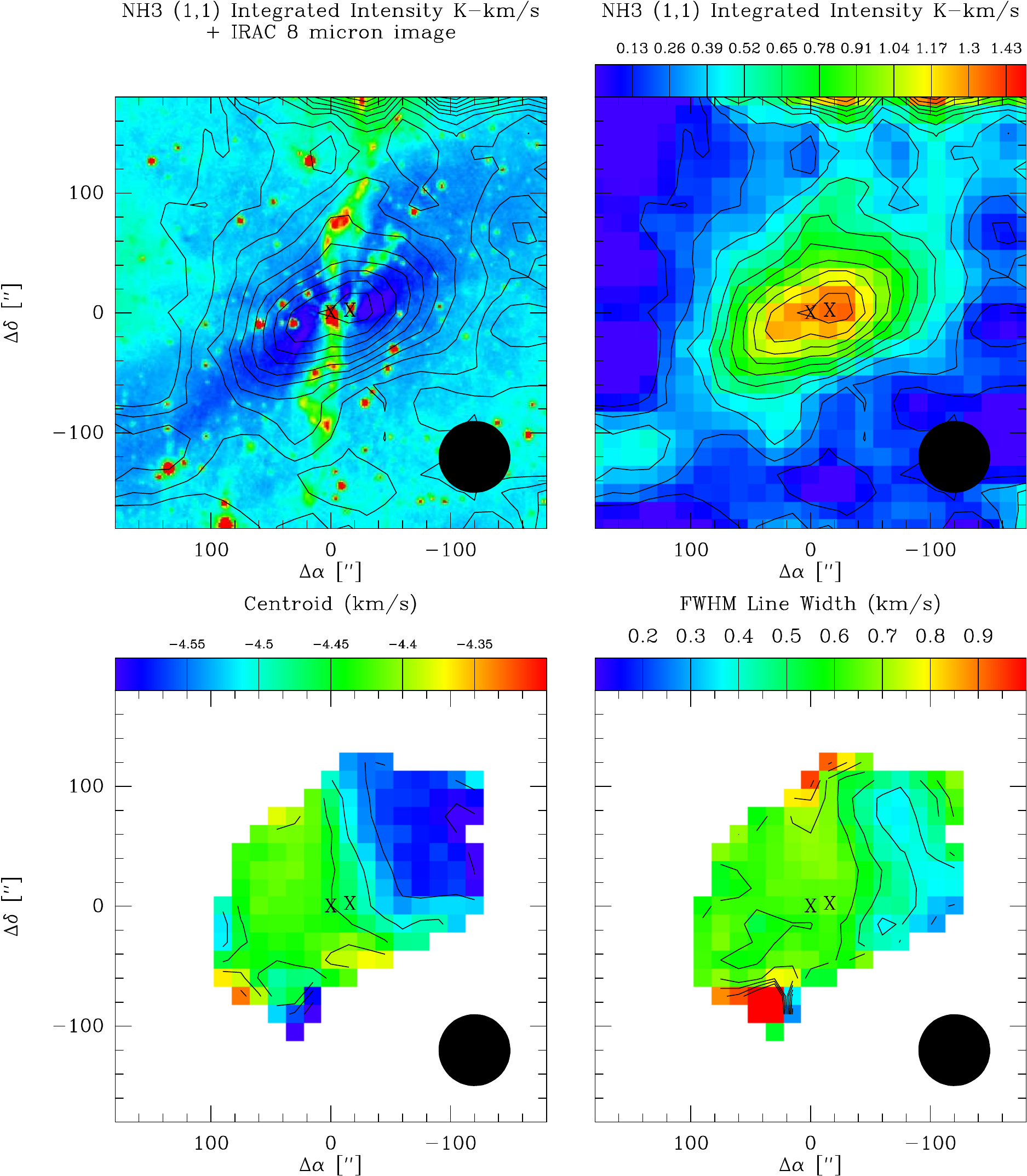}
\end{center}
\caption{Parkes single-dish NH$_3$ (1,1) maps. The upper left panel shows the integrated
intensity of the main NH$_3$ (1,1) hyperfine lines overlaid on the \textit{Spitzer} 8~\micron\
image (color scale). The upper right panel shows just the integrated
intensity of the main NH$_3$ (1,1) hyperfine lines (color scale and contours). The lower
left panel shows the line center velocity derived from a fit to all the NH$_3$ (1,1) 
hyperfine lines, and the lower right panel shows the linewidth also derived from fitting all 
hyperfine components. The integrated intensity peaks toward the position of IRS2, and
the velocity map shows a very small gradient across the source, only about a 0.15~\kms\
differential. The linewidth map shows that the lines are narrow, at most only about 0.6~\kms\
wide. The positions of the IRS1 and IRS2 are marked with X's. The intensity levels
in the top panels are from the integrated intensity map of the main hyperfine lines between -6.5 to -2.2~\kms;
the intensity levels start at and increase on 3$\sigma$ intervals, where $\sigma$=0.045~K~\kms.
}
\label{parkes-nh3}
\end{figure}

\begin{figure}
\begin{center}
\includegraphics[scale=0.5]{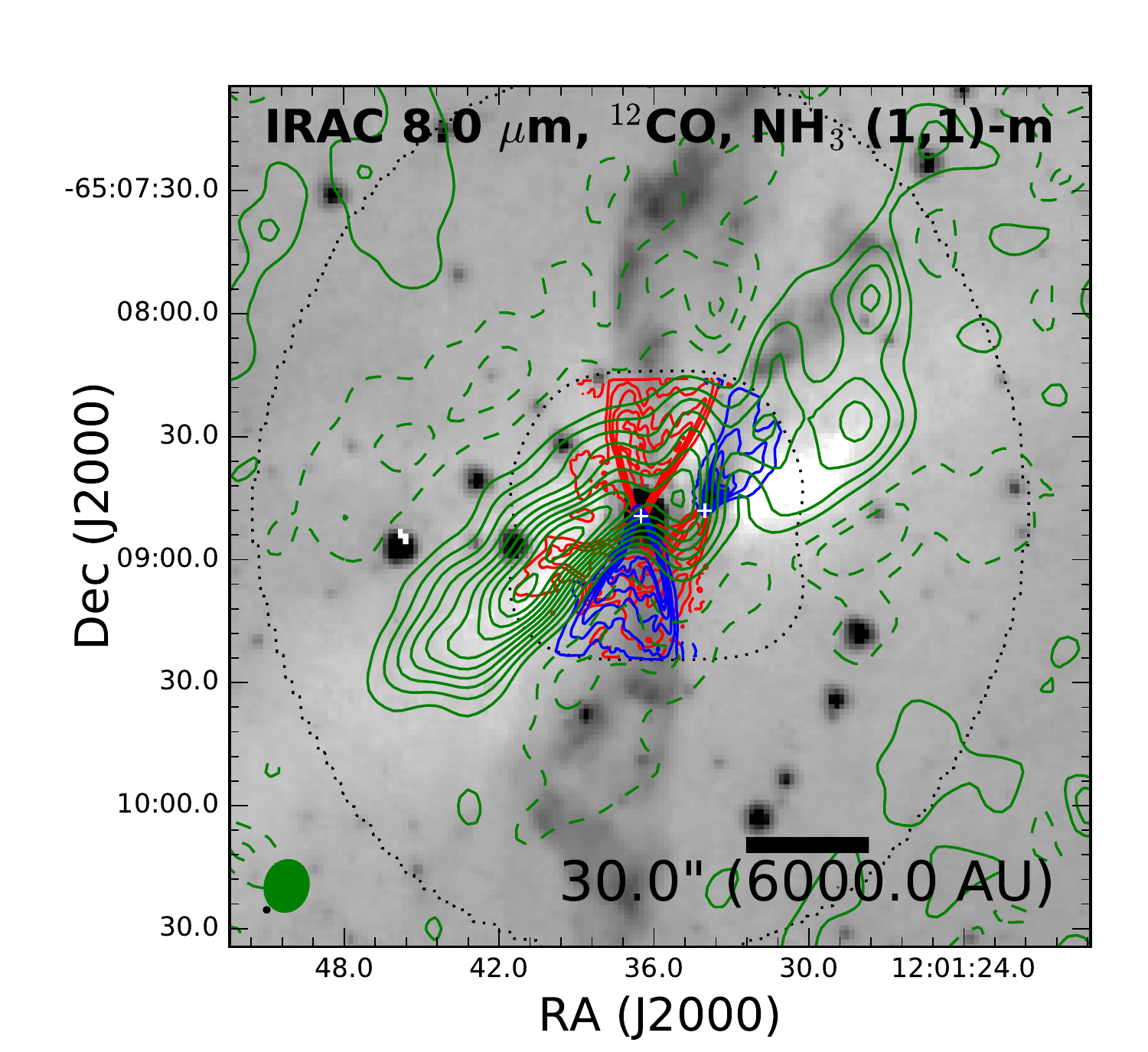}
\includegraphics[scale=0.5]{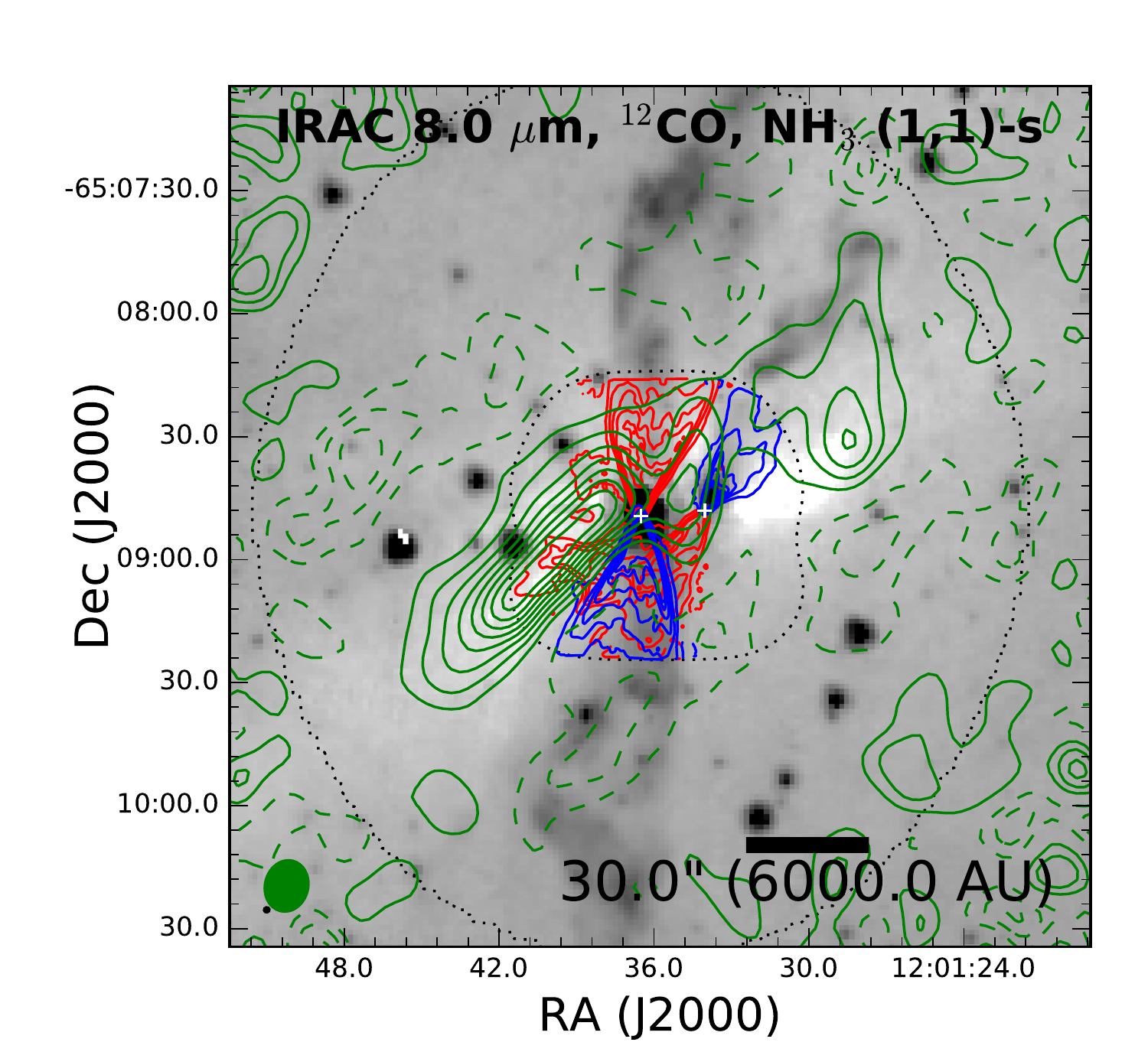}
\end{center}
\caption{The integrated intensity contours from the ATCA NH$_3$ (1,1) 
observations (green) are overlaid on the \textit{Spitzer} 8~\micron\ image (inverse grayscale).
The integrated intensity of the main \nht\ lines is shown in the left panel
 and the satellite lines in the right panel (between -12.6 to -11.4~\kms).
We also overlay the \twco\ red and blue-shifted integrated 
intensity contours (with corresponding colors) from Figure \ref{co-isotopes}, showing the relationship
of the outflow with the dense gas.
The \nht\ (1,1) emission at this higher resolution shows excellent correspondence with 
the 8~\micron\ absorption feature of the envelope on the background, and
the \nht\ maps exhibit a cavity and depressions toward the location of IRS1. IRS2 is located
directly beside a \nht\ peak. The \nht\ main and (satellite)
contours start at 5$\sigma$ (3$\sigma$) and increase on 5$\sigma$ (3$\sigma$) intervals, 
where $\sigma$=0.075~\kkms\ (0.072~\kkms), a -5$\sigma$ (-3$\sigma$) contour is also drawn.}
\label{atca-8micron}
\end{figure}

\begin{figure}
\begin{center}
\includegraphics[scale=0.8]{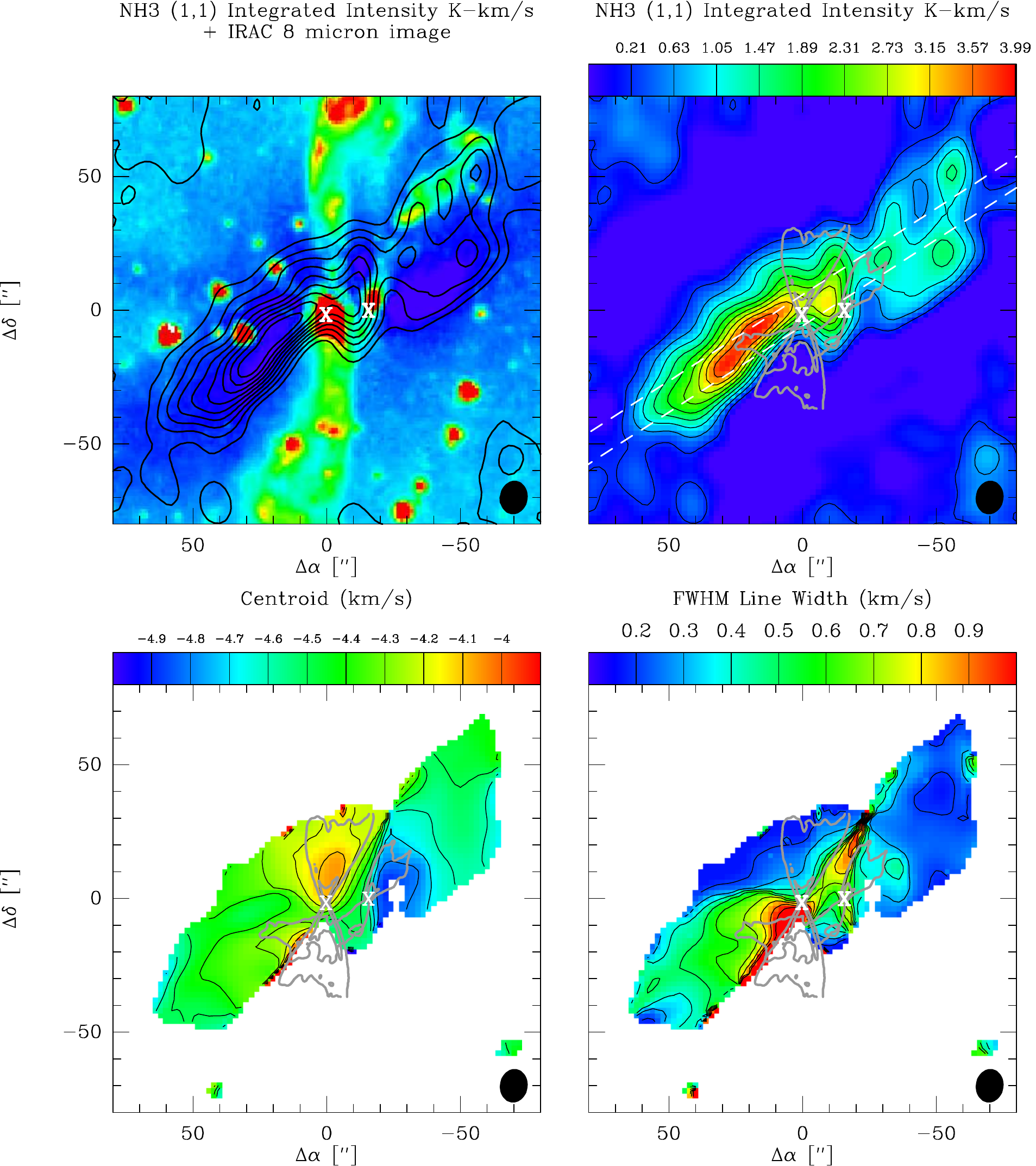}

\end{center}
\caption{Same as for Figure \ref{parkes-nh3}, but for the ATCA \nht\ observations. 
The spatial extents of the outflows probed by ALMA in \twco\ are marked 
with the gray contours in all panels but the top left. The 
velocity map shows considerably more structure than the Parkes map. There is a red-shifted
component that appears to reflect the red-shifted outflow cavity of IRS1, then there is
an abrupt velocity gradient between IRS1 and IRS2. Some of the velocity gradient is likely related 
to the outflow influence, but even without the additional velocity structure
of the outflow, the gradient is quite strong. Furthermore, the gradient also appears as
a region of large linewidth which results from the line velocity changing on the scale of
one beamwidth. There is also some evidence of outflow influence on the south side where there 
is increased linewidth where the eastern edge of the IRS1 outflow cavity is 
located, in addition to the red-shifted side of the IRS2 outflow. However, there is
no increased linewidth toward the red-shifted side of the outflow where there appears to be a shift
in the line-center map. The intensity levels
in the top panels are from the integrated intensity map of the main hyperfine lines between -5.0 to -3.5~\kms;
the intensity levels start at 3$\sigma$ and increase on 6$\sigma$ intervals, where $\sigma$=0.07~\kkms.  The dashed lines
in the upper right panel marks the region of position-velocity (PV) 
extraction.
}
\label{atca-nh3}
\end{figure}

\begin{figure}
\begin{center}
\includegraphics[scale=0.8]{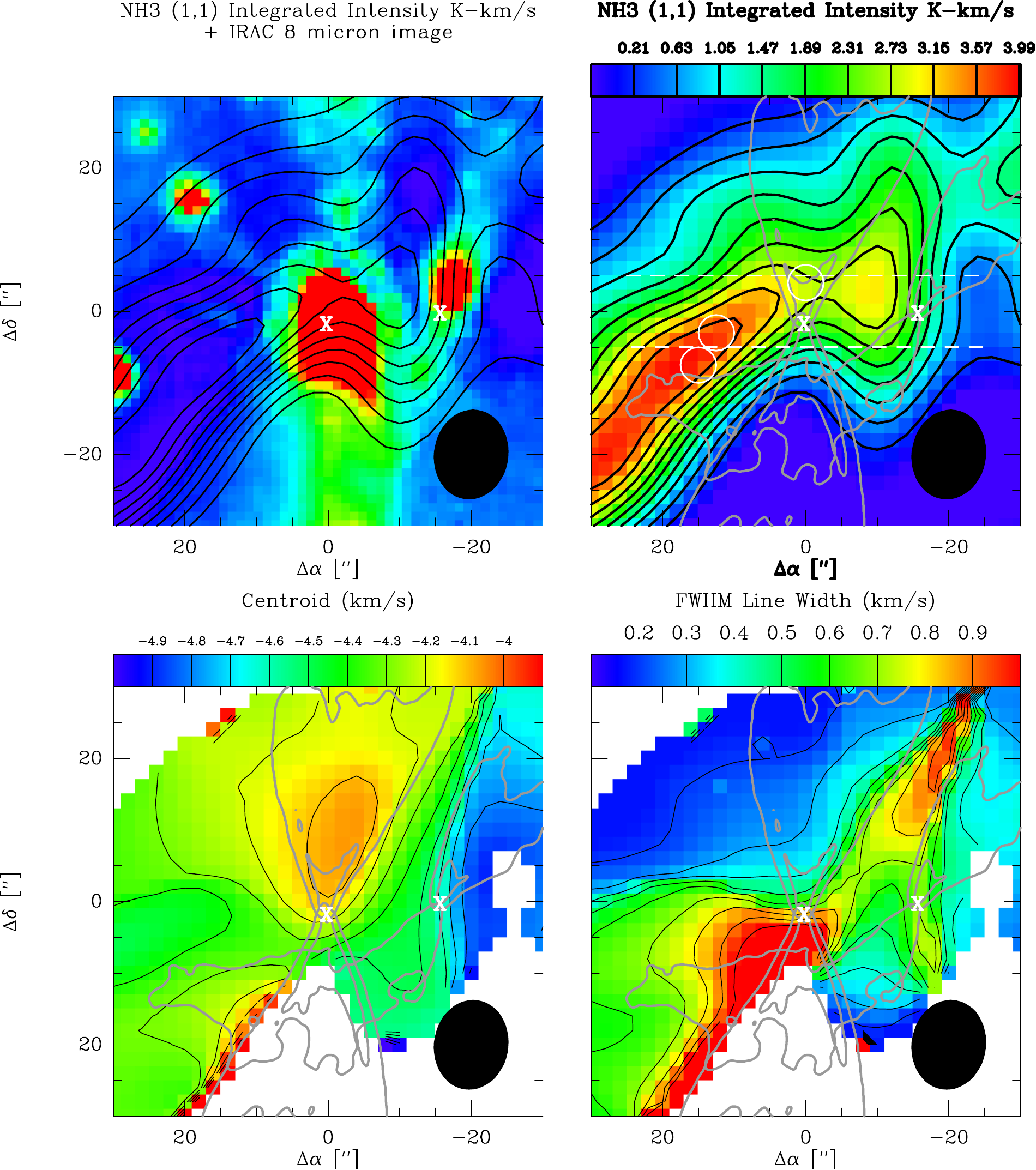}
\end{center}
\caption{Same as Figure \ref{atca-nh3} but zooming in.
The dashed circles in the upper right panel correspond to regions of spectral 
extraction to examine the quality of hyperfine fitting, see Appendix A.
}
\label{atca-nh3-zoom}
\end{figure}

\begin{figure}
\begin{center}
\includegraphics[scale=0.5]{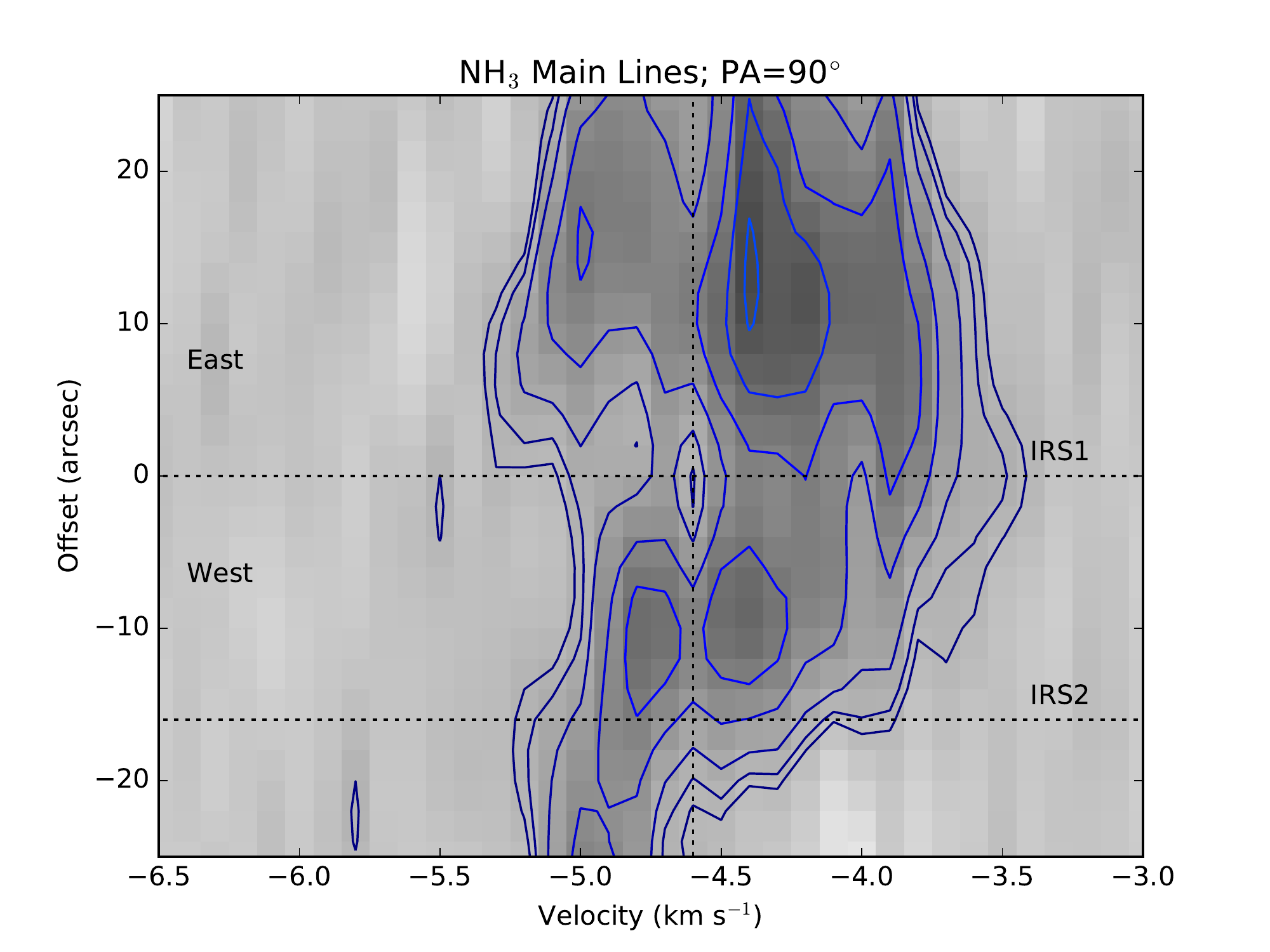}
\includegraphics[scale=0.5]{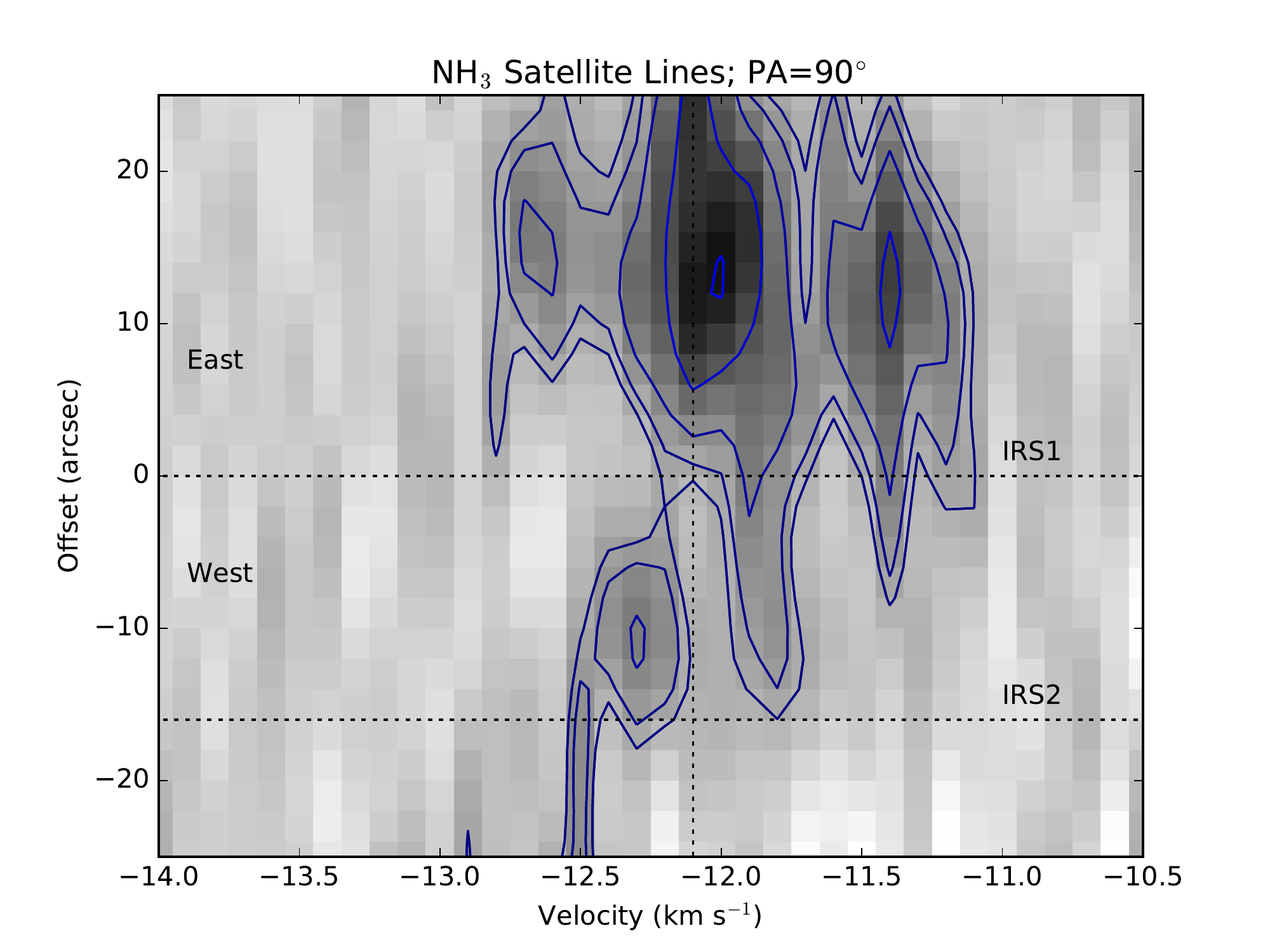}
\end{center}
\caption{Position-velocity diagrams of \nht\ taken in a 10\arcsec\ wide strip in the
east-west direction, centered on IRS1. The region extracted is bounded
by the dashed lines in the upper right panel of Figure \ref{atca-nh3-zoom}.
The top panel shows the PV diagram of the
main \nht\ hyperfine lines, and the bottom panel shows a PV diagram of a
set of \nht\ satellite hyperfine lines that are not as closely spaced as the main lines.
}
\label{atca-nh3-pv}
\end{figure}

\begin{figure}
\begin{center}
\includegraphics[scale=0.5]{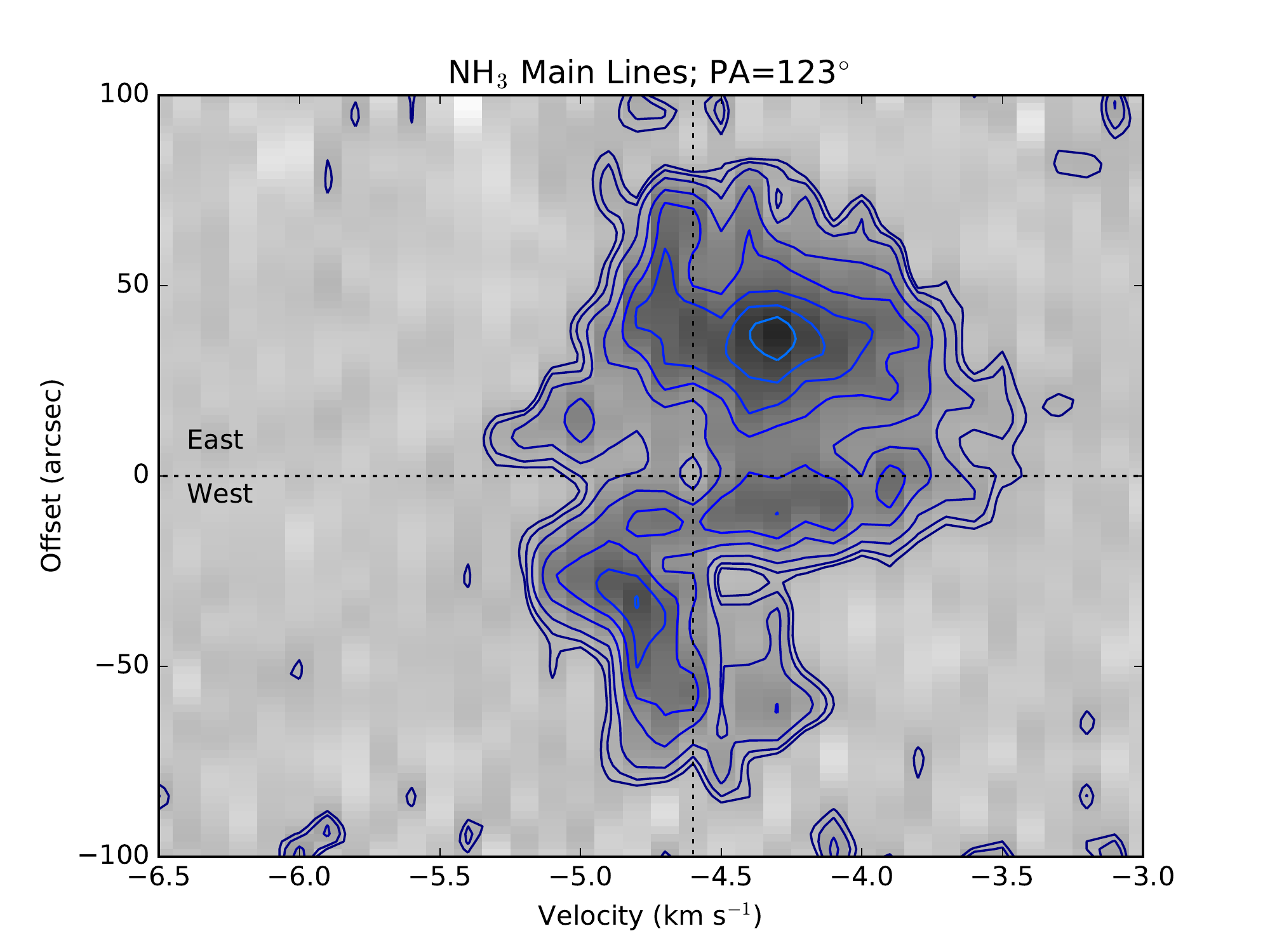}
\includegraphics[scale=0.5]{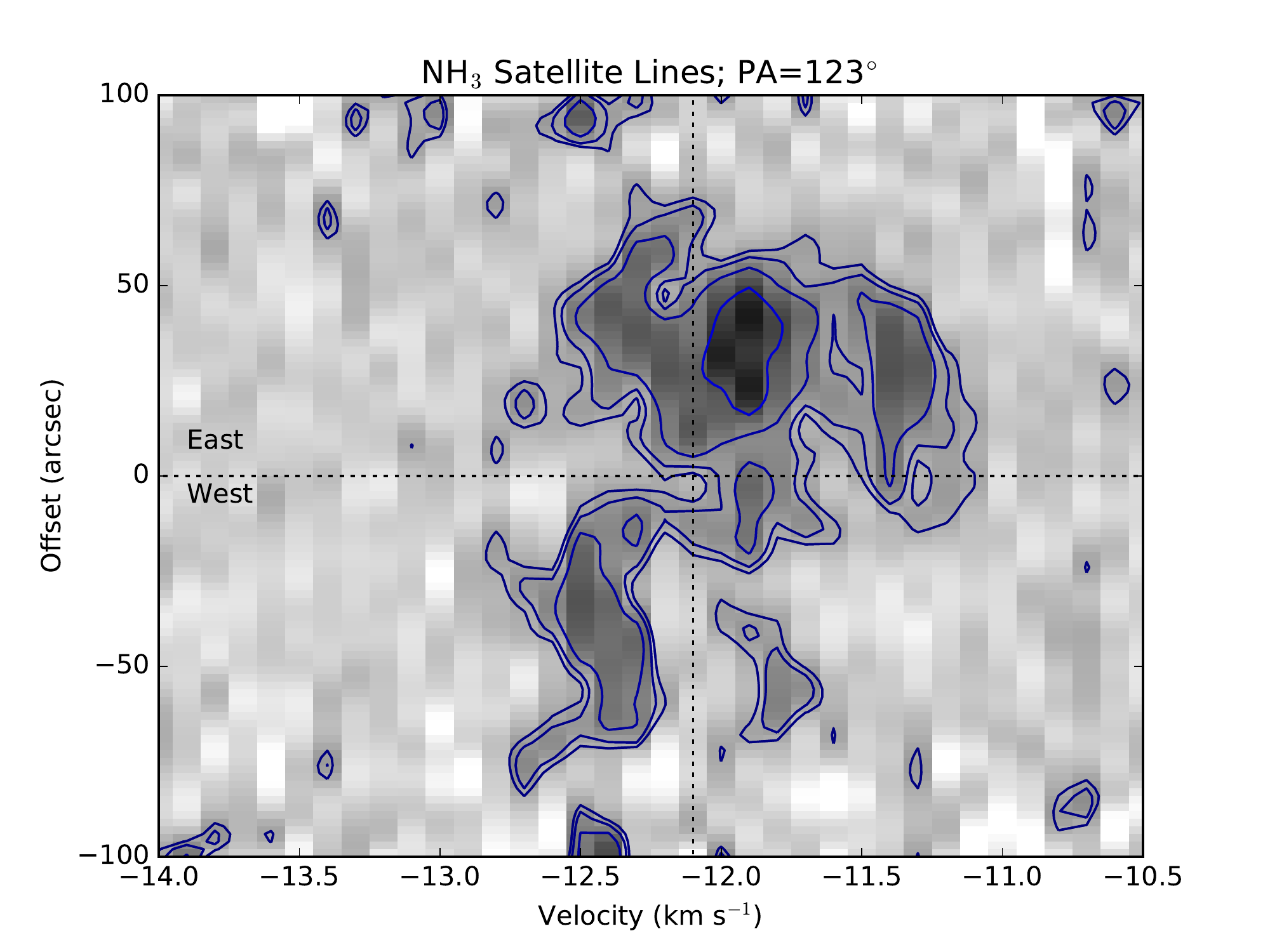}
\end{center}
\caption{Position-velocity diagrams of \nht\ taken in a 10\arcsec\ wide strip 
along a position angle of 123\degr\ east of north, centered on IRS1.
The region extracted is bounded
by the dashed lines in the upper right panel of Figure \ref{atca-nh3}.
The top panel shows the PV diagram of the
main \nht\ hyperfine lines, and the bottom panel shows a PV diagram of a
set of \nht\ satellite hyperfine lines that are not as closely spaced as the main lines.
}
\label{atca-nh3-pv-diag}
\end{figure}

\begin{figure}
\begin{center}
\includegraphics[scale=0.5]{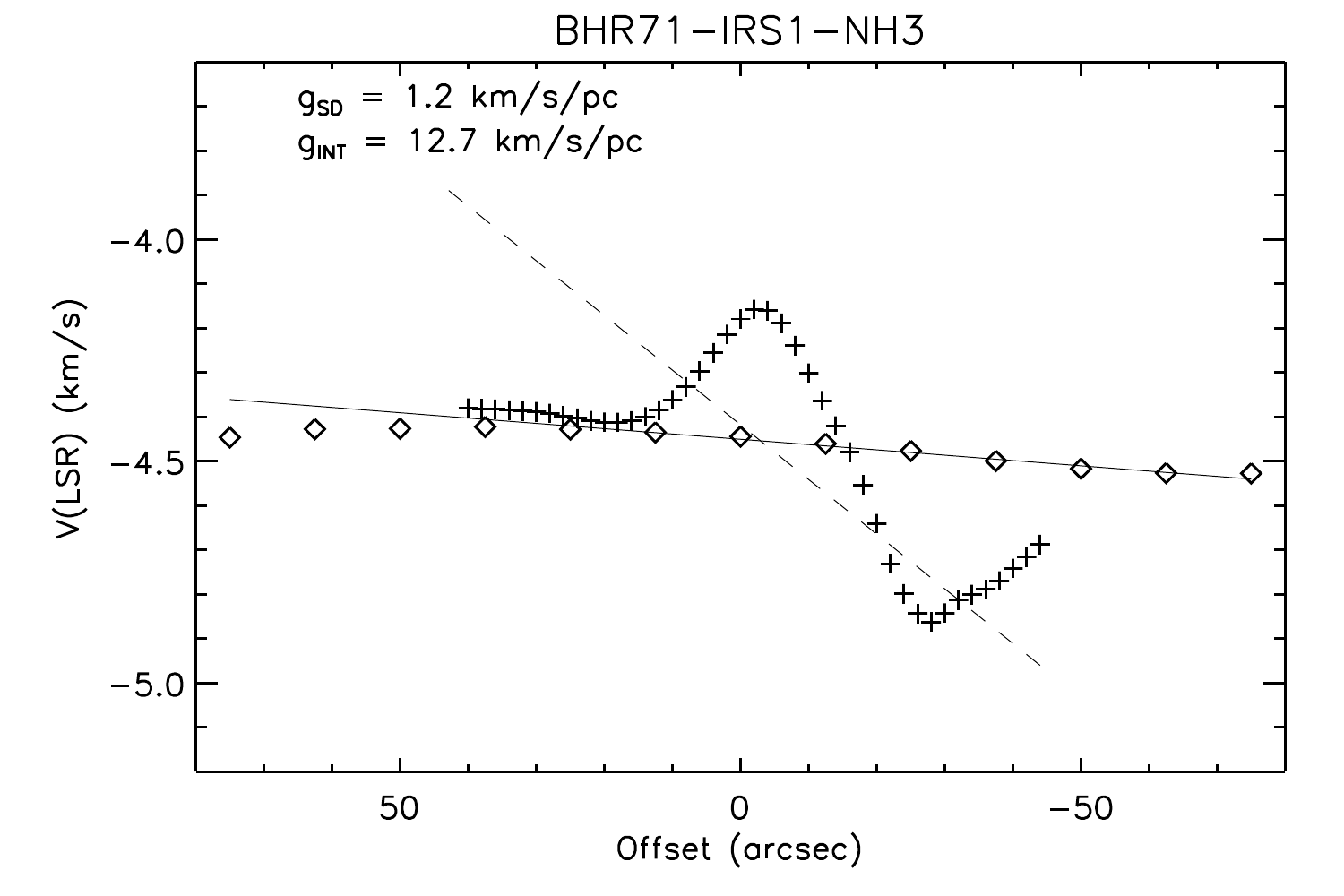}
\includegraphics[scale=0.5]{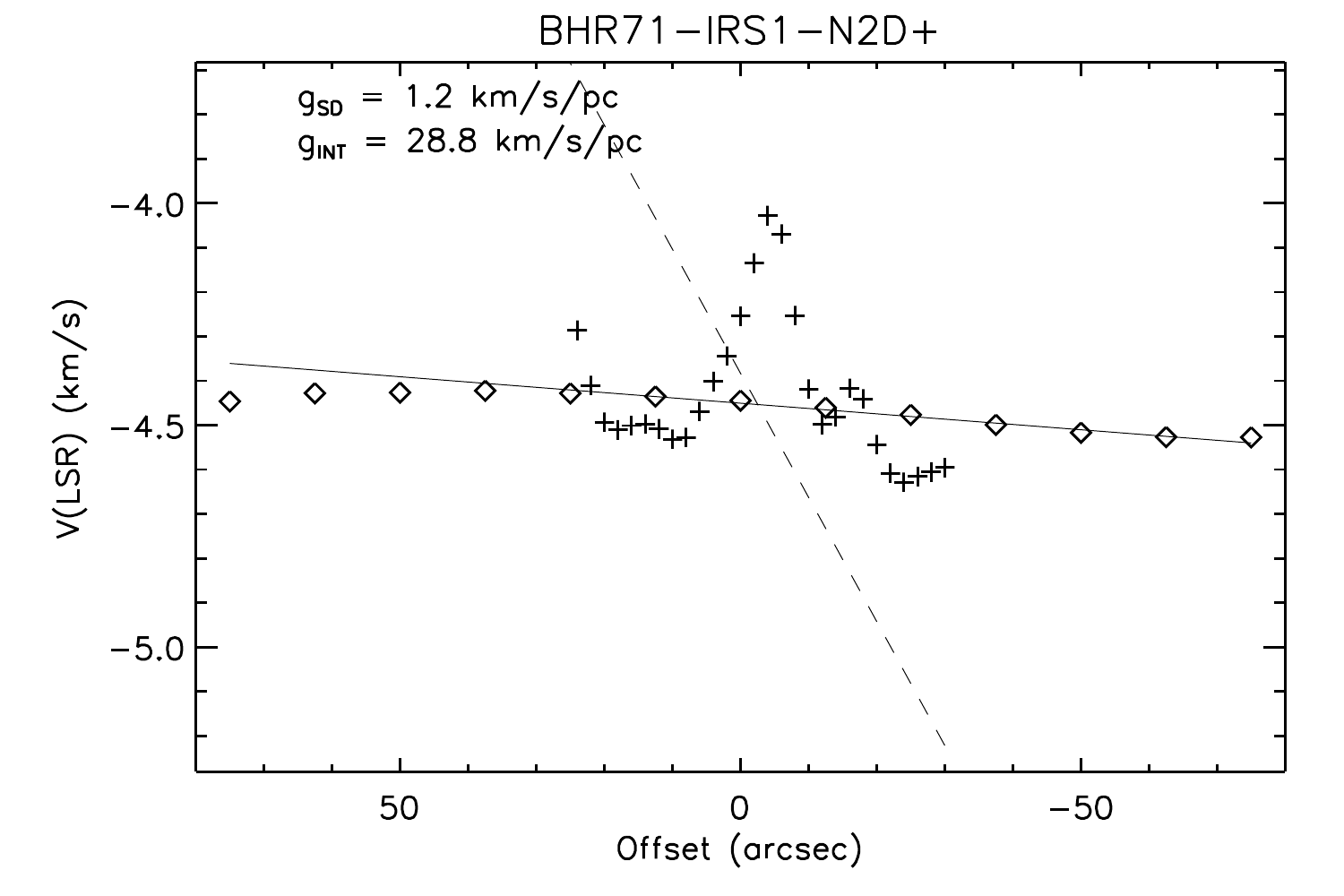}
\includegraphics[scale=0.5]{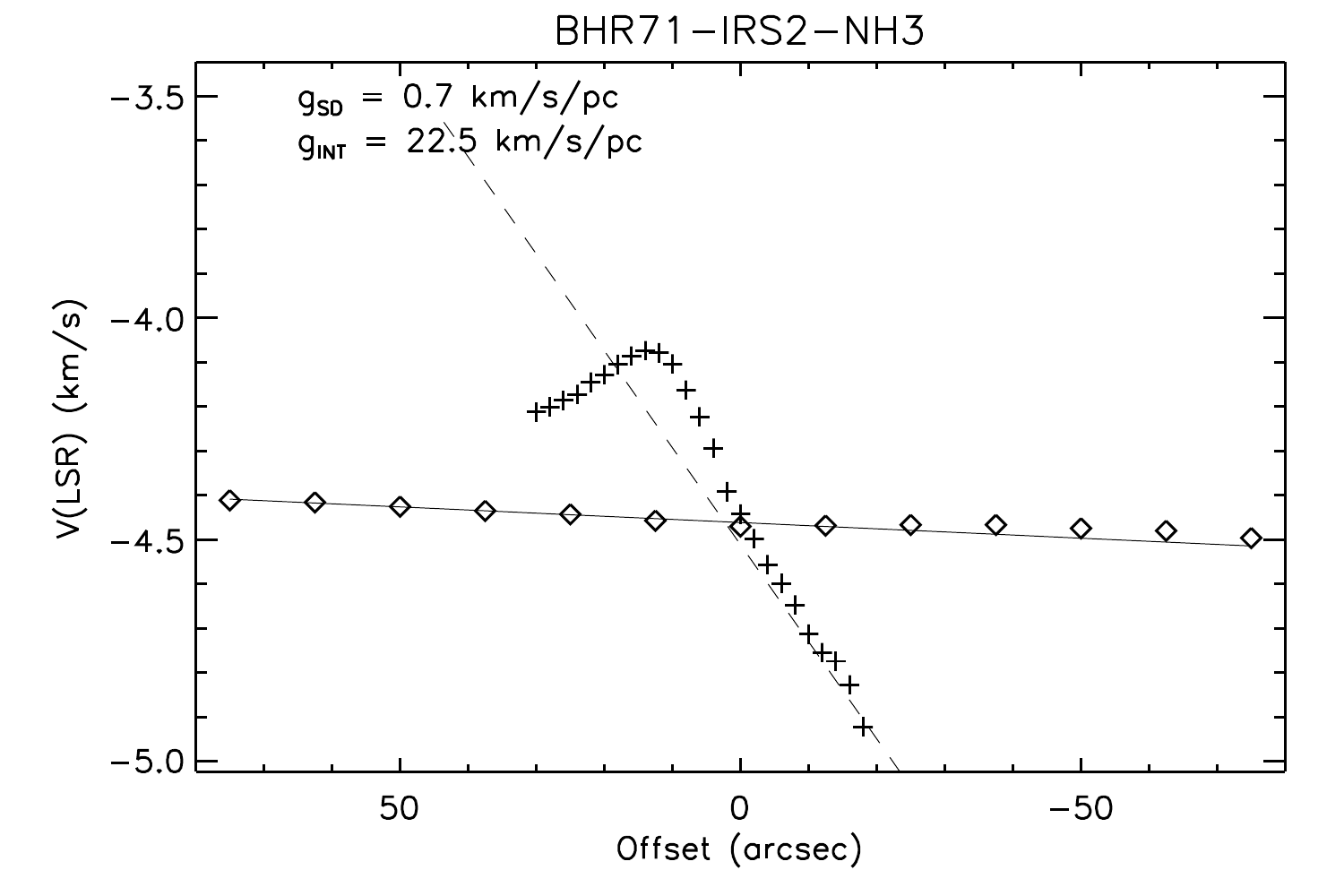}
\includegraphics[scale=0.5]{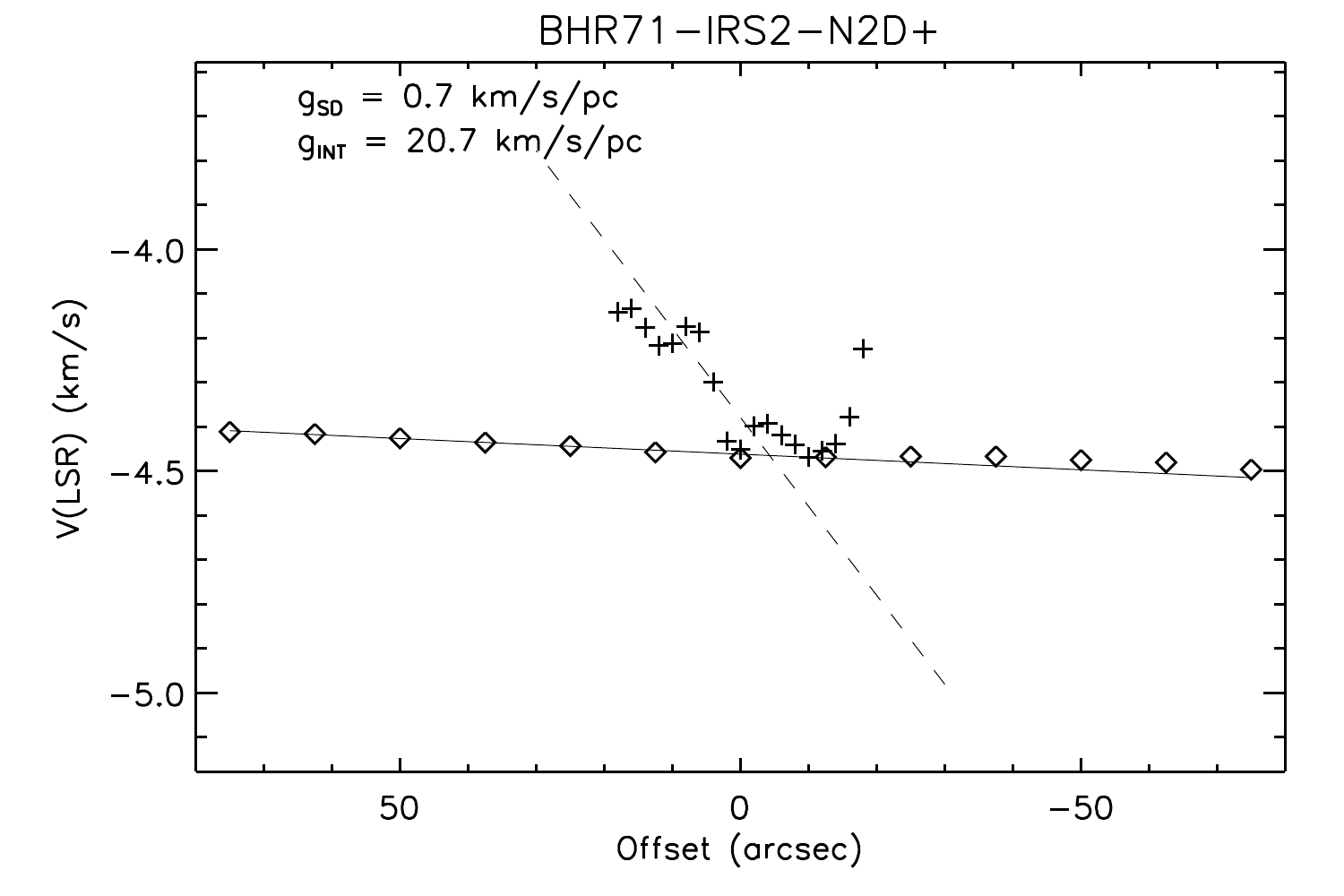}
\end{center}
\caption{Velocity profiles extracted orthogonal to the outflow directions for
IRS1 (top row) and IRS2 (bottom row). 
Velocity profiles are shown for both \nht\ (1,1) (left panels) and \ntdp\ (right panels). All panels also show the profile
extracted from the Parkes \nht\ data (diamonds) which is fit with a line (solid line). The single-dish data show a gradual
velocity gradient and the interferometer data (plus signs) show more structure. The \ntdp\ 
data may have different centroid velocities relative
to the \nht. This could be due to \ntdp\ tracing a different region of the cloud
due to chemistry. The velocity gradients in the interferometer data are fit (dashed lines), but
their structure limits the utility of the values derived.
}
\label{1d-profiles}
\end{figure}

\begin{figure}
\begin{center}
\includegraphics[scale=0.5]{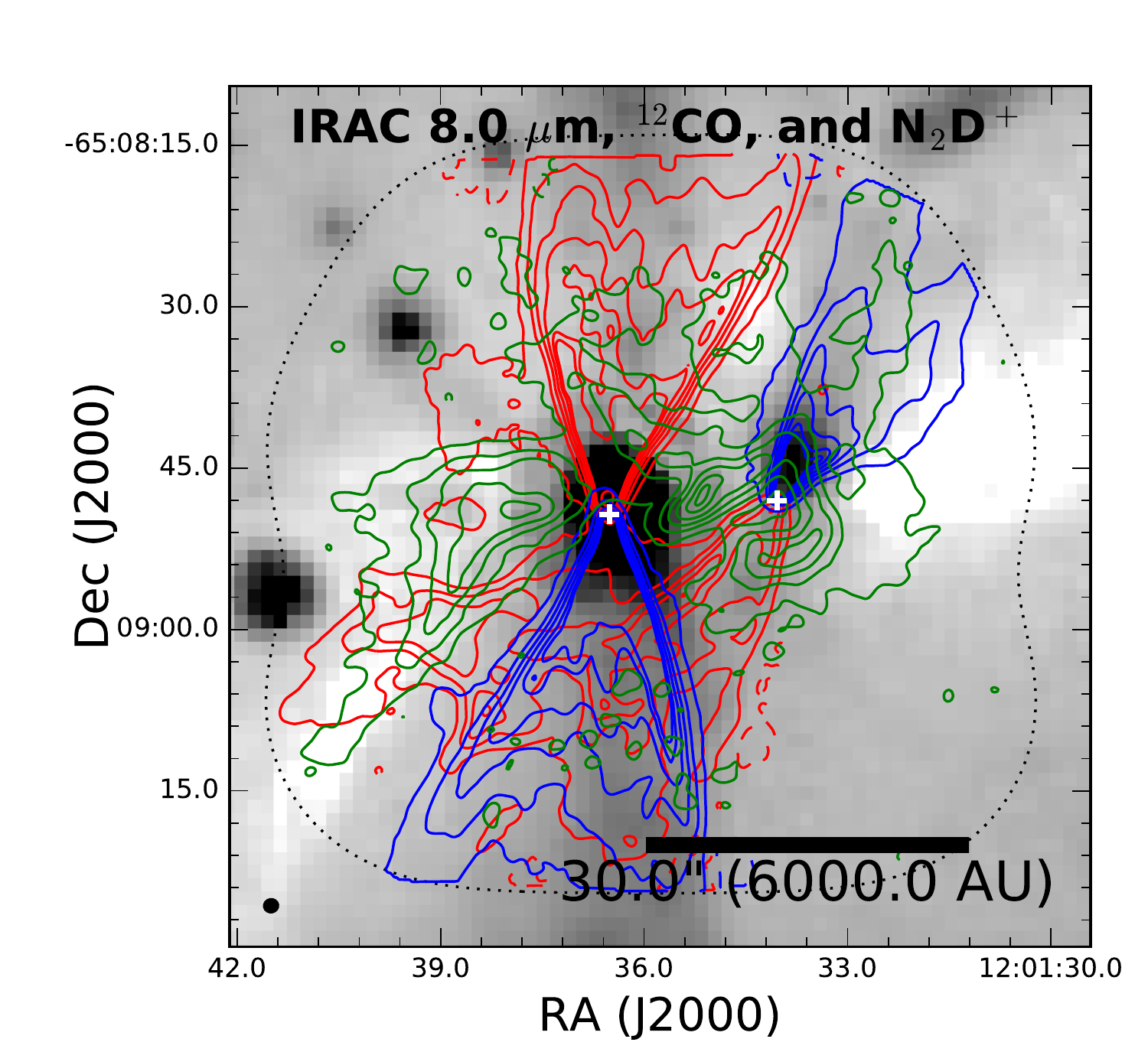}
\includegraphics[scale=0.5]{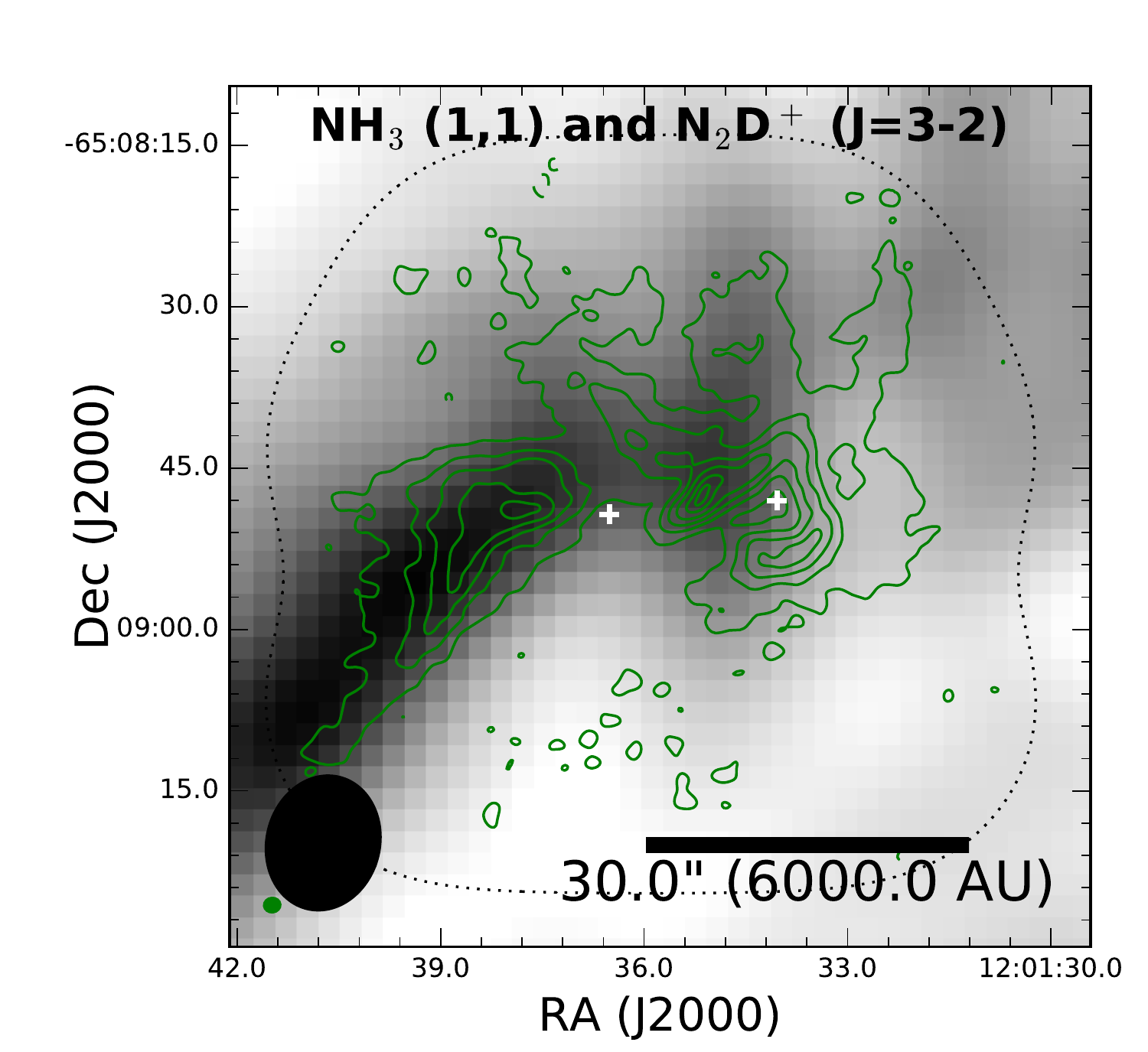}
\end{center}
\caption{The integrated intensity contours from the ALMA \ntdp\ ($J=3\rightarrow2$) observations
are overlaid (green) on the \textit{Spitzer} 8.0~\micron\ image (inverse grayscale) 
in the left panel and the \ntdp\ is overlaid (green) on the ATCA \nht\ (1,1) 
integrated intensity map (inverse grayscale) in the right panel. We also overlaid the \twco\ blue- and red-shifted 
integrated intensity contours (with corresponding colors) from Figure \ref{co-isotopes}.
The extent of C$^{18}$O shown in Figure \ref{co-isotopes}
is consistent with the size of the \ntdp\ deficit regions. There is some correspondence
on the east side of the \ntdp\ and the 8~\micron\ absorption of the envelope, 
but less than the ATCA \nht\ due to the smaller area mapped by ALMA. The ALMA \ntdp\ corresponds 
well with the \nht\ on the eastern side
to the edge of the map and the emission toward IRS1 is lower in both \nht\ and \ntdp. Near IRS2,
the correspondence of \ntdp\ and \nht\ is not as close, the peak of \nht\ east of IRS2 corresponds
with the \ntdp\ peak, but the \ntdp\ east and south of IRS2 has lower \nht\ intensity.
The \ntdp\ contours start at 5$\sigma$ and increase on 10$\sigma$ intervals, where $\sigma$=0.098~\kkms.
}
\label{atca-alma-8micron}
\end{figure}

\begin{figure}
\begin{center}
\includegraphics[scale=0.8]{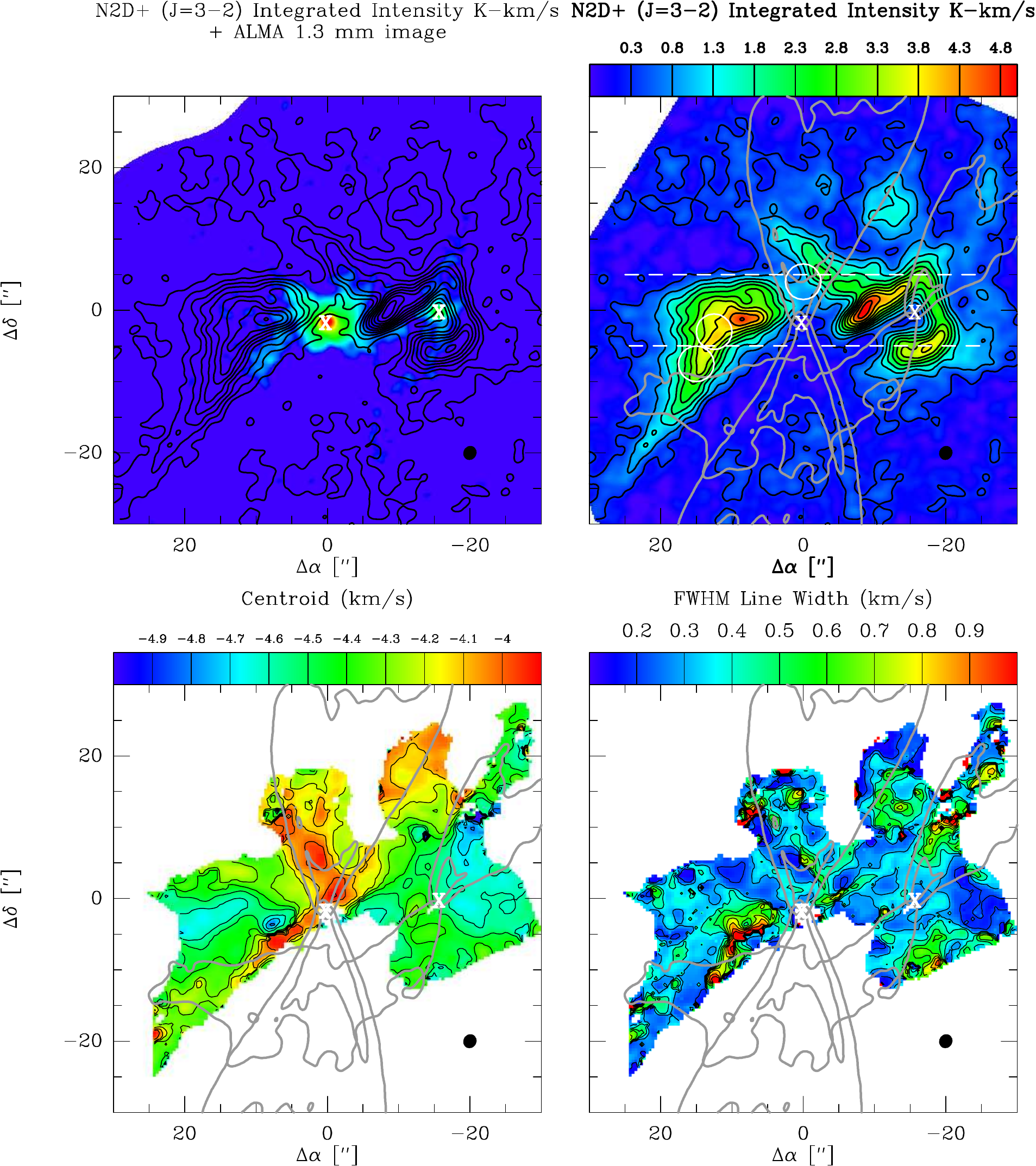}
\end{center}
\caption{Same as for Figure \ref{atca-nh3} but for the ALMA \ntdp\ ($J=3\rightarrow2$) 
transition and the data are overlaid on the ALMA 1.3~mm image instead of the \textit{Spitzer} 8~\micron. 
The spatial extents of the outflows probed by ALMA in \twco\ are marked with the gray contours in all panels
but the top left.
The integrated intensity maps show clear deficits of \ntdp\ toward the protostar locations,
as expected from protostellar heating. However, toward IRS2, the \ntdp\ is found in a 
horseshoe-shaped feature. The velocity structure is markedly different,
going to lower velocities on the eastern side of
the envelope in \ntdp\ vs. \nht. Some of this is likely due to the higher resolution of the
ALMA data and there is not as much blending of line components. Here the more red-shifted features
to the north 
appear to occupy more area than the outflow region of influence delineated by the \twco\ 
contours.
Ignoring the features on the north side of the 
envelope, the east and west side of the envelope are at
practically the same velocity. The linewidth map
is very noisy due to the low S/N in some parts of the data, but linewidths are all 
quite narrow, much more narrow than \nht\ in most places. The intensity levels
in the top panels are from the integrated intensity map of the main hyperfine lines between -5.5 to -3.7~\kms;
the intensity levels start at 3$\sigma$ and increase on 5$\sigma$ intervals, where $\sigma$=0.1~\kkms. The dashed lines in the upper right panel denote the
region of PV extraction, and the dashed circles in the upper 
right panel correspond to regions of spectral 
extraction to examine the quality of hyperfine fitting, see Appendix A.
}
\label{alma-n2dp}
\end{figure}

\begin{figure}
\begin{center}
\includegraphics[scale=0.5]{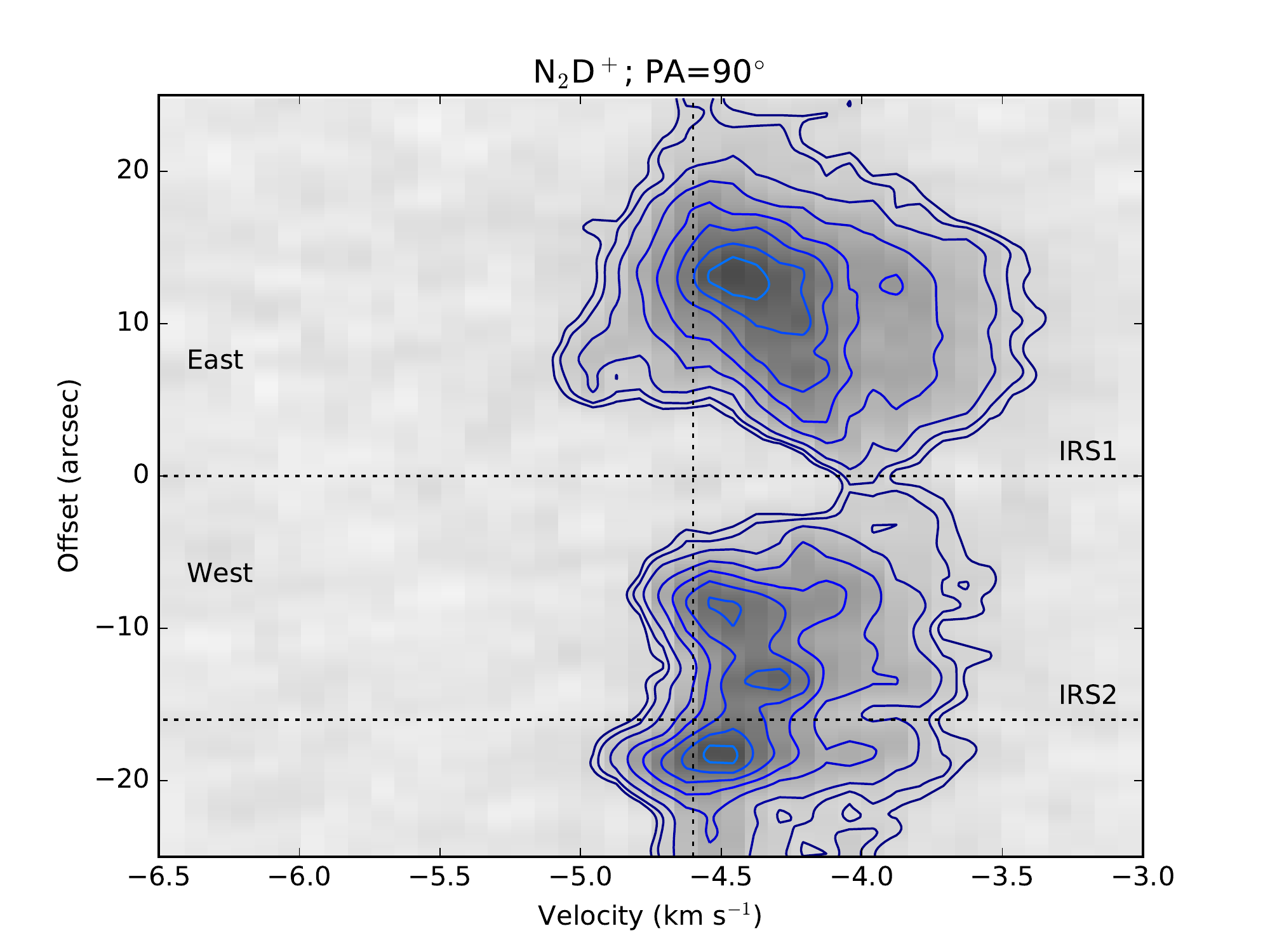}
\end{center}
\caption{Position-velocity diagram of \ntdp\ emission toward IRS1 and IRS2 taken
in a 10\arcsec\ cut across the equatorial plane of the envelope (PA = 90\degr), centered
on IRS1. The region extracted is bounded
by the dashed lines in the upper right panel of Figure \ref{alma-n2dp}.
The second velocity component shown in Figure \ref{atca-nh3-pv}
is difficult to pick out because the hyperfine structure of \ntdp\ ($J=3\rightarrow2$) is 
comprised of the closely spaced lines with comparable relative intensities. But, the overall
emission is more broad on the east side of the envelope than on the west due to the additional 
velocity component. The red-shifted velocities related to the outflow are clear
at the position of IRS1 where the only detectable emission is red-shifted. The overall line profiles
are more shifted toward the red because the rest frequency of our observation was centered
on the brightest hyperfine line of \ntdp\ ($J=3\rightarrow2$), which located at a slightly
higher frequency that most of the other lines having comparable relative intensity.
}
\label{alma-n2dp-pv}
\end{figure}

\begin{figure}
\begin{center}
\includegraphics[scale=0.8]{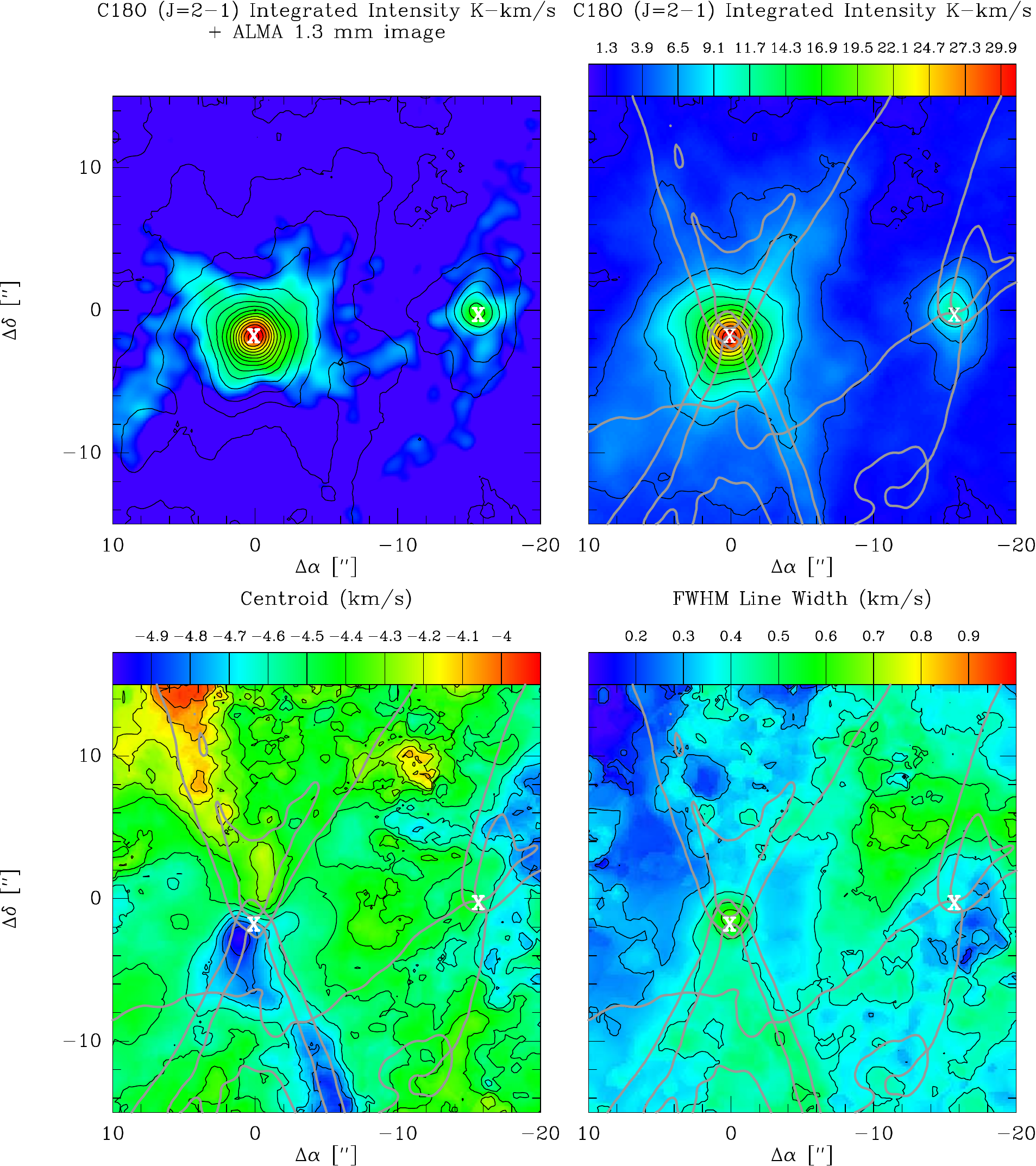}
\end{center}
\caption{Same as for Figure \ref{alma-n2dp} but for the ALMA \cateo\ ($J=2\rightarrow1$) and
zoomed-in on a smaller area centered on the two protostars. The line center velocity map in 
the lower left shows the clear influence of the outflow on the \cateo\ kinematics, but there
is a twist near the location of the protostar that is inconsistent with the outflow. The 
linewidth map for \cateo\ shows enhanced linewidth toward IRS1, indicative of more rapid
motion along the line of sight; IRS2 does not show an increased line width.
The intensity levels shown are the same as in Figure \ref{co-isotopes}, using the 12m+ACA+TP data.
}
\label{alma-c18o}
\end{figure}

\begin{figure}
\begin{center}
\includegraphics[scale=0.4]{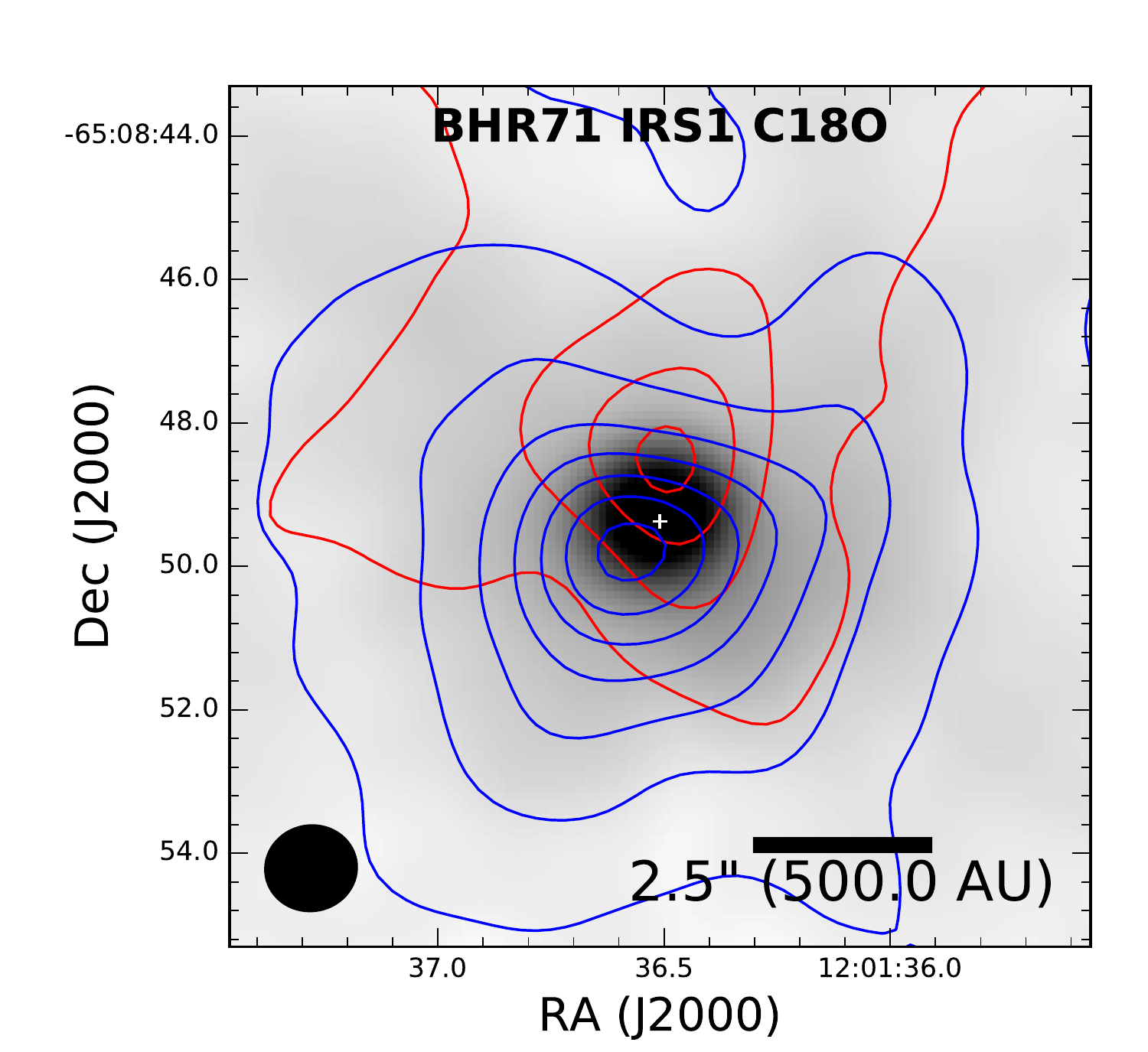}
\includegraphics[scale=0.4]{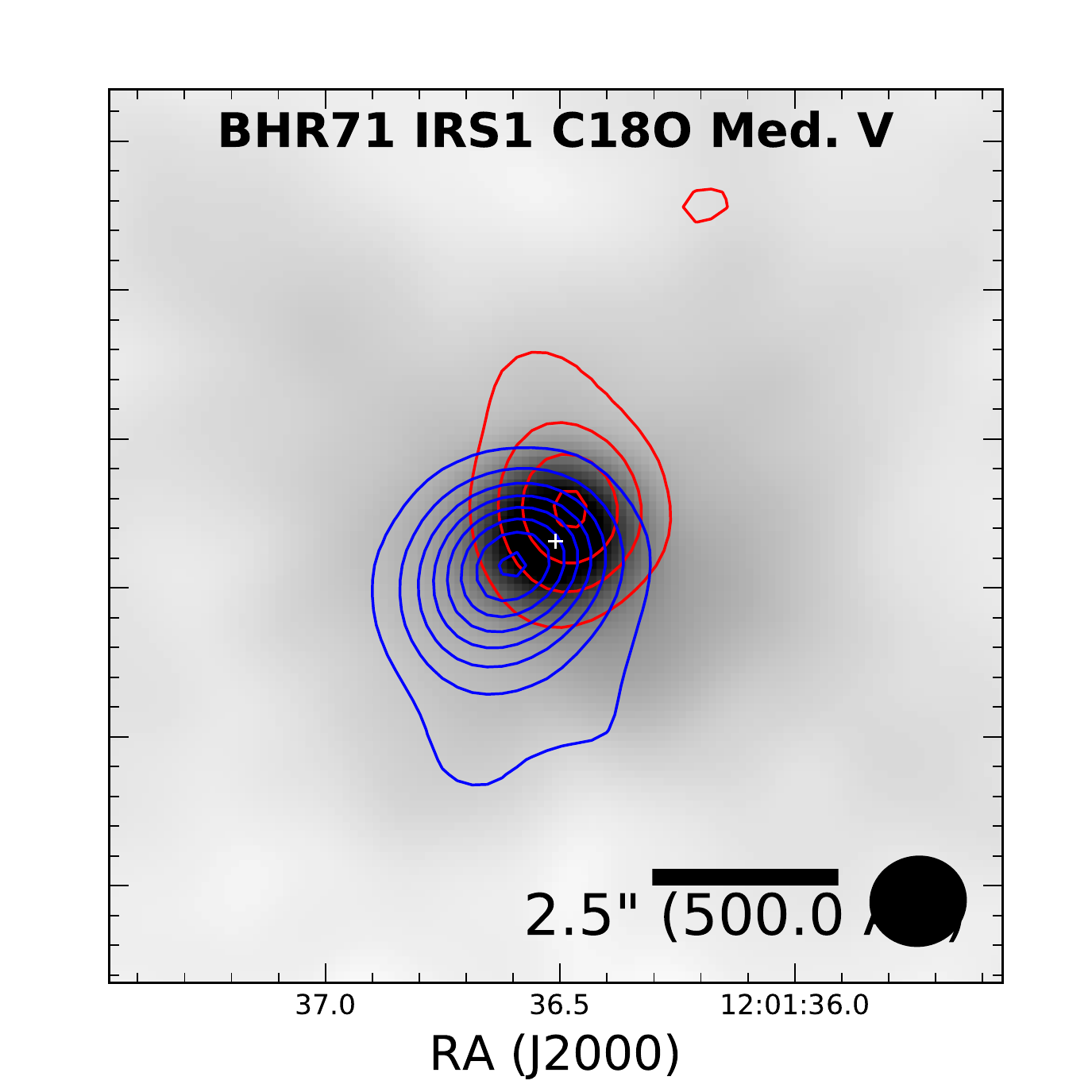}
\includegraphics[scale=0.4]{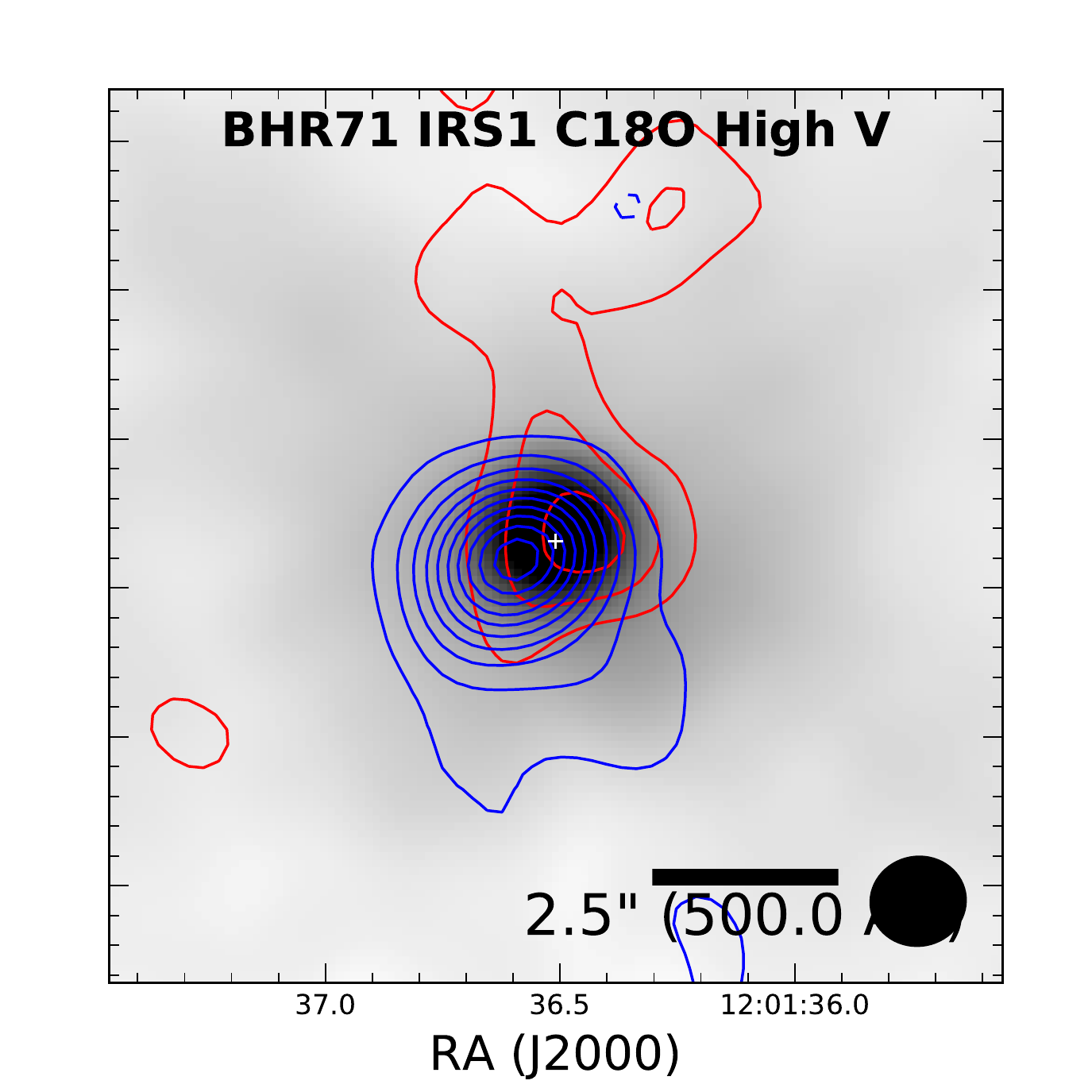}
\end{center}
\caption{C$^{18}$O red and blue-shifted integrated intensity maps 
toward IRS1 (using 12m+ACA+TP data) shown in different velocity ranges and overlaid on the 1.3~mm continuum
image. The low-velocity
C$^{18}$O clearly is influenced by the outflow, but at moderate and higher velocities,
the C$^{18}$O begins to trace a velocity gradient orthogonal to the outflow, the 
expected velocity gradient direction for rotation.
The low-velocity emission is integrated between -5.8 to -4.4~\kms\ and -4.4 to -3.2~\kms,
and $\sigma_{red,blue}$ = 0.09, 0.1~\kkms\ with the contours are drawn starting at 30$\sigma$ and increasing on intervals of 30$\sigma$.
The moderate-velocity emission is integrated between -6.2 to -5.6~\kms\ and -3.6 to -2.8~\kms,
and $\sigma_{red,blue}$ = 0.08, 0.07~\kkms\ with the contours are drawn starting at 10$\sigma$ and increasing on intervals of 10$\sigma$.
The high-velocity emission is integrated between -6.3 to -5.8~\kms\ and -3.0 to -2.5~\kms,
and $\sigma_{red,blue}$ = 0.06, 0.06~\kkms\ with the contours are drawn starting at 3$\sigma$
(5$\sigma$) and increasing on intervals of 3$\sigma$ (5$\sigma$); the values in parentheses
correspond to the blue contours. The 12m+ACA+TP data were used to generate this Figure.
}
\label{IRS1-c18o}
\end{figure}

\begin{figure}
\begin{center}
\includegraphics[scale=0.5]{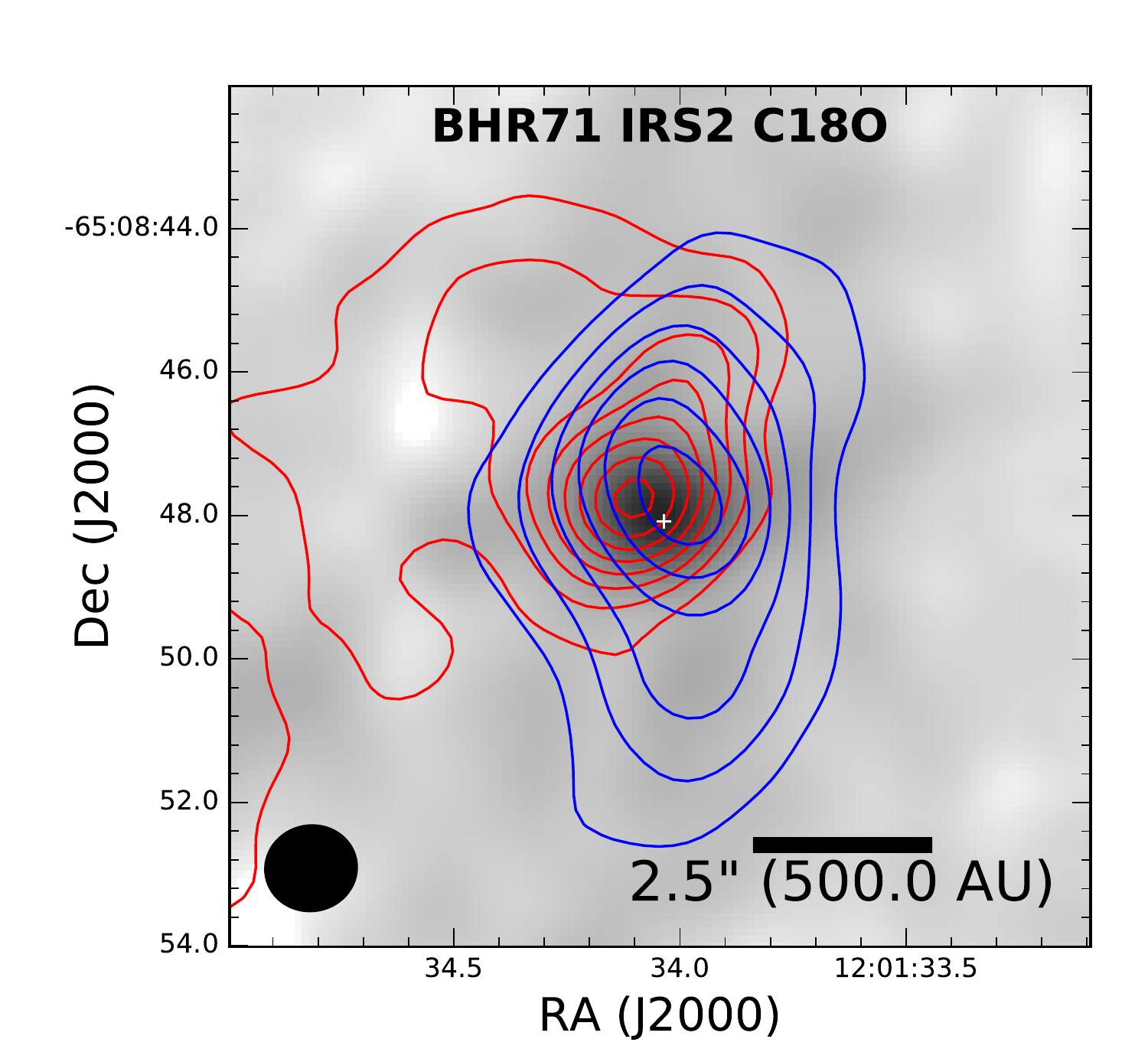}
\end{center}
\caption{C$^{18}$O red and blue-shifted integrated intensity maps 
toward IRS2 (using 12m+ACA+TP data). A velocity gradient is evident, in a direction that is
near orthogonal to the outflow. However, the velocity gradient direction is in
\textit{opposite} sense with respect to the gradient direction in IRS1.
The \cateo\ emission is integrated between -5.1 to -4.4~\kms\ and -4.4 to -3.5~\kms,
and $\sigma_{red,blue}$ = 0.08, 0.074~\kkms. The red and (blue) contours are drawn starting at 15$\sigma$ (35.0$\sigma$) 
and increasing on intervals of 5$\sigma$ (10$\sigma$). 
}
\label{IRS2-c18o}
\end{figure}

\begin{figure}
\begin{center}
\includegraphics[scale=0.5]{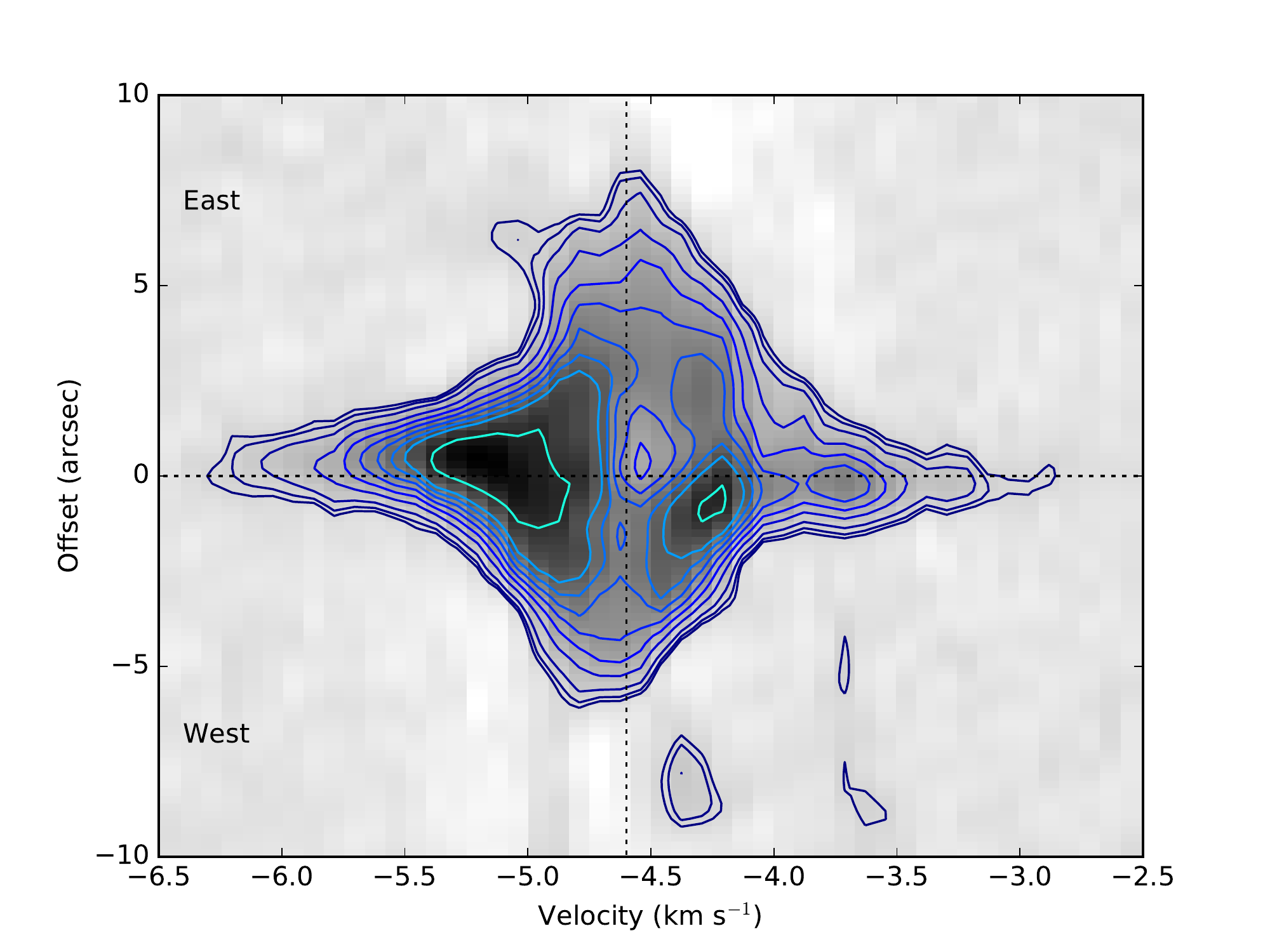}
\includegraphics[scale=0.5]{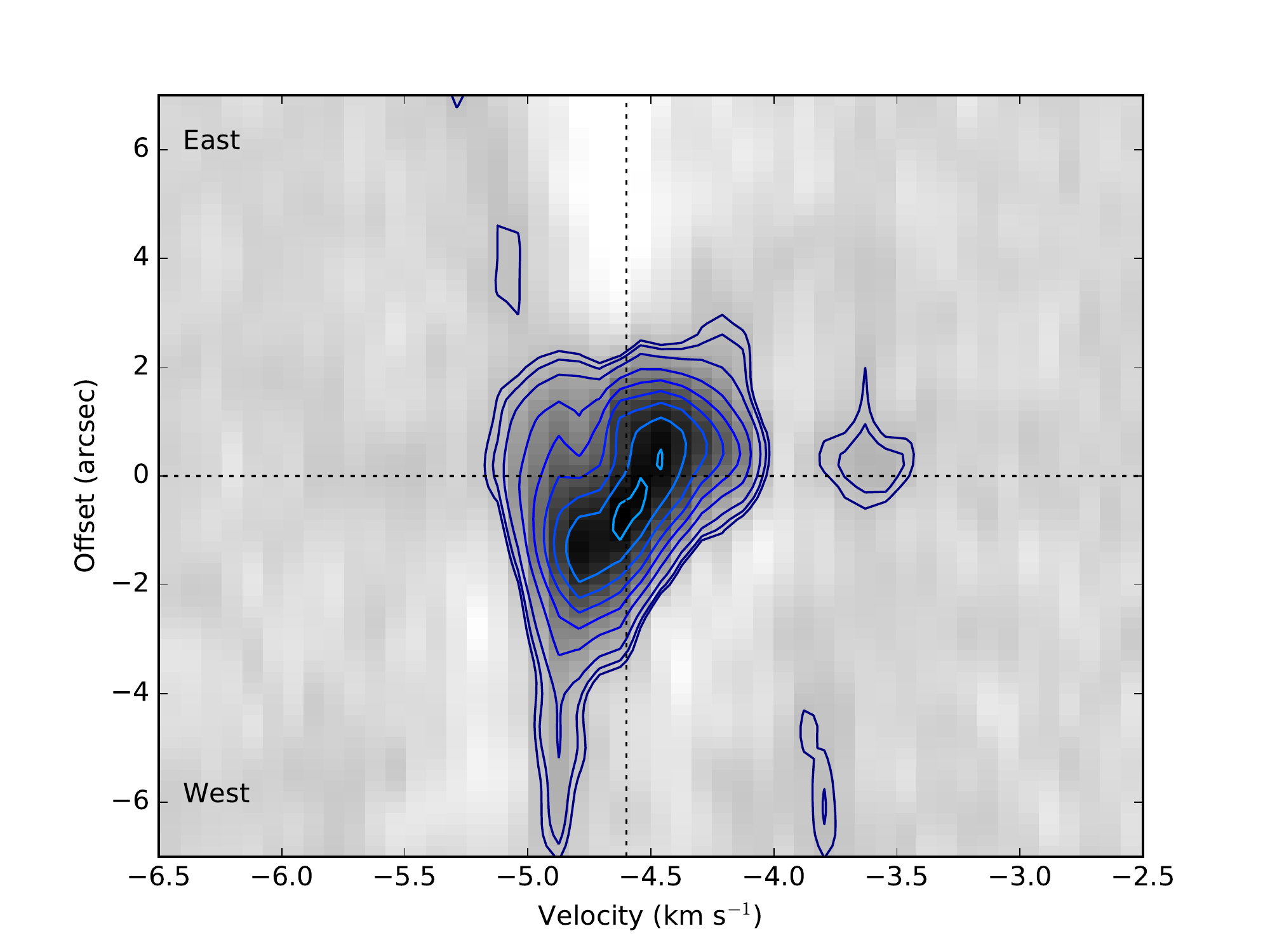}
\end{center}
\caption{Position-velocity diagrams of C$^{18}$O emission toward IRS1 (top) and IRS2 (bottom)
taken in cuts orthogonal to the outflow directions, 6\farcs2 in width.
The velocity gradient in IRS1 
appears to go from the upper left quadrant to the lower right quadrant at the highest 
velocities, and the velocity gradient in IRS2 appears to go from the lower left quadrant 
to the upper right quadrant.
The two PV diagrams are generated using only the data from the ALMA 12m array, excluding the 
ACA and Total Power. This limits confusion from the inclusion of the extended \cateo\ emission.
}
\label{PV-c18o}
\end{figure}

\begin{figure}
\begin{center}
\includegraphics[scale=0.5]{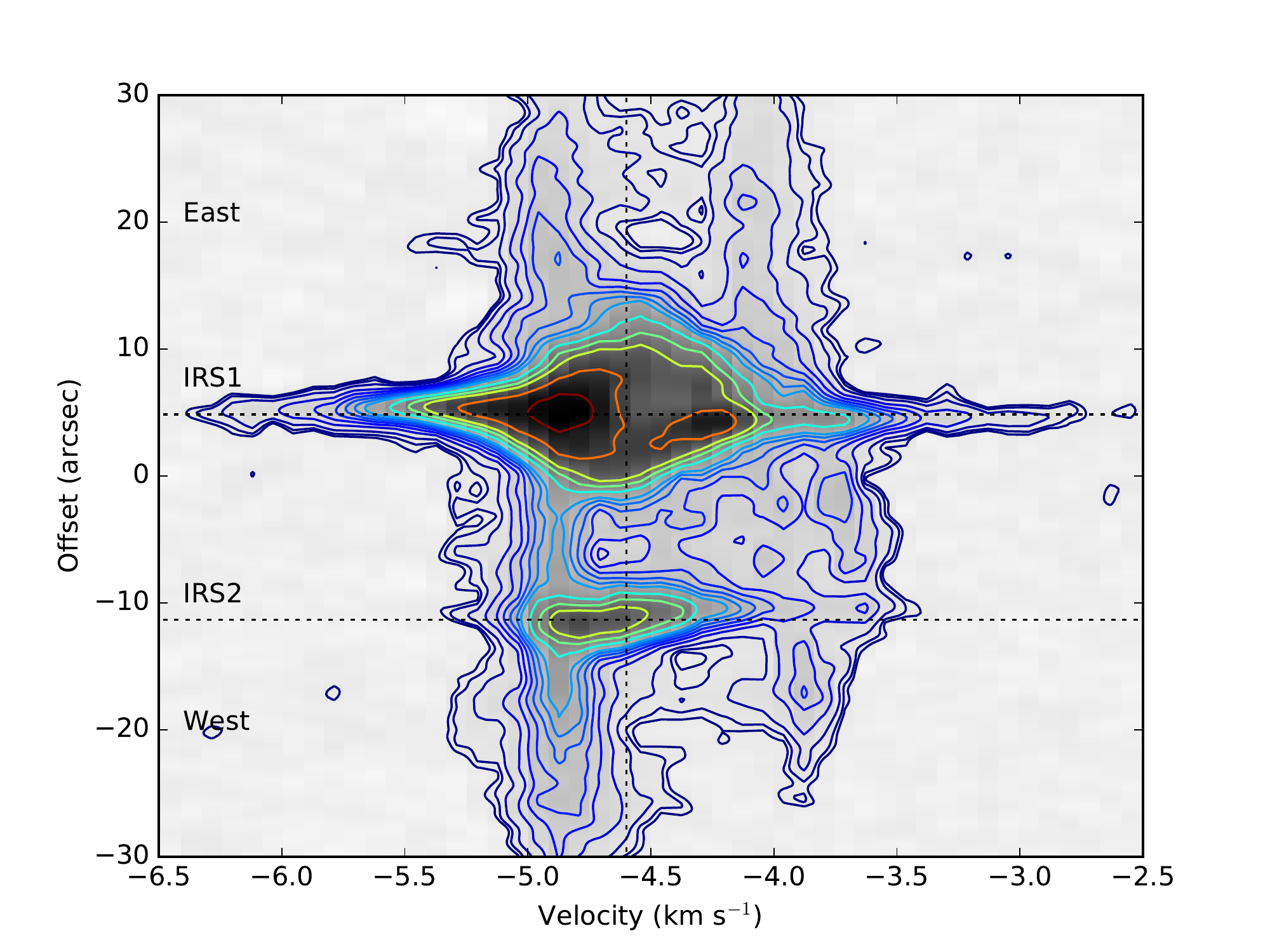}
\end{center}
\caption{Position-velocity diagrams of C$^{18}$O emission encompassing both IRS1 and IRS2 
taken in the same cut as the \nht\ and \ntdp\ PV diagrams shown in Figures \ref{atca-nh3-pv} and \ref{alma-n2dp-pv}.
This PV diagram includes the 12m, ACA, and Total Power data to recover the extended structure from the core.
Similar to the \nht\ and \ntdp, the \cateo\ also shows two velocity components, one at about -4.9~\kms\ and the
other at about -4.0~\kms. The \cateo\ emission concetrated around
the protostars themselves is a bit between the two velocity components, but more closely
associated with the blue-shifted velocity component, having
their velocity centers at about -4.6~\kms.
}
\label{PV-c18o-ext}
\end{figure}

\begin{figure}
\begin{center}
\includegraphics[scale=0.5]{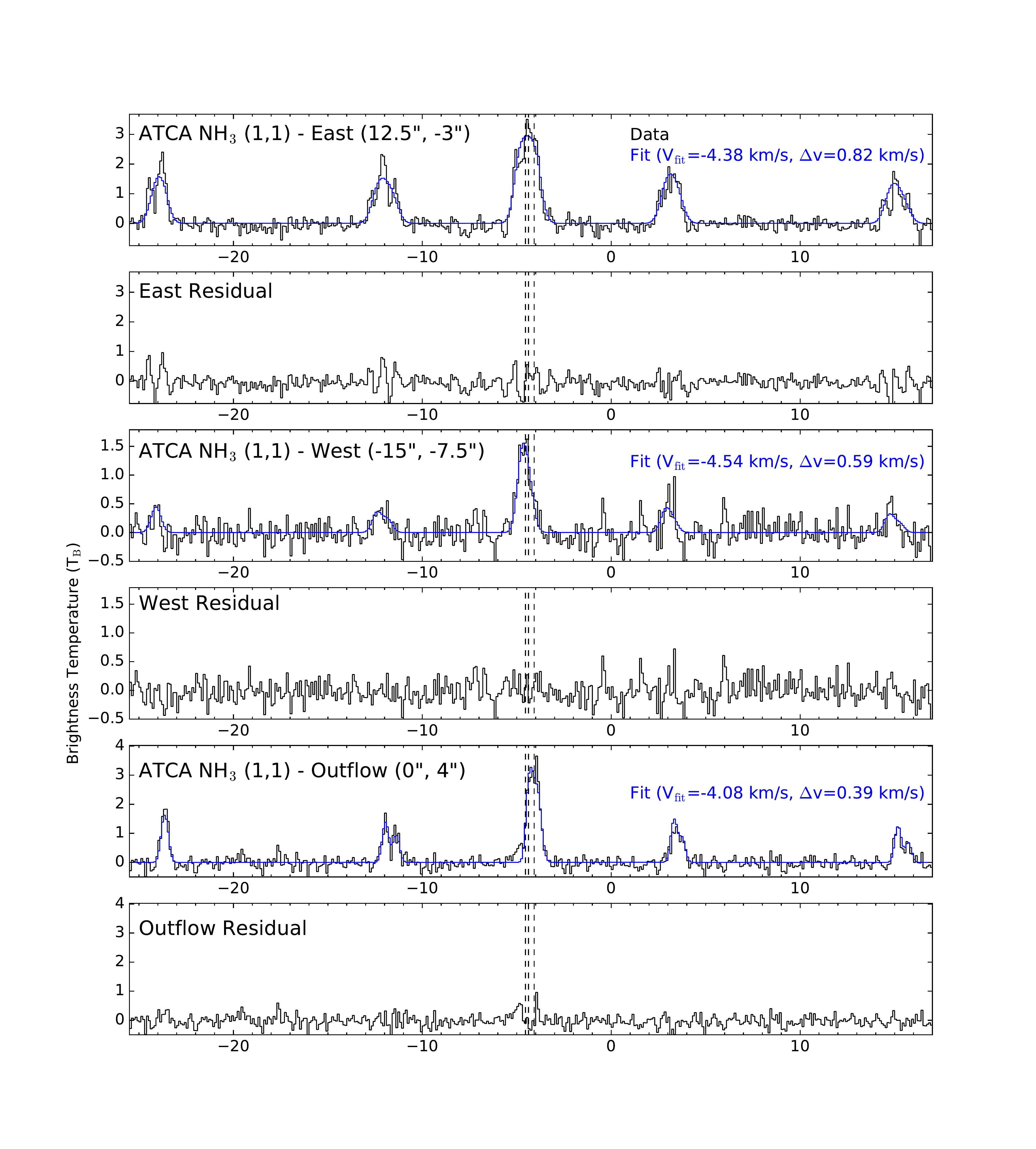}
\end{center}
\caption{ATCA \nht\ spectra taken toward three positions, showing the spectrum
and hyperfine fit in one panel and the residual of the fit below. 
The coordinates denoted for East, West, and
Outflow are relative to the position of IRS1 in Figure \ref{atca-nh3},
and the extraction regions are also drawn as circles in the upper
right panel of Figure \ref{atca-nh3-zoom}. The
East spectrum has the largest residual which is likely due to the two velocity
components that are apparent in Figure \ref{atca-nh3-pv}. 
The two blended
components in the East spectrum results in the fit having a broader linewidth.
The West and Outflow
positions are dominated by a single velocity component; however, the Outflow
position has a small blue-shifted residual.
}
\label{nh3-hyperfine}
\end{figure}

\begin{figure}
\begin{center}
\includegraphics[scale=0.5]{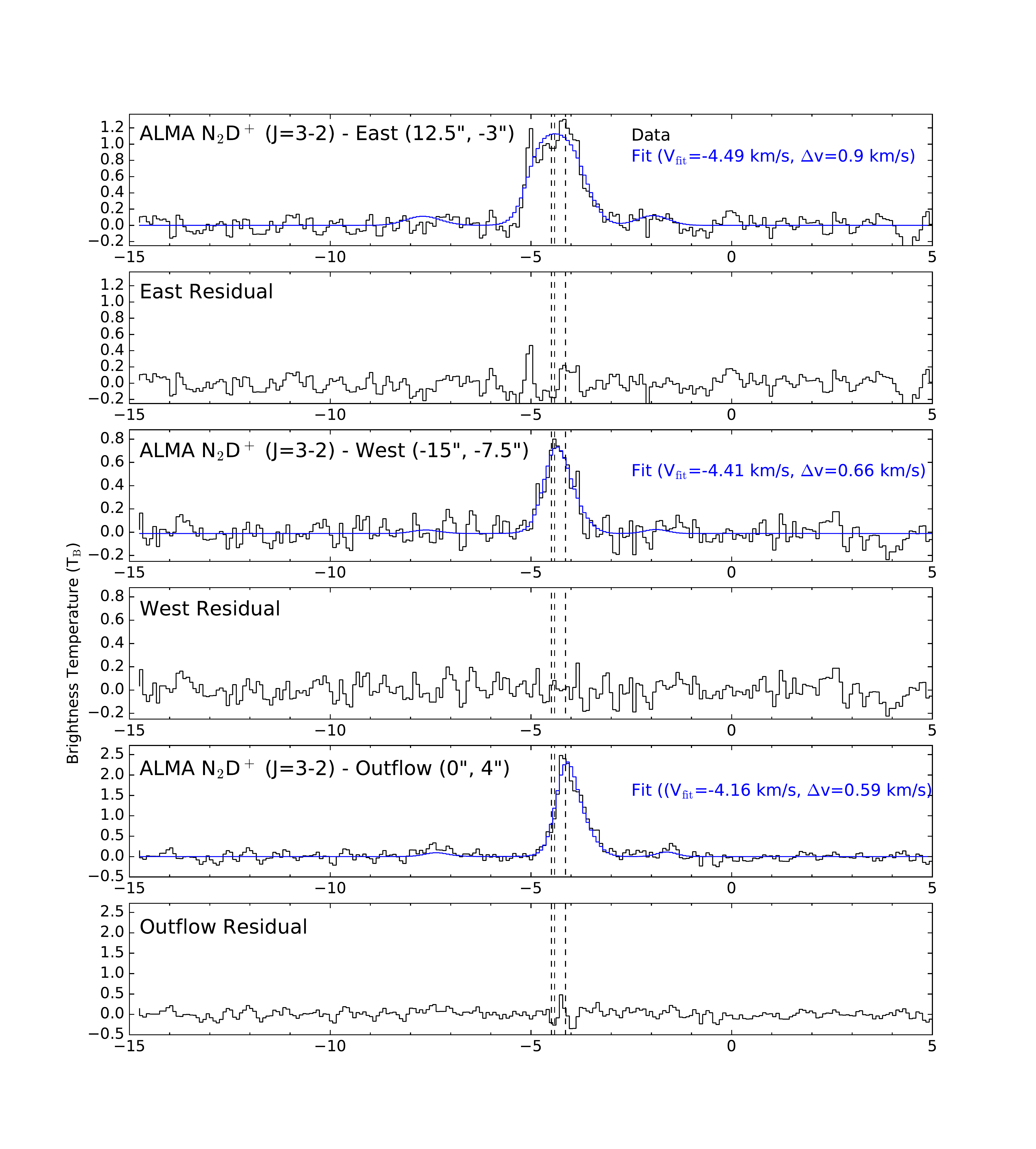}
\end{center}
\caption{ALMA \ntdp\ spectra taken toward three positions, showing the spectrum
and hyperfine fit in one panel and the residual of the fit below.
The coordinates denoted for East, West, and
Outflow are relative to the position of IRS1 in Figure \ref{alma-n2dp},
and the extraction regions are also drawn as circles in the upper right
panel of that Figure. The
East spectrum has the largest residual which is likely due to the two velocity
components that are apparent in Figure \ref{alma-n2dp-pv}. The two blended
components in the East spectrum results in the fit having a broader linewidth.
The West position
is dominated by a single velocity component, and the Outflow
position has a small residual.
}
\label{n2dp-hyperfine}
\end{figure}

\begin{figure}
\begin{center}
\includegraphics[scale=0.5]{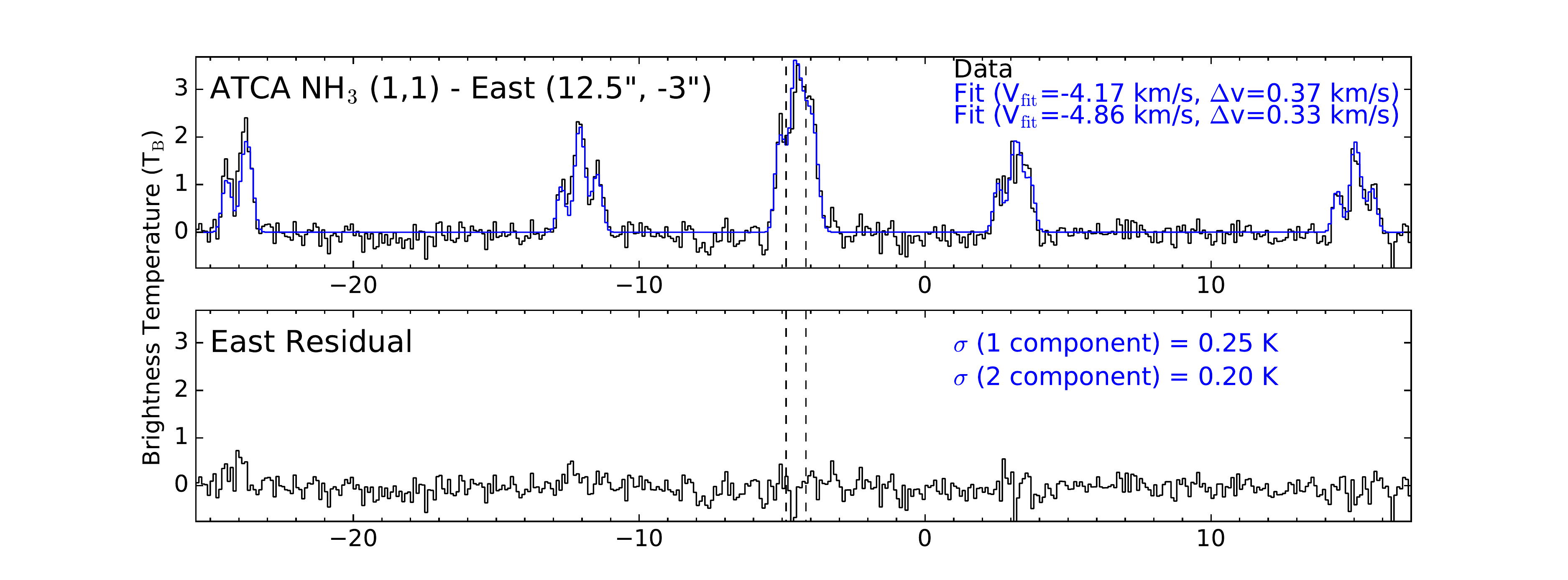}
\includegraphics[scale=0.5]{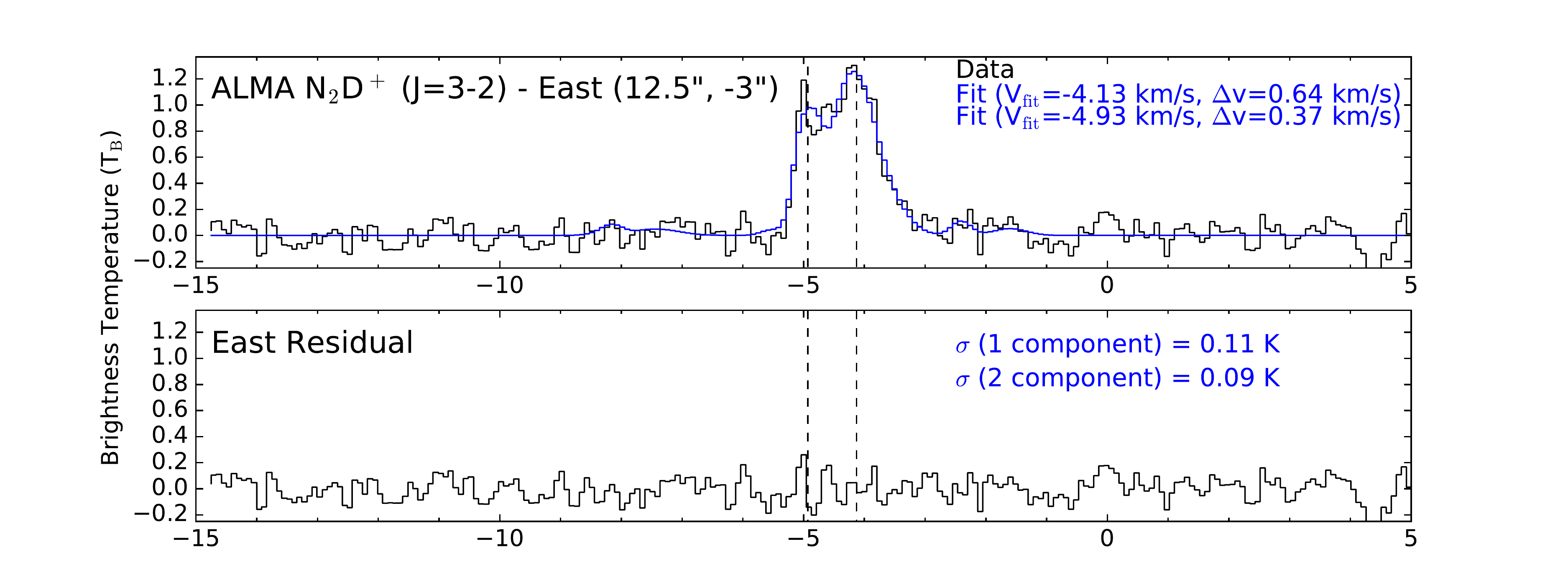}
\end{center}
\caption{ALMA \nht\ and \ntdp\ spectra taken at the same East envelope
position as in Figures \ref{nh3-hyperfine} and \ref{n2dp-hyperfine}. However,
instead of a single component fit, we attempt a two-component fit at this position
to disentangle and characterize the two velocity components. The two-component fit
is visually and statistically better than a single component fit. The vertical dashed
lines in both the upper and lower panels correspond to the central velocities of each
component fit.
}
\label{nh3-n2dp-hyperfine}
\end{figure}

\clearpage

\begin{deluxetable}{llllllll}
\tabletypesize{\scriptsize}
\tablewidth{0pt}
\tablecaption{Molecular Line Setups}
\tablehead{\colhead{Telescope} & \colhead{Molecule} & \colhead{Frequency} & \colhead{Raw Spectral Resolution} & \colhead{Map Spectral resolution} & \colhead{Map RMS} & \colhead{Beam} & \colhead{Detected?}\\
                                &                    & \colhead{(GHz)}     & \colhead{(kHz, km~s$^{-1}$)}      & \colhead{(km~s$^{-1}$)}           & \colhead{(Jy~beam$^{-1}$ or K)} & \colhead{(\arcsec)} &
}
\startdata
Parkes & NH$_3$ (1,1) & 23.694495 & 1.0, 0.012 & 0.05 & 0.1~K & 60 & Y\\
ATCA  & NH$_3$ (1,1) & 23.694495 & 0.5, 0.006 & 0.1 & 0.015~Jy~beam$^{-1}$, 0.19~K & 9.1$\times$7.8 & Y\\
      & NH$_3$ (2,2) & 23.722633 & 0.5, 0.006 & 0.2 & 0.012~Jy~beam$^{-1}$ & 10.8$\times$9.2 & M\\
      & NH$_3$ (3,3) & 23.870129 & 0.5, 0.006 & 0.2 & 0.011~Jy~beam$^{-1}$ & 10.8$\times$9.2 & N\\
      & Continuum    & 23.160    & 1.0, 13.0  & \nodata & 0.000048~Jy~beam$^{-1}$  & 9.4$\times$8.1& Y\\
ALMA  & C$^{18}$O ($J=2\rightarrow1$) &219.560354 & 61.035, 0.083 & 0.083 & 0.029~Jy~beam$^{-1}$& 1.7$\times$1.5 & Y\\
      & $^{13}$CO ($J=2\rightarrow1$) &220.398684 & 61.035, 0.083  & 0.083 & 0.035~Jy~beam$^{-1}$ & 1.7$\times$1.4 & Y\\
      & $^{12}$CO ($J=2\rightarrow1$) & 230.538   & 61.035, 0.079 & 0.166 & 0.029~Jy~beam$^{-1}$ & 1.5$\times$1.3 & Y\\
      & N$_2$D$^+$ ($J=3\rightarrow2$) & 231.321828 & 61.035, 0.079 & 0.083 & 0.034~Jy~beam$^{-1}$, 0.23~K & 2.0$\times$1.8 & Y\\
      & Continuum                     & 232.5    & 14.6, 18.8 & \nodata & 0.00087~Jy~beam$^{-1}$ & 1.3$\times$1.2& Y\\
\enddata
\tablecomments{
The final column, with the label `Detected?' denotes whether the molecule or continuum was detected with high-confidence (Y),
a marginal detection (M), or a non-detection (N).
}

\end{deluxetable}

\begin{deluxetable}{llllllll}
\tabletypesize{\scriptsize}
\tablewidth{0pt}
\tablecaption{Photometry}
\tablehead{\colhead{Wavelength} & \colhead{IRS1 Flux Density} & \colhead{IRS2 Flux Density} & \colhead{Total Flux Density} & \colhead{Instrument} & \colhead{Resolution} & \colhead{Aperture Radius\tablenotemark{a}} & \colhead{Reference} \\
          \colhead{(\micron)}  & \colhead{(Jy)}              & \colhead{(Jy)}              & \colhead{(Jy)}               &                      & \colhead{(\arcsec)}  & \colhead{(\arcsec)} & 
}

\startdata
1.25 & 0.0048$\pm$0.0005  & $<$1.4e-4  & 0.017$\pm$0.0017 & CTIO/ISPI & 0.9 & 30, 5, 50 & 1\\
1.66 & 0.025$\pm$0.006  & $<$2.6e-4  & 0.072$\pm$0.007 & CTIO/ISPI & 0.9 & 30, 5, 50 & 1\\
2.15 & 0.14$\pm$0.014  & $<$2.5e-3     & 0.33$\pm$0.03 & CTIO/ISPI & 0.9 & 30, 5, 50 & 1\\
3.6  & 0.073$\pm$0.07  & 0.0045$\pm$0.0004  & 0.15$\pm$0.02& IRAC  & 1.7 & 30, 5, 50  & 1\\
4.5  & 0.15$\pm$0.02  & 0.012$\pm$0.007 & 0.27$\pm$0.015 & IRAC  & 1.7 & 30, 5, 50  & 1\\
5.8  & 0.17$\pm$0.02  & 0.015$\pm$0.007  & 0.27$\pm$0.004& IRAC  & 1.8 & 30, 5, 50   & 1\\
8.0  & 0.14$\pm$0.03  & 0.009$\pm$0.0006  & 0.17$\pm$0.08 & IRAC  & 1.9 & 30, 5, 50  &1\\
24.0  &  5.0$\pm$0.3 & 0.09$\pm$0.003 & 5.1$\pm$0.3 & MIPS & 6 & PSF, PSF, 50 & 2\\
60.0  & \nodata & \nodata &    77.4$\pm$15.5 & IRAS & 72 & 300 &  3\\
70.0  & 127.2$\pm$12.7 & 8.8$\pm$1.0  & 145.7$\pm$14.5 & PACS & 5.6 & PSF, PSF,50 & 1\\
100.0 & 230.9$\pm$23.0 & 22.8$\pm$4.1  & 292.7$\pm$30.0 & PACS & 6.8 & PSF, PSF,50 & 1\\
160.0 & 248.4$\pm$25.0 & 37.6$\pm$8.0  & 353.8$\pm$35.0 & PACS & 10.7 & PSF, PSF, 50 & 1\\
250.0 & 157.5$\pm$30.0 & 36.1$\pm$6.0  & 213.4$\pm$22.0 & SPIRE & 18 & PSF, PSF, 22 & 1\\
350.0 & \nodata        & \nodata       & 127.0$\pm$9.1 & SPIRE & 24 & 30 & 1\\
500.0 & \nodata        & \nodata       & 64.8$\pm$4.5 & SPIRE  & 35 & 40 & 1\\
1300.0 & \nodata        & \nodata      &  3.8$\pm$0.72 & SEST & 23 & 60 & 4\\
1300.0 & 1.28$\pm$0.01  &  0.12$\pm$0.005 & 1.41$\pm$0.016   & ALMA & 1.25 & 12.5, 5, 17.5 & 1\\
3200.0 &  0.140$\pm$0.028 & 0.0028$\pm$0.0021 &  0.143$\pm$0.028 & ATCA & 3.5 & Gaussian & 2\\
12960.0 &  0.0027$\pm$0.0002 & 0.00022$\pm$0.0001 & 0.0029$\pm$0.0002  &     ATCA & 9 & Gaussian & 1\\
\enddata
\tablecomments{
The flux density for IRS1 includes a minor contribution from IRS2 due to the 30\arcsec\ (6000~AU)
radius aperture coming close to encompassing the system. The total flux density refers to the flux measured 
in the third aperture listed in the Aperture column. 
References: (1) This work; (2) \citet{chen2008}; (3) IRAS Photometry, (4) \citep{launhardt2010}.
}
\tablenotetext{a}{
The comma separated values in this column represent the aperture used for IRS1, IRS2, and
the their combined flux densities, respectively. The aperture for the combined flux densities of IRS1 and IRS2 is chosen
to encompass the emission of both sources. The numerical sizes denote aperture photometry, PSF denotes photometry by 
point spread function fitting and Gaussian denotes photometry by Gaussian fitting.
}

\end{deluxetable}

\end{document}